\documentclass[rmp,aps]{revtex4-1}
\pdfoutput=1
\usepackage{latexsym}
\usepackage[pdftex]{graphicx}
\usepackage{amsmath}
\usepackage{natbib}

\begin{document}
\title{Landau-Stark states and cyclotron-Bloch oscillations of a quantum particle}
\author{Andrey R. Kolovsky}
\affiliation{Kirensky Institute of Physics, 660036 Krasnoyarsk Russia}
\affiliation{Siberian Federal University, 660041 Krasnoyarsk Russia}
%{\small \em andrey.r.kolovsky@gmail.com}

\author{Giorgio Mantica}
\affiliation{Center for Nonlinear and Complex Systems, University of Insubria 22100 Como Italy}
\affiliation{CNISM unit\`a di Como and  I.N.F.N. sezione di Milano \mbox{,} Italy}
%{\small \em giorgio@uninsubria.it}
\date{\today}

\begin{abstract}
Recent experimental progress in the creation of synthetic electric and magnetic fields, acting on cold atoms in a two-dimensional lattice, has attracted renewed interest to the problem of a quantum particle in the Hall configuration. The present work contains a detailed analysis of the eigenstates of this system, called Landau-Stark states, and of the associated dynamical phenomenon of cyclotron-Bloch oscillations. It is shown that Landau-Stark states and cyclotron-Bloch oscillations crucially depend on two factors. The first is the orientation of the electric field relative to the primary axes of the lattice. The second is ratio between the frequencies of Bloch and cyclotron oscillations, that is also the ratio between the magnitudes of electric and magnetic fields. The analysis is first carried out in the tight-binding approximation, where the magnetic field is characterized by the Peierls phase entering the hopping matrix elements. Agreement of this analysis with the full quantum theory is also studied.
\end{abstract}
%\pacs{05.60.Gg}{Quantum transport}
%\pacs{72.10.Bg}{General formulation of transport theory}
%\pacs{73.43-f}{Quantum Hall effects}
%\pacs{05.45.-a}{Nonlinear dynamics and chaos}
\maketitle

\tableofcontents
%%%%%%%%%%%%%%%%%%%%%%%%%%%%%%%%%%%%%%%%%%%%%%%
\section{Introduction}
%%%%%%%%%%%%%%%%%%%%%%%%%%%%%%%%%%%%%%%%%%%%%%%
%

It is not uncommon, in Physics, that certain problems suddenly arise to a revived interest, after having been studied, and at least partially solved, in a distant past and henceforth disregarded. A bright example of such a revival is given by Bloch oscillations of a quantum particle in a periodic potential. Predicted by Bloch and Zener around 1930 for electrons in a solid crystal subject to an electric field \cite{Bloc28,Zene32,Zene34}, this phenomenon has been experimentally detected only sixty years later in semiconductor superlattices \cite{Feld92}. However, only after its observation via cold atoms in optical lattices \cite{Daha96, Mors01} and via light in photonic  crystals \cite{Pert99,Mora99} new research in this direction blossomed.  Currently, the field of Bloch oscillations of cold atoms is very diverse, including phenomena like super Bloch oscillations \cite{Hall10} and new applications, like precision measurement of gravity \cite{gravity}, and more: in this review we elect to limit our citations to a few seminal and typical papers, out of an impressive total list.

During this revival, experimental achievements has gone hand by hand with progress in the theory of Wannier-Stark states, {\em i.e.} the eigenstates of a quantum particle in a tilted lattice, that are the spectral counterpart of Bloch oscillations. New analytical methods have been developed, which allow one to calculate Wannier-Stark states for arbitrary values of the system parameters \cite{53}. This review is particularly focused on spectral theory: in this respect Wannier-Stark states constitute the second item in the list of the four fundamental quantum states in the problem of a particle in a lattice---the first place being obviously occupied by Bloch waves.

The third element in this list are Landau states. Originally, the term was meant to indicate the eigenstates of a charged particle, in free space, subject to a magnetic field. However, it is nowadays also used to denote the eigenstates in the presence of an additional periodic potential. The combined action of the magnetic field and the periodic potential yields the extremely rich physics of Landau states, which is reflected in the notions of magnetic bands \cite{Azbe64,Zak64}, Hofstadter's butterfly \cite{Hofs76}, topological invariants \cite{Kohm85}, edge states \cite{Hats93}, {\em et cetera}. Recently, following a number of theoretical suggestions \cite{Jaks03,84}, experimentalists have been able to simulate magnetic fields acting on {\em neutral} atoms in optical  lattices \cite{Aide11,Aide13,Miya13,Stru13}. Precisely,  the physical systems so constructed are described by a tight-binding Hamiltonian, formally identical to that of a charged particle in a lattice, subject to a magnetic field. This achievement opens wide prospects for experimental studies of Landau states and  related phenomena \footnote{One must also mention similar studies for photonics crystals \cite{Rech13,Hafe13}.}.

The fourth kind of fundamental states is related to the case of a quantum particle in the Hall configuration. This is defined by a  two-dimensional lattice, an in-plane (real or synthetic) electric field, and a normal to the lattice plane (real or synthetic) magnetic field. It will be shown in the subsequent sections that the eigenstates of a quantum particle in the Hall configuration share some features with Wannier-Stark states. For this reason we call them {\em Landau-Stark} states and the associated dynamical problem {\em cyclotron-Bloch  oscillations}. Some aspects of Landau-Stark states and cyclotron-Bloch oscillations have been addressed earlier \cite{Naka95,Bare99,Kuno00,Naza01,Muno05}. Having been written in the period before revival, these papers did not attract the deserved attention. In addition, they were lacking a systematic approach. This is partly due to the fact that in the laboratory the Hall configuration was exclusively realized by solid state systems, where the electric field can be treated as a perturbation. This is definitely not the case for the new experimental systems based on cold atoms and photonic crystals, that require a complete approach, in which electric and magnetic fields (more generally, potential and gauge fields) are treated on equal footing from the very beginning. The aim of the present review, which is largely based on recent works \cite{85,87,90,91,93,94,96,98,99}, is to fill gaps in the theory of Landau-Stark states and cyclotron-Bloch oscillations, by systematically considering the system in all regions of its parameter space.

The structure of this work is as follows. In Sec.~\ref{secA} we introduce our main model -- the tight-binding Hamiltonian of a quantum particle in the Hall configuration. In this Hamiltonian the magnetic field is characterized by the Peierls phase $\alpha$, the electric field by its magnitude $F$ and orientation $\beta=F_x/F_y$. We then recall the main properties of this system: for $\alpha=0$, $F\ne0$, Wannier-Stark states are introduced; the case $\alpha\ne0$, $F=0$, yields Landau states; finally, when $\alpha\ne0$, $F\ne0$ we describe the important construction of Landau-Stark states. Global properties of these states are discussed in Sec.~\ref{secA4}. We show that, unlike the case of Landau states, where rationality of the Peierls phase $\alpha$ plays the crucial role, the structure of Landau-Stark states is principally determined by the parameter $\beta$. Namely, for rational $\beta$ Landau-Stark states are extended functions with continuous spectrum, while for irrational $\beta$ they are localized states with discrete spectrum.

Extended and localized Landau-Stark states and their spectra are further discussed in Sec.~\ref{secB} and Sec.~\ref{secC}, where we focus on two limiting values of the Peierls phase: $\alpha=1/2$, in Sec.~\ref{secB}, and $\alpha \ll 1$,  in Sec.~\ref{secC} \footnote{Going ahead we mention that within the tight-binding approximation one can restrict $\alpha$ to the interval $|\alpha|\le1/2$ without any loss of generality.}. In Sec.~\ref{secB} we also analyze the case of a {\em staggered} magnetic field, which provides a relatively easy experimental realization of the Hall configuration with cold atoms \cite{Aide11}. Important results of these sections are the estimates of the localization length of Landau-Stark states for irrational $\beta$ in different situations. In particular, for $\alpha\ll 1$, the localization length is shown to diverge exponentially, when the electric field decreases below a critical value.

The next two sections of the review are devoted to the analysis of two difficult problems of mathematical and physical relevance, that broaden the scope of the investigation and more closely address real experimental situations. In Sec.~\ref{secD} we consider finite systems, with specified boundary conditions, while in Sec.~\ref{secE} we study infinite lattices, beyond the single-band approximation. Finite systems can show features, like edge states, that do not appear in infinite systems. We analyze two types of boundary conditions: harmonic confinement, which is the default boundary condition for cold atoms in optical lattices, and   Dirichlet boundary conditions, which are realized in photonic and solid state crystals. The analysis of the system beyond the single-band approximation is aimed to answer the fundamental question whether Landau-Stark states are in fact metastable states, {\em i.e.}, quantum resonances, as it is known to be the case of Wannier-Stark states.

We finally end the review by listing open problems in the concluding Sec.~\ref{secEE}.

Before proceeding with the main part we find useful to list the abbreviations used in the text. These are:
% \begin{itemize}
$\bullet$ 1D: one-dimensional
$\bullet$ 2D: two-dimensional
$\bullet$ WS-states: Wannier-Stark states
$\bullet$ BOs: Bloch oscilllations
$\bullet$ LS-states: Landau-Stark states
$\bullet$ LZ-tunneling: Landau-Zener tunneling
$\bullet$ BC: boundary condition.
% \end{itemize}

%%%%%%%%%%%%%%%%%%%%%%%%%%%%%%%%%%%%%%%%%%%%%%
\section{Preliminaries}
\label{secA}
%%%%%%%%%%%%%%%%%%%%%%%%%%%%%%%%%%%%%%%%%%%%%%
\subsection{The model}
\label{secA1}

In this section we introduce the physical model under investigation. Although the theory is applicable to different physical systems, we shall use the standard terminology of solid state physics. Let us therefore consider a quantum particle of mass $M$ and charge $e$, moving in a 2D square lattice of side $a$ in the $(x,y)$ plane, created by a periodic potential $V(x,y)$. The particle is also under the action of an electric and a magnetic field. The former lies in the $(x,y)$ plane, while the latter is normal to it. This particle is described by the Hamiltonian
%***************************************************
\begin{equation}
\label{b1}
\widehat{H}=\frac{1}{2M}\left(\hat{\bf{p}}-\frac{e}{c}{\bf A}\right)^2+V({\bf r}) +e\; {\bf F} \cdot {\bf r} \;,\quad
V(x+la,y+ma)=V(x,y) \;,
\end{equation}
where ${\bf A}$ is the vector potential of the magnetic field $B$ and ${\bf F}$ denotes the electric field. Standard choices of the vector potential are ${\bf A}=B(0,x)$ or ${\bf A}=B(-y,0)$, known as the Landau gauge, and  ${\bf A}=B(-y/2,x/2)$, known as the symmetric gauge. A rigorous analysis of the system (\ref{b1}) is a difficult, unsolved problem. Therefore, one deals with this problem by introducing suitable approximations, among which the most common is the tight-binding approximation.

%%%%%%%%%%%%%%%%%%%%%%%%%%%%%%%%%%%%%%%%%%%%%%%
\subsubsection{The tight-binding Hamiltonian}
\label{secA1a}

The starting point of the tight-binding approximation is the Bloch spectrum of the quantum particle when $B$ and $F$ are null. This spectrum consists of the ground energy band, $E=E(\kappa_x,\kappa_y)$, separated by a finite gap $\Delta$ from higher energy bands, that may overlap. Using the Bloch states associated with the ground band, one constructs the localized Wannier functions $w_{l,m}({\bf r})$, where the indexes $l$ and $m$ label the local wells of the periodic potential.  Since Wannier functions are related to each other by elementary translations, $w_{l,m}(x,y)=w_{0,0}(x-la,y-ma)$, the single function $w_{0,0}({\bf r})$ is sufficient to construct all of them. On these premises the single-band approximation amounts to the following ansatz for the wave function,
%***************************************************
\begin{equation}
\label{b2}
\Psi({\bf r})=\sum_{l,m} \psi_{l,m} w_{l,m}({\bf r}) \;.
\end{equation}
We thus truncate the Hilbert space of the system to the subspace spanned by the Wannier functions $w_{l,m}({\bf r})$. We shall discuss the validity of this approximation in some detail later on, in Sec.~\ref{secE}.

Once Wannier functions have been determined, we can calculate the matrix elements of the Hamiltonian (\ref{b1}) in their basis: $H_{l,m}^{l',m'}=\int  w_{l',m'}({\bf r})\widehat{H}w_{l,m}({\bf r}) {\rm d}{\bf r}$. If the potential $V({\bf r})$ is deep enough (and, hence, Wannier functions are well localized within one well) we may neglect matrix elements with $|l'-l|>1$ and $|m-m'|>1$. After some elementary algebra we obtain:
%***********************************************
\begin{equation}
\label{b3}
\left(\widehat{H} \psi\right)_{l,m}= -\frac{J_x}{2}\left(\psi_{l+1,m}+\psi_{l-1,m}\right)
-\frac{J_y}{2}\left(e^{i2\pi\alpha l} \psi_{l,m+1}+e^{-i2\pi\alpha l} \psi_{l,m-1}\right)
+ea(F_x l + F_y m)\psi_{l,m} \;,
\end{equation}
where we used the Landau gauge ${\bf A}=B(0,x)$ and we omitted a constant term in the energy of Wannier states. The parameter $\alpha$ in this Hamiltonian is the famous Peierls phase,
%***************************************************
\begin{equation}
\label{b4}
\alpha=\frac{eBa^2}{hc} \;,
\end{equation}
which physically corresponds to the magnetic flux through the elementary cell, in units of the flux quantum $hc/e$. The tight-binding Hamiltonian (\ref{b3}) will be our model in the main part of this review.
% In the other words, we  shall consider it irrespective to the original Hamiltonian (\ref{b1}).
Then, without any loss of generality, we can restrict the value of the parameter $\alpha$ to the interval $-1/2<\alpha\le 1/2$.

%%%%%%%%%%%%%%%%%%%%%%%%%%%%%%%%%%%%%%%%%%%%%%
\subsubsection{Characteristic frequencies}
\label{secA1b}

Two characteristic frequencies determine the dynamics of the tight-binding Hamiltonian (\ref{b3}). They can be determined by alternatively letting $B$ and $F$ to be null.

Firstly, consider the case $F=0$. Within the tight-binding approximation, the ground Bloch band is approximated by the cosine dispersion relation
%*******************************************
\begin{equation}
\label{b5}
E(\kappa_x,\kappa_y)=-J_x\cos(a\kappa_x)-J_y\cos(a\kappa_y) \;.
\end{equation}
Thus, the hopping matrix elements $J_x$ and $J_y$ in Eq.~(\ref{b3}) are related to the effective mass of the particle, $M^*_{x,y}=\hbar^{-2}{\rm d}^2 E/{\rm d}\kappa_{x,y}^2$, via $J_{x,y}=\hbar^2/a^2 M^*_{x,y}$. Using the notion of effective mass we can introduce the cyclotron frequency  $\omega_c=eB/c\sqrt{M_x^*M_y^*}$. Next, expressing the magnetic field $B$ through the Peierls phase $\alpha$ in Eq. (\ref{b4}) and the effective masses $M^*_{x,y}$ through the hopping matrix elements $J_x$ and $J_y$, we can also write
%***************************************************
\begin{equation}
\label{b6}
\omega_c=2\pi\alpha(J_xJ_y)^{1/2}/\hbar  \;.
\end{equation}
%
%It should be noted, however, that Eq.~(\ref{b6}) is a meaningful quantity only for $|\alpha|\ll 1$.

Secondly, let $B=0$, to define the other characteristic frequency of the system, the Bloch frequency $\omega_B$:
%***************************************************
\begin{equation}
\label{b7}
\omega_B=\sqrt{\omega_x^2 +\omega_y^2}  \;,\quad \omega_x=eaF_x/\hbar \;,\quad \omega_y=eaF_y/\hbar \;,
\end{equation}
that characterizes the so--called Bloch oscillations experienced by a particle when the electric field is switched on. Observe that we can also define two different frequencies, $\omega_x$ and $\omega_y$, one for each of the spatial directions.

Generally, when both fields are present, one expects qualitatively different dynamics of the system (\ref{b3}) depending on relative size of cyclotron and Bloch frequencies. In what follows, to simplify equations, we set the charge $e$, the lattice period $a$ and all fundamental constants to unity. In so doing, the Bloch frequencies are given by $F_x$ and $F_y$. Also, if not stated otherwise, we also set $J_x=J_y=J=1$, which means that energy is measured in units of the hopping energy $J$ and time in units of the tunneling period $T_J=2\pi\hbar/J$. %Then the cyclotron frequency is given by the Peierls phase multiplied by $2\pi$.

%%%%%%%%%%%%%%%%%%%%%%%%%%%%%%%%%%%%%%%%%%%%%
\subsection{Vanishing magnetic field}
\label{secA2}

Let us now describe the two limiting cases of null magnetic and electric field, starting from the former.

%%%%%%%%%%%%%%%%%%%%%%%%%%%%%%%%%%%%%%%%%%%%%
\subsubsection{Wannier-Stark states}
\label{secA2a}
For vanishing magnetic field, the eigenstates of the Hamiltonian (\ref{b3}) are the localized WS-states $\Psi^{(n,k)}$, for integer $n$ and $k$, whose spectrum is given by the sum of two Wannier-Stark ladders,
%*******************************************
\begin{equation}
\label{c1}
E=F_x n + F_y k \;.
\end{equation}
One can easily prove Eq.~(\ref{c1}) and find the explicit form of WS-states by noticing that for $\alpha=0$ the  Hamiltonian (\ref{b3}) is separable and, hence, the 2D eigenvalue problem reduces to two independent 1D problems: one in the $x$ direction,
%*******************************************
\begin{equation}
\label{c2}
-\frac{J_x}{2}(b_{l+1}+b_{l-1}) + F_x lb_l=Eb_l  \;,
\end{equation}
and a fully similar one for the second degree of freedom. The solutions of Eq.~(\ref{c2}) can be expressed in terms of Bessel functions ${\cal J}_l$ of the first kind, of argument $z=J_x/F_x$. This can be seen in a variety of ways. Let us introduce a technique that will be used in the following: define the generating function
%*******************************************
\begin{equation}
\label{c2a}
Y(\vartheta)=\frac{1}{2\pi}\sum_l b_l \exp(il\vartheta).
\end{equation}
In terms of the latter, the system of algebraic equations (\ref{c2}) reduces to the ordinary differential equation
%*******************************************
\begin{equation}
   \label{c2b}
iF_x\frac{{\rm d}Y}{{\rm d}\vartheta}=-(J_x\cos\vartheta + E) Y \;.
\end{equation}
The solution of this equation reads
%*******************************************
\begin{equation}
\label{c2c}
Y(\vartheta)=\exp\left(-i \frac{J_x}{F_x} \sin\vartheta + i\frac{E}{F_x}\vartheta\right) \;,
\end{equation}
where, to satisfy the periodicity/quantization condition $Y(\vartheta+2\pi)=Y(\vartheta)$, the energy $E$ must be a multiple of $F_x$, {\em i.e.}, $E=nF_x$. Using the Fourier transformation, we compute $b_l^{(n)}={\cal J}_{l-n}(J_x/F_x)$. Thus, the explicit form of  the 2D WS-states reads
%*******************************************
\begin{equation}
\label{c3}
\Psi^{(n,k)}_{l,m}={\cal J}_{l-n}\left(\frac{J_x}{F_x}\right){\cal J}_{m-k}\left(\frac{J_y}{F_y}\right) \;.
\end{equation}
It should be observed that Eq.~(\ref{c1}) and Eq.~(\ref{c3}) do not apply to the case where either $F_x$ or $F_y$ are null. In fact, when $F_x=0$, WS-states are extended functions in the $x$ direction, $\Psi^{(k,\kappa)}_{l,m}\sim \exp(i\kappa l){\cal J}_{m-k}(J_y/F)$, and the spectrum is given by a ladder of energy bands,
%*******************************************
\begin{equation}
\label{c5}
E_\nu(\kappa)=F \nu-J_x\cos(\kappa) \;, \quad \nu=0,\pm1,\ldots \;.
\end{equation}
Clearly, a similar equation holds when $F_y=0$.

%%%%%%%%%%%%%%%%%%%%%%%%%%%%%%%%%%%%%%%%%%%%%
\subsubsection{Bloch oscillations}
\label{secA2b}
From the form of WS-states we can predict the dynamics of an initially localized wave packet in the $(x,y)$ plane. In the general case, $F_x,F_y\ne0$, the motion is a superposition of harmonic 1D Bloch oscillations.
%where the mean velocity and the mean coordinate of a quantum particle change according to the cosine and sine laws, respectively.
As a result, a coherent wave packet  follows a Lissajous-like trajectory \cite{58}.  Observe again that, in the case when one field component is null, say $F_x=0$, the packet oscillates in the $y$ direction and spreads ballistically in the $y$ direction.

A second, specific yet important initial condition is the Bloch wave
%*******************************************
\begin{equation}
   \label{c6a}
\Phi^{(\kappa)}_{l,m} \sim \sum_{l,m} e^{i\kappa_x l} e^{i\kappa_y m}  \;,
\end{equation}
that is characterized by the quasimomentum ${\bf \kappa}=(\kappa_x,\kappa_y)$. Clearly, in this case one can only speak of oscillations of the expectation value of the velocity. To describe these oscillations, it is convenient to move from the Wannier-Stark to the  Bloch picture. In fact, using
the Bloch acceleration theorem, {\em i.e.} replacing $\kappa$  by $\kappa-{\bf F}t$, we find
%*******************************************
\begin{equation}
\label{c6b}
\psi(t)=A(t)\Phi^{(\kappa-{\bf F}t)} \;, \quad A(t)=\exp\left[-i\int_0^t E(\kappa-{\bf F}t'){\rm d} t'\right] \;,
\end{equation}
where $E(\kappa)$ is the Bloch dispersion relation, Eq.~(\ref{b5}). In the following we shall consider the projection of the mean velocity on the direction of the electric field,
%*******************************************
\begin{equation}
   \label{c6c}
v(t)=v_x(t)\sin\beta +  v_y(t)\cos\beta  \;,\quad v_{x,y}(t)=\langle\psi(t) |\hat{v}_{x,y} |\psi(t)\rangle \;,
\end{equation}
where $\hat{v}_{x,y}$ are the velocity operators. These operators should not be confused with the momentum operators. In fact, they are given by the expressions
%***************************************************
\begin{eqnarray}
\nonumber
(\hat{v}_x\psi)_{l,m}=\frac{J_x}{2i}\left(\psi_{l+1,m} - \psi_{l-1,m}\right) \;, \\
   \label{c6d}
(\hat{v}_y\psi)_{l,m}=\frac{J_y}{2i}\left(\psi_{l,m+1}e^{i2\pi\alpha l} - \psi_{l,m-1}e^{-i2\pi\alpha l} \right) \;,
\end{eqnarray}
from where one notices that they coincide with the momentum operators only if $\alpha=0$ \footnote{These equations follow from the definition
$\hat{v}_{x}=-i[\widehat{H}_0,\hat{x}]$ where $\widehat{H}_0$ is the Hamiltonian (\ref{b3}) for $F=0$ and $(\hat{x}\psi)_{l,m}=l \psi_{l,m}$. For the operator $\hat{v}_y$ one  has similar expressions.}. Using the solution (\ref{c6b}), it is easy to prove that the projection oscillates as
%*******************************************
\begin{equation}
\label{c7}
v(t)=E'(\kappa-{\bf F}t) \;,
\end{equation}
where the prime denotes the first derivative of the dispersion relation along the line $\kappa(t)=\kappa-{\bf F}t$.

%%%%%%%%%%%%%%%%%%%%%%%%%%%%%%%%%%%%%%%%%%%%%%
\subsubsection{Double-periodic lattices}
\label{secA2c}
The dynamics of Bloch oscillations becomes much richer if we break separability of the tight-binding Hamiltonian. This can be done, for example, by introducing checkerboard onsite energies: $V_{l,m}=\delta (-1)^{l+m}$. This problem was considered in \cite{58}. The on-site potential doubles the lattice period and splits the Bloch band (\ref{b5}) into two subbands
%*******************************************
\begin{equation}
\label{c8}
E_{1,2}(\kappa_x,\kappa_y)=\pm \sqrt{J^2\cos^2(d\kappa_x)\cos^2(d\kappa_y) +\delta^2} \;,
\end{equation}
where $d=\sqrt{2}$ and $J_x=J_y=J$. Notice that in Eq.~(\ref{c8}) the quasimomenta $\kappa_x$ and $\kappa_y$ refer to the coordinate system defined by the primary axes of the double-periodic lattice, which are rotated by $\pi/4$ with respect to the primary axes of the simple square lattice.

As soon as the Bloch band is split into subbands a new effect comes into play: interband LZ-tunneling.  To take LZ-tunneling into account, Eq.~(\ref{c6b}) must be modified as follows:
%*******************************************
\begin{equation}
\label{c9}
\psi(t)=c_1(t)\Phi_1^{(\kappa-{\bf F}t)} + c_2(t)\Phi_2^{(\kappa-{\bf F}t)}  \;,
\end{equation}
where $c_1(t)$ and $c_2(t)$ are the complex amplitudes of the lower and upper subband, respectively. Recently, LZ-tunneling has been intensively studied in 1D double-periodic lattices \cite{Breid06,Drei09,Klin10}. When moving to 2D systems, a crucial difference appears: the dynamics of LZ-tunneling  not only depends on the field electric magnitude $F$, but also on its orientation, relative to the primary axes of the lattice. We shall discuss this orientation effect in detail in Sec.~\ref{secB}, when considering the case of $\pi$-flux,  where the particle dispersion relation also consists of two subbands.

The complex Landau-Zener dynamics of double-periodic lattices (or, more generally, lattices with two sublattices) is reflected in the complicated structure of WS-states and of their spectrum. This problem was considered in \cite{92,96}: it was found that WS-states are localized for `irrational' directions of the electric field, but extended for `rational' directions, defined by the straight lines connecting any two sites of the lattice. In a lattice with square symmetry the last conditions means that $F_x/F_y$ is a rational number. Anticipating results to be detailed later, let us mention that the Landau-Stark states ($\alpha\ne0$) are also extended or localized, depending on the rationality of the parameter $\beta=F_x/F_y$. This is just one of the many similarities existing between lattices with non-trivial geometry and the square lattice in the presence of a magnetic field.

%this spectrum is continues for `rational' orientations of the field ($F_x/F_y$ is a rational number) and pure point for `irrational' orientations. Generally, the approach based on the notion of WS-states is more constructive because the interband  LZ-tunneling is somehow encoded in WS-states.  However, the approach based on the notion of LZ-tunneling gives more transparent physical picture. In what follows we shall use both approaches in parallel that provides physical interpretation of analytical results.

%%%%%%%%%%%%%%%%%%%%%%%%%%%%%%%%%%%%%%%%%%%%%
\subsection{Vanishing electric field}
When the electric field is null, $F=0$, the key notion of the theory is that of magnetic bands. We discuss this notion within the framework of the tight-binding Hamiltonian (\ref{b3}), although it has a more general meaning, that can be applied to the original Hamiltonian (\ref{b1}) as well, see \cite{Azbe64,Zak64}.

\subsubsection{Magnetic bands}
\label{secA3}
It is easy to see that a non-zero Peierls phase breaks separability of the 2D Hamiltonian and, hence, the single Bloch band (\ref{b5}) splits into subbands -- known as magnetic bands. To find these bands one uses a formal substitution in which the wave function is a plane wave in the $y$ direction\footnote{If one uses the gauge ${\bf A}=B(-y,0)$ the substitution must rather be a plane wave in the $x$ direction.},
%***************************************************
\begin{equation}
\label{d1}
\Psi_{l,m}= \frac{e^{i\kappa_y m}}{\sqrt{L_y}}  b_l \;,\quad \kappa_y=\frac{2\pi}{L_y}k \;.
\end{equation}
Notice that in Eq.~(\ref{d1}) one assumes periodic boundary conditions, of period $L_y$, that in a successive step tends to infinity. The substitution (\ref{d1}) results in the celebrated Harper equation of \cite{Harp55},
%**************************************************
\begin{equation}
\label{d2}
-\frac{J_x}{2}(b_{l+1}+b_{l-1}) -J_y\cos(2\pi\alpha l+\kappa_y)b_l =E b_l \;,
\end{equation}
which is parametrized by the quasimomentum $\kappa_y$. If $\alpha$ is a rational number, $\alpha=r/q$, we can apply Bloch theorem to Eq.~(\ref{d2}), thereby introducing the quasimomentum $\kappa_x$, defined in the reduced Brillouin  zone, $-\pi/q<\kappa_x\le \pi/q$. Then, the spectrum of the system (\ref{b3}) consists of $q$ energy bands.

\begin{figure}[t]
\center
\includegraphics[width=8.0cm, clip]{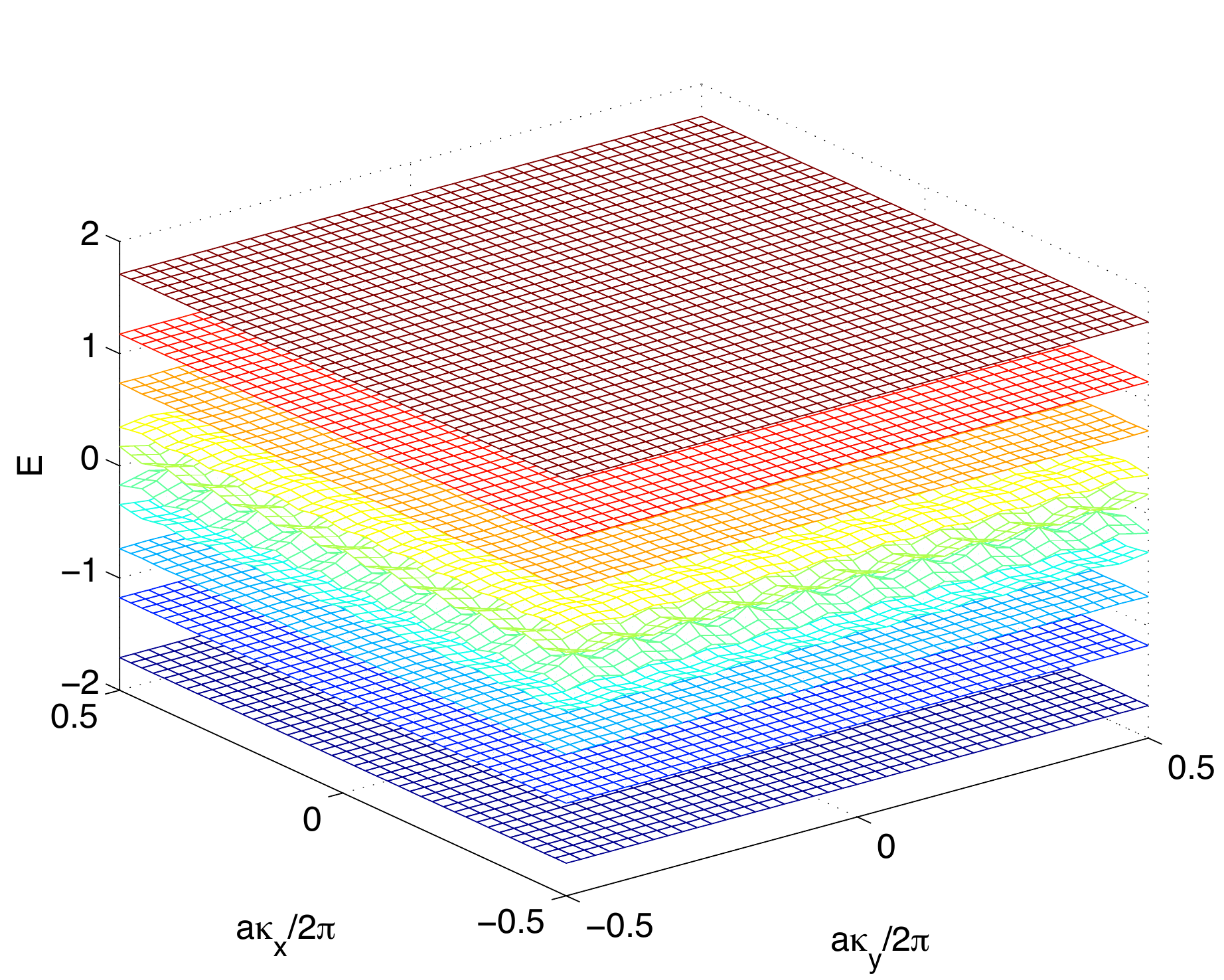}
\includegraphics[width=8.0cm, clip]{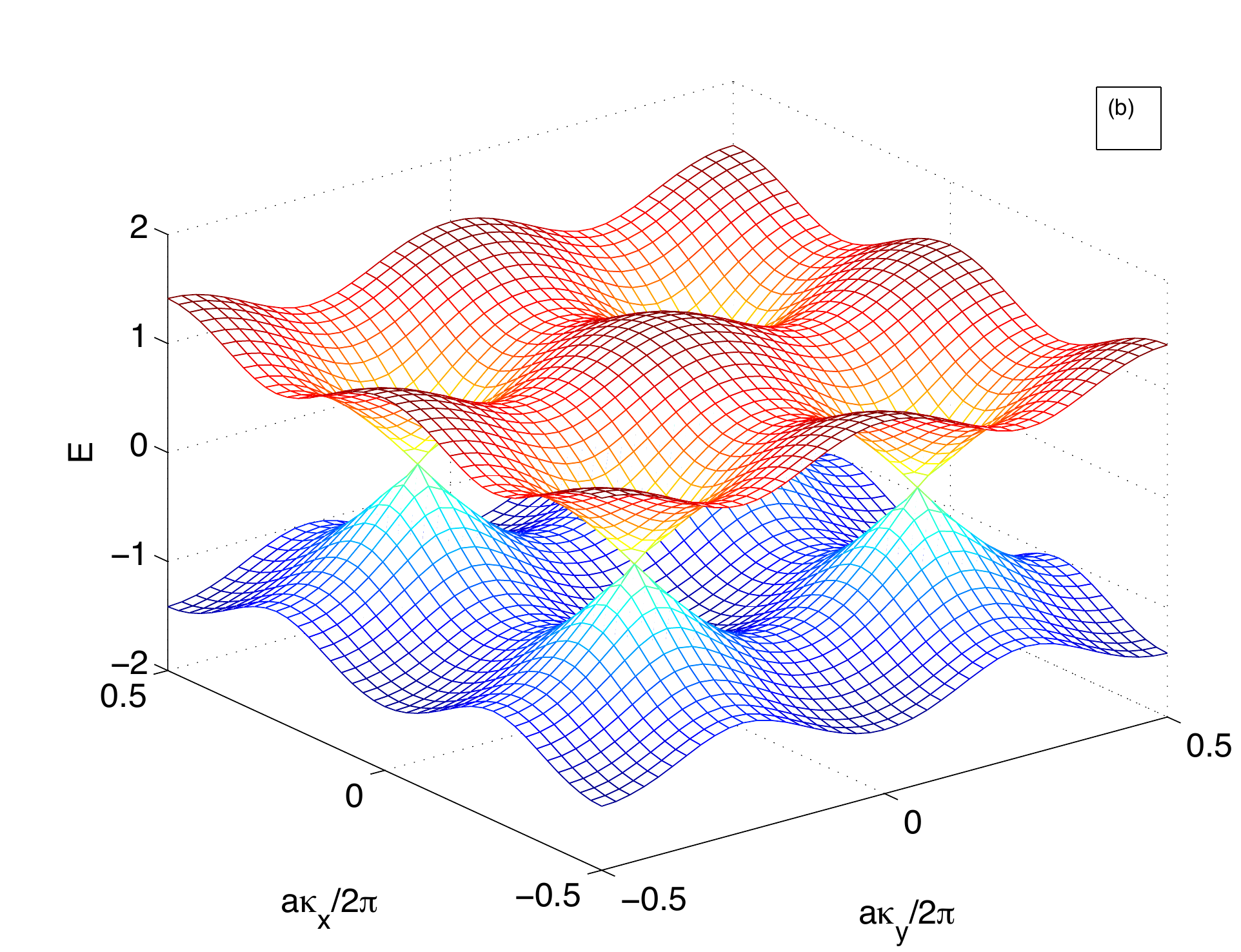}
\caption{Magnetic bands for $\alpha=1/10$ (a) and $\alpha=1/2$ (b). The tunneling amplitudes are $J_x=J_y=1$.}
\label{figA1}
\end{figure}

As an example, Fig.~\ref{figA1}(a) shows  this spectrum for $\alpha=1/10$. It is seen that the low and high-energy bands are practically flat. Moreover, the energy of these band is approximately given by the equation
%**************************************************
\begin{equation}
   \label{d3}
E=
\begin{cases}
-(J_x+J_y)+\omega_c(n+1/2) \;,& n\ll q \\
(J_x+J_y)-\omega_c(q-n+1/2) \;,& q-n \ll q \;,
\end{cases}
\end{equation}
as it is expected on the basis of the effective mass approximation. Thus, for $\alpha\ll 1$ low and high--energy bands can be viewed as multi-degenerate Landau levels of a quantum particle with positive and negative mass, respectively.

Figure \ref{figA1}(b) shows  the spectrum for $\alpha=1/2$. This particular value of $\alpha$ may be considered as exceptional, since for $\alpha=1/2$ the Hamiltonian is real and its spectrum is known analytically, see Eq.~(\ref{g7}) in Sec.~\ref{secB4}. Nonetheless, this case is physically important, in relation to doubly-periodic lattices. %As it was briefly mentioned in Sec.~\ref{secA2c}, Bloch dynamics in double-periodic lattices is far from trivial.  The presence of the Dirac cones in the dispersion relation makes this dynamics even more complex.

In what follows, when analyzing the general situation of non-zero electric and magnetic field, we shall focus separately on the cases of small $\alpha\ll 1$ and large $\alpha=1/2$. Although having some common features, these two cases appear to be quite different and require different theoretical approaches. By understanding these two limiting cases, we shall be able to make reliable predictions for arbitrary values of $\alpha$.
%#############################################################

%%%%%%%%%%%%%%%%%%%%%%%%%%%%%%%%%%%%%%%%%%%%%%%
\section{Landau-Stark states}
\label{secA4}
%%%%%%%%%%%%%%%%%%%%%%%%%%%%%%%%%%%%%%%%%%%%%%%

We now proceed to the general case of non-zero electric and magnetic fields. As mentioned in the introduction, in this case we call Landau-Stark states the eigenstates of the Hamiltonian (\ref{b3}). In this section we describe a fundamental property of LS-states that is valid for arbitrary $\alpha$, whether large or small, rational or irrational. This property is the sensitivity of the spectrum to the orientation of the electric field, relative to the primary axes of the lattice. Namely, for rational values of  the parameter $\beta$,
%*******************************************
\begin{equation}
   \label{e1}
\beta=F_x/F_y \;,
\end{equation}
the spectrum of LS-states is continuous (band structured) and, correspondingly, LS-states are extended, Bloch-like states. To the contrary, when $\beta$ is an irrational number, the spectrum is discrete and LS-states are localized. In the next two subsection we describe analytical and numerical approaches to compute these spectra.

%%%%%%%%%%%%%%%%%%%%%%%%%%%%%%%%%%%%%%%%%%%%%%%%
\subsection{Localized Landau-Stark states}
\label{secA4a}

\begin{figure}[t]
\center
\includegraphics[width=8.5cm, clip]{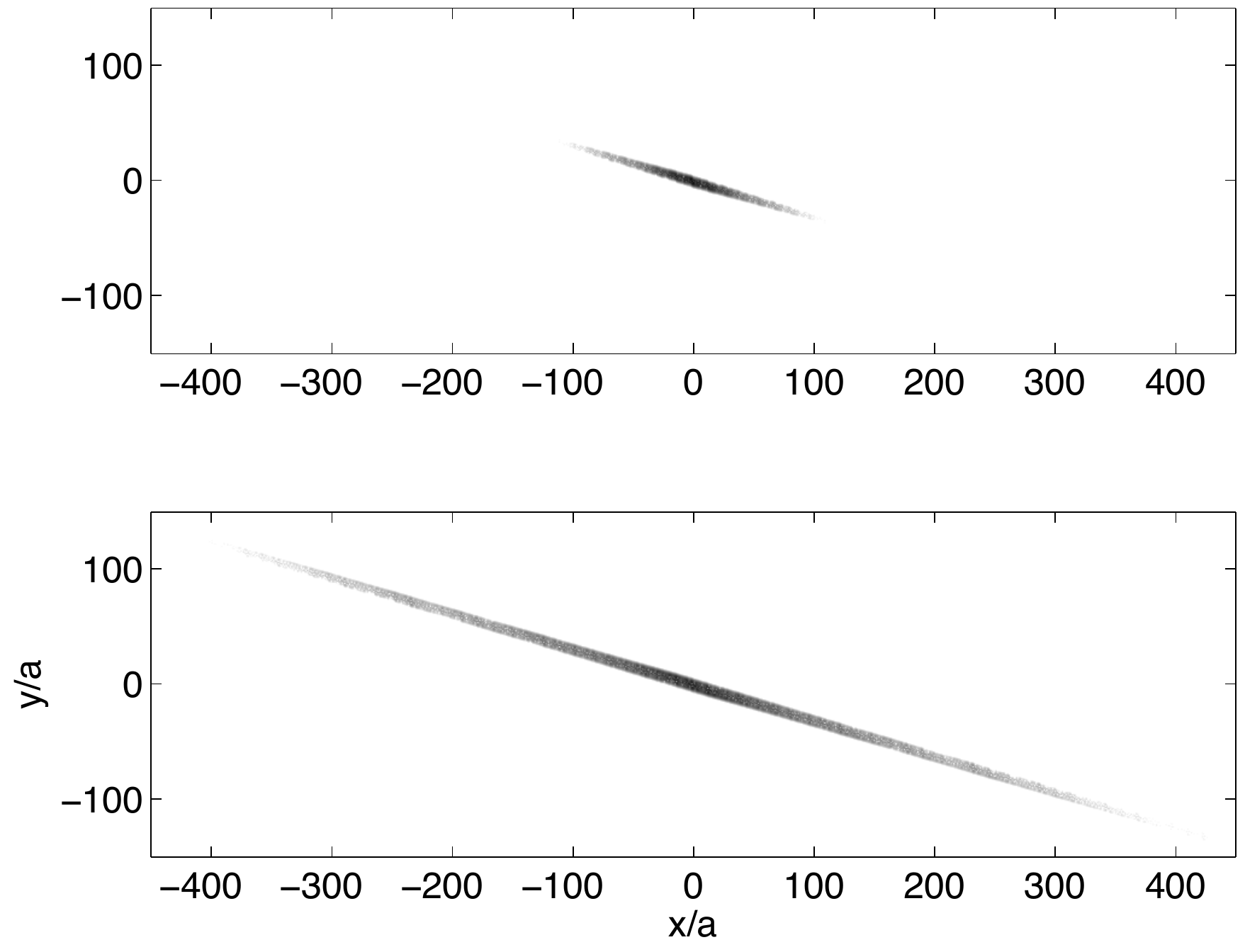}
\caption{Example of a localized LS-state for $\beta=F_x/F_y=(\sqrt{5}-1)/4\approx1/3$ and $F=0.7$ (a) and $F=0.5$ (b). The Peierls phase is $\alpha=(\sqrt{5}-1)/8\approx1/6$.}
\label{figA2}
\end{figure}

To facilitate the discussion, Fig.~\ref{figA2} shows examples of LS-states for irrational $\beta=(\sqrt{5}-1)/4$, a value which is close to the rational $\beta=1/3$. More precisely, the figure shows a single LS-state with the energy $E=0$, for two values of the electric field magnitude: $F=0.7$ and $F=0.5$. It is evident in Fig.~\ref{figA2} that these states stretch in the direction orthogonal to the vector ${\bf F}$, that is a common property of LS-states. It is also seen that the localization length of the state increases when $F$ is decreased. We will discuss the scaling law for the localization length in Sec.~\ref{secB3a} and Sec.~\ref{secC3}, let us focus here on the spectrum. It can be deduced from the following simple theorem.

Let $\Psi_{l,m}$ be an eigenstate of the Hamiltonian (\ref{b3}), of energy $E$. Then the state
%*********************************************************************
\begin{equation}
\label{e2}
\tilde{\Psi}_{l,m}=\Psi_{l-n,m-k}e^{-i2\pi\alpha nm}
\end{equation}
is also an eigenstate of (\ref{b3}), of energy $\tilde{E}=E+(F_x n+F_y k)$. It follows from the last equation that every LS-state can be labeled by two integer numbers, $n$ and $k$, so that the spectrum of localized LS-states coincides with that of WS-states,{\em i.e.},
%*********************************************************************
\begin{equation}
   \label{e3}
E_{n,k}=F_x n + F_y k \;\quad \mbox{for  } F_x/F_y\ne r/q \;.
\end{equation}
Although the spectra of WS-states ($\alpha=0$) and LS-states ($\alpha\ne0$)  coincide, the states themselves are completely different. There are two approaches to find  LS-states. The first is straightforward diagonalization of the 2D Hamiltonian (\ref{b3}).  Notice that it suffices to find only one LS-state: the other states (which, of course, form a complete basis in the Hilbert space) are obtained by translating this state across the lattice and imprinting on it a gauge-dependent phase according to Eq.~(\ref{e2}).  The second approach, while more involved, opens the way for an analytical analysis of LS-states. We discuss it in the next subsection.
%#############################################################

%%%%%%%%%%%%%%%%%%%%%%%%%%%%%%%%%%%%%%%%%%%%%%%%
\subsubsection{Evolution operator approach}
\label{secA4b}

Let us construct, from the complete set of LS-states $\Psi^{(n,k)}$, a set of Bloch-like functions,
%********************************************************************
\begin{equation}
\label{e4}
\Phi^{(n,\kappa)}=\sum_k e^{i\kappa k} \Psi^{(n,k)} \;,
\end{equation}
which are labeled by the integer index $n$ and the real quasimomentum $\kappa$. Observe that the spatial $l,m$ dependence is left implicit in the above equation and below, not to overburden notation. It is easy to prove that the Bloch-like states (\ref{e4}) are eigenstates of the evolution operator $\widehat{U}$ over the period $T_y=2\pi/F_y$,
%********************************************************************
\begin{equation}
   \label{e5}
\widehat{U}\Phi^{(n,\kappa)}=\lambda_n \Phi^{(n,\kappa)} \;, \quad \widehat{U}=\exp(-i\widehat{H}T_y)
\end{equation}
where $\lambda_n=\exp(-i F_x n T_y)=\exp(-i2\pi \beta n)$. Notice that the eigenphases $\lambda_n$ do not depend on $\kappa$ (contrary to the eigenstates, see below). The trick is now to find a different representation of the states  $\Phi^{(n,\kappa)}$. To do this, we consider the  time-dependent Schr\"odinger equation with the Hamiltonian (\ref{b3})
%********************************************************************
\begin{equation}
   \label{e6}
i\dot{\psi}_{l,m}=-\frac{J_x}{2}(\psi_{l+1,m}+\psi_{l-1,m})
-\frac{J_y}{2}(\psi_{l,m+1}e^{i2\pi\alpha l}+\psi_{l,m-1}e^{-i2\pi\alpha l})+(F_x l+F_y m)\psi_{l,m} \;.
\end{equation}
Using the substitution (compare with Eq.~(\ref{d1}))
%******************************************************************
\begin{equation}
   \label{e7}
\psi_{l,m}(t)=\frac{1}{\sqrt{L_y}} e^{i(\kappa-F_y t) m} b_l(t)
\end{equation}
we get the following evolution equation for the amplitudes $b_l(t)$, that can be interpreted as generated by a time dependent Hamiltonian $\widehat{H}(t)$:
%****************************************************************
\begin{equation}
\label{e8}
i\dot{b}_{l}=-\frac{J_x}{2}(b_{l+1}+b_{l-1}) -J_y \cos(2\pi\alpha l +\kappa-F_y t) b_{l} + F_x l b_{l}
\equiv \left(\widehat{H}(t){\bf b}\right)_l \;.
\end{equation}
The solution of Eq.~(\ref{e8}), in which $\kappa$ is a parameter, is equivalent to the construction of the $\kappa$-fiber evolution operator $\widehat{U}^{(\kappa)}$, over the period $T_y=2\pi/F_y$. This latter is defined formally as the time-ordered integral
%*************************************************************
\begin{equation}
\label{e10}
\widehat{U}^{(\kappa)}=\widehat{\exp}\left[-i\int_0^{T_y}\widehat{H}(t) {\rm d}t \right] \;.
\end{equation}

Note that if we set the hopping matrix element $J_x$ to zero, the evolution operator (\ref{e10}) is a diagonal matrix with entries $U_{l,l}^{(\kappa)}=\exp(-i2\pi \beta l)$. For a finite $J_x$ it becomes (rather, it can be approximated by) a banded matrix, with band width proportional to $J_x/F_x$. At this point, irrationality of the parameter $\beta$ is responsible for the point spectrum of the evolution operator and for the localization of its eigenstates ${\bf b}^{(n)}(\kappa)$ \footnote{We mention an analogy with the paradigmatic model of quantum chaos -- the kicked rotor \cite{Casa79}. For vanishing kick amplitude, the evolution operator of the kicked rotor is a diagonal matrix with matrix elements $U_{l,l}=\exp(-i2\pi \xi l^2)$, which is a periodic or aperiodic function of $l$ according to rationality of the parameter $\xi$. The number theoretic characteristics of $\xi$ are crucial \cite{Casa84}, since they determine whether eigenfunctions of the kicked rotor are extended or localized.}. We use these  eigenstates to construct an alternative representation of the 2D Bloch-like states $\Phi^{(n,\kappa)}$ as follows:
%******************************************************************
\begin{equation}
\label{e12}
\Phi^{(n,\kappa)}_{l,m}=\frac{1}{\sqrt{L_y}} e^{i\kappa m} b^{(n)}_l(\kappa) \;.
\end{equation}
(Clearly, a large limit process over $L_y$ is implicit here.) Finally, using the definition (\ref{e4}), we find the localized LS-states by Fourier transforming Eq.~(\ref{e12}):
%*************************************************************
\begin{equation}
\label{e13}
\Psi_{l,m}^{(n,k)}=\frac{1}{2\pi}\int_0^{2\pi}  \Phi^{(n,\kappa)}_{l,m} e^{-i\kappa k} {\rm d}\kappa
=\frac{1}{2\pi} \int_0^{2\pi}  b_l^{(n)}(\kappa) e^{-i\kappa(m- k)} {\rm d}\kappa\;.
\end{equation}
In essence, the evolution operator approach reduces the original 2D eigenvalue problem to a set of 1D problems, parameterized by the quasimomentum $\kappa$.
%#############################################################
\begin{figure}[t]
\includegraphics[width=8.0cm]{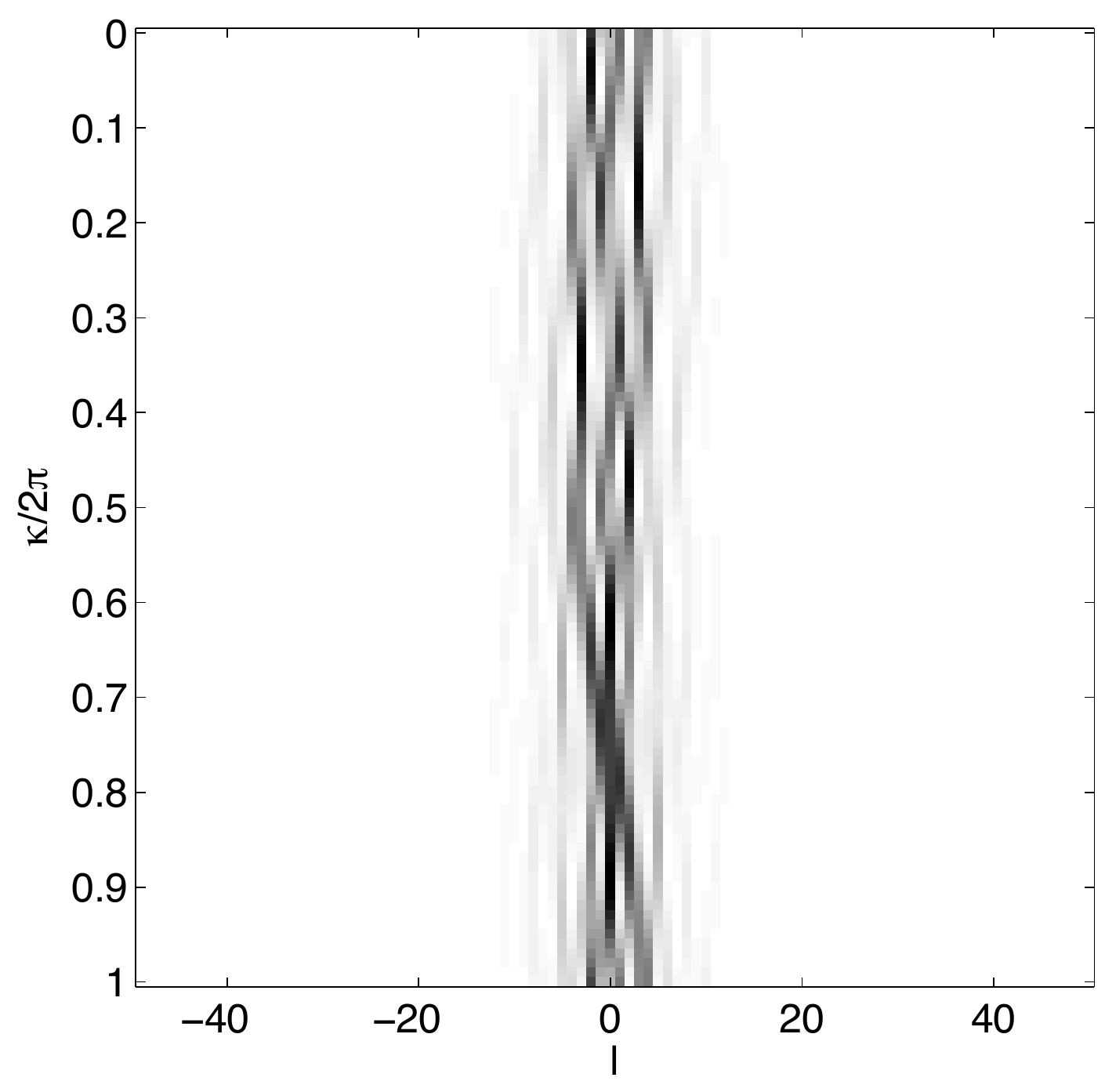}
\caption{Eigenstate of the 1D evolution operator (\ref{e10}) as a function of the quasimomentum $\kappa$. Squared amplitudes $|b_l^{(n)}(\kappa)|^2$ are shown as a grey-scaled map. Parameters are $\alpha=1/3$, $F=0.5$,  $\beta=(\sqrt{5}-1)/4$ and $n=0$.}
\label{figA3}
\end{figure}

We conclude this subsection by two remarks. The most delicate step in the numerical implementation of the described procedure is matching the eigenstates obtained for different values of $\kappa$, so that the functions  $b_l^{(n)}(\kappa)$ in Eq.~(\ref{e13}) become analytic functions of $\kappa$. Fortunately, for some purposes like, for example, studying the localization length of LS-states, this step can be omitted. In fact, let us consider the projection of the localized state $\Psi^{(n,k)}$ onto the $x$ axis,
%***********************************************************
\begin{equation}
   \label{e14}
\rho^{(n)}_l=\sum_m |\Psi_{l,m}^{(n,k)}|^2 \;.
\end{equation}
Using Eq.~(\ref{e12}) we have
%***********************************************************
\begin{equation}
\label{e15}
\rho^{(n)}_l=\int_0^{2\pi} {\rm d}\kappa \; |b_l^{(n)}(\kappa)|^2 \;.
\end{equation}
Implementation of Eq.~(\ref{e15}) does not require any phase matching. The typical behavior of  the squared amplitude $|b_l^{(n)}(\kappa)|^2$ is shown in Fig.~\ref{figA3} where $n=0$. It is seen that one can sample a few values of $\kappa$, using a Monte-Carlo approach, to obtain a reliable estimate for extension of LS-states in the $x$ direction.  Similarly,  we can find the extension of LS-states in the $y$ direction. To do this, one should consider evolution operator over the period $T_x=2\pi/F_x$ and repeat all steps described above.
%
%Needless to say that in this case the Hamiltonian of the driven Harper model reads
%%**********************************************************
%\begin{equation}
%\label{e16}
%(\widehat{H}(t) {\bf b})_m=-\frac{J_y}{2}(b_{m+1}+b_{m-1}) + F_y m b_{m} -J_x \cos(2\pi\alpha m +\kappa+F_x t) b_{m}  \;.
%\end{equation}

%%%%%%%%%%%%%%%%%%%%%%%%%%%%%%%%%%%%%%%%%%%%%%%%
\subsection{Extended Landau-Stark states}
\label{secA4c}

If the parameter $\beta$ is a rational number, LS-states are extended states in the direction orthogonal to the vector ${\bf F}$ and their spectrum consists of energy bands. To find these bands we follow the approach introduced in \cite{Naka95} and further developed in \cite{91}. This approach involves several steps. The first step consists of choosing a convenient gauge for the magnetic field. Letting $\beta=r/q$, where $r$ and $q$ are relatively prime integers, the electric field vector can be written as:
%*************************************************
\begin{equation}
   \label{f1}
{\bf F}= \frac{F}{\sqrt{N}}\left( r,q \right) \;, \quad N=r^2+q^2 \;.
\end{equation}
We choose the gauge
%*****************************************************
\begin{equation}
   \label{f2}
{\bf A}=B\left( -\frac{q(rx+qy)}{N},\frac{r(rx+qy)}{N}\right)  \;,
\end{equation}
which reflects the geometry induced by the electric field. Within this gauge, the tight-binding Hamiltonian (\ref{b3}) becomes
%*****************************************************
\begin{eqnarray}
\label{f3}
(\widehat{H}\psi)_{l,m}=
-\frac{J_x}{2}\left( \exp\left[-i2\pi\alpha \frac{q}{N}(rl+qm)\right] \psi_{l+1,m}  +  h.c. \right) +
\nonumber \\
-\frac{J_y}{2}\left( \exp\left[i2\pi\alpha \frac{r}{N}(rl+qm)\right]\psi_{l,m+1} +h.c. \right)
+ F\frac{(rl+qm)}{\sqrt{N}} \psi_{l,m} \;,
\end{eqnarray}
where $h.c.$ denotes the terms required to render the Hamiltonian Hermitian.

\begin{figure}[t]
\includegraphics[width=8.0cm]{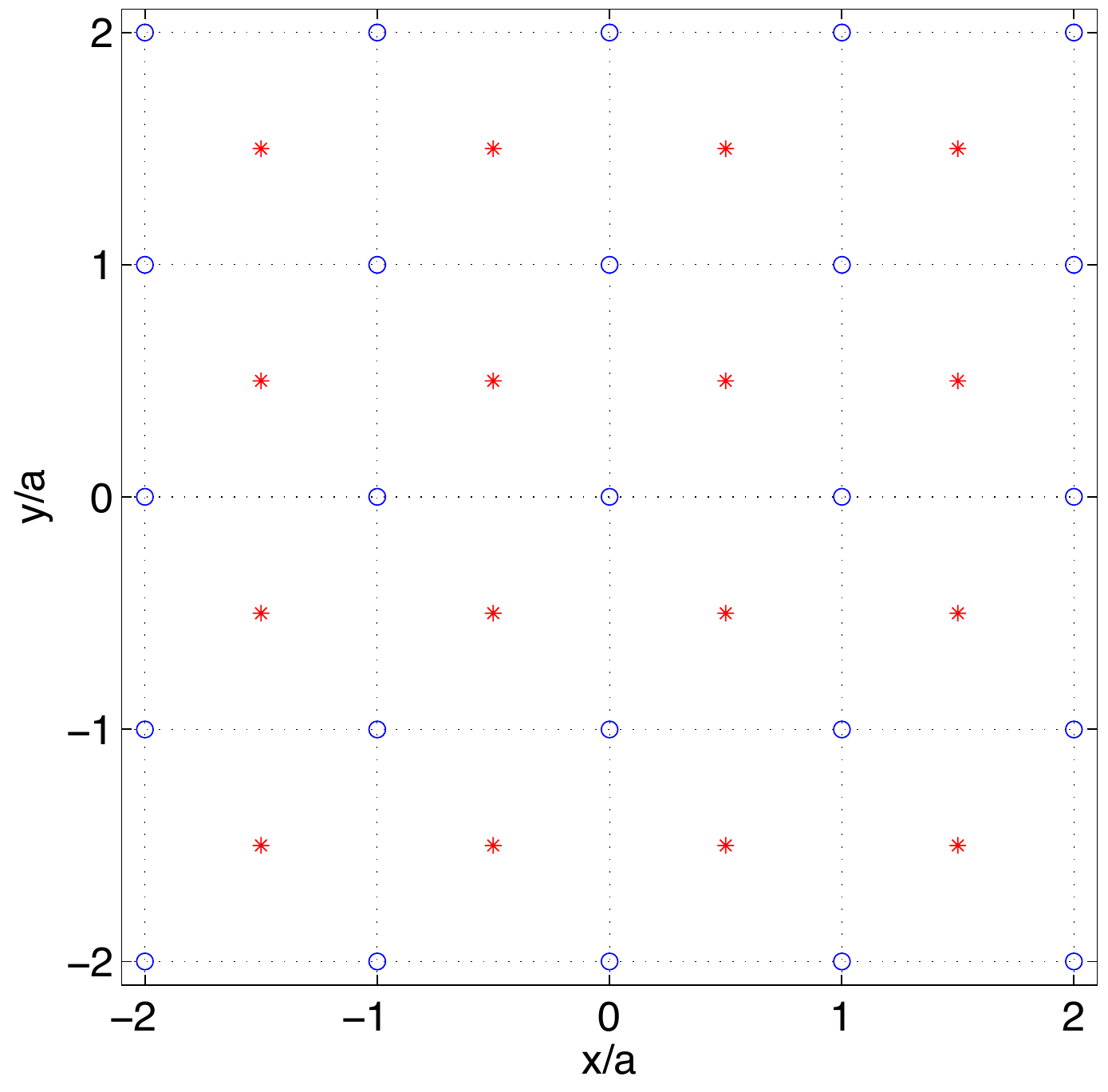}
\caption{Extended lattice for the field direction $(r,q)=(1,1)$, where it consists of two sublattices, marked by stars and circles, respectively.}
\label{figA4}
\end{figure}

The second step is to simplify the Hamiltonian (\ref{f3}), by rotating coordinates so to align the electric field with the vertical axis $\xi$ of a new
coordinate frame $(\eta,\xi)$:
%***************************************************
\begin{equation}
   \label{f4}
\eta=\frac{qx-ry}{\sqrt{N}}  \;,\quad  \xi=\frac{rx+qy}{\sqrt{N}}  \;.
\end{equation}
In rotated coordinates, the original lattice sites $(l,m)$ (recall that the lattice period is set to unity) appear to lie on a sublattice, embedded into a new square lattice of spacing $d$,
%***************************************************
\begin{equation}
   \label{f5}
d=1/\sqrt{N}  \;,
\end{equation}
whose sites $(sd,pd)$ can be labeled by pair of integer indexes $(s,p)$. Note hat this new lattice actually consists of $N$ independent sublattices, only one of which coincides with the original lattice, see Fig.~\ref{figA4}.

Letting now $\phi_{s,p}$ denote the wave-function amplitude at site $(sd,pd)$ in the rotated frame of reference $(\eta,\xi)$, and using the fact that $(rl+qm) = p$, we can write the Hamiltonian action as
%*************************************************
\begin{equation}
   \label{f6}
(\widehat{H}\phi)_{s,p} = -\frac{J_x}{2}\left(e^{-i2\pi\alpha q  p/ N} \phi_{s+q,p+r}  +  h.c. \right)
-\frac{J_y}{2}\left(e^{ i2\pi\alpha r   p/ N} \phi_{s-r,p+q} + h.c. \right) + dFp \phi_{s,p}  \;.
\end{equation}
Observe that, coherently with Eq.~(\ref{f3}), the $N$ sub-lattices described above are uncoupled, and that the shift $s \rightarrow s+N$ is an invariant transformation. Nevertheless, it is convenient to solve the eigenvalue problem for all sublattices simultaneously. We therefore consider the stationary Schr\"odinger equation for the complex amplitudes $\phi$, $(\widehat{H}\phi)_{s,p} = E \phi_{s,p}$. Following \cite{Naka95} we use the plane wave basis
%***************************************************
\begin{equation}
   \label{f7}
\phi_{s,p}=\frac{e^{id\kappa s}}{\sqrt{L}} b_p(\kappa) \;,
\end{equation}
where $L$ eventually tends to infinity. We finally arrive at the following equation, %where we have put $\theta = 2\pi\alpha  N^{-1}$:
%%***************************************************
%\begin{equation}
%\label{f8}
%-\frac{J_x}{2}\left(e^{-i \theta q p+iq d\kappa} b_{p+r}  + e^{i \theta q (p-r)-iq d\kappa} b_{p-r}  \right)
%-\frac{J_y}{2}\left(e^{i \theta  r p -i r d\kappa} b_{p+q} + e^{-i \theta  r (p-q) +i r d\kappa} b_{p-q} \right)
%+ dFp b_{p}=E b_{p} \;.
%\end{equation}
%%
%***************************************************
\begin{eqnarray}
\label{f8}
-\frac{J_x}{2}\left[Q(p) e^{iq d\kappa} b_{p+r}  + Q(p-r)e^{-iq d\kappa} b_{p-r}  \right]
-\frac{J_y}{2}\left[R(p) e^{-ir d\kappa} b_{p+q} + R(p-q)e^{ir d\kappa} b_{p-q} \right]
+ dFp b_{p}=E b_{p} \;, 
\end{eqnarray}
where
%************************************************
\begin{eqnarray}
Q(p)=\exp\left(-i2\pi\alpha \frac{qp}{N}\right) \;,\quad R(p)=\exp\left(i2\pi\alpha \frac{rp}{N}\right) \;.
\end{eqnarray}
The left hand side of Eq.~(\ref{f8}) is a banded matrix of bandsize five, hence,  it can be numerically diagonalized rather easily. This yields the energy bands
%***************************************************
\begin{equation}
E=E_\nu(\kappa) \;,\quad E_\nu(\kappa+2\pi/\sqrt{N})=E_\nu(\kappa) \;.
\end{equation}
In the next section we discuss the asymptotic behavior of these bands in the limit $F\rightarrow\infty$.
%#############################################################

%%%%%%%%%%%%%%%%%%%%%%%%%%%%%%%%%%%%%%%%
\subsubsection{Perturbation theory}
\label{secA4d}

The asymptotic behavior of the energy bands crucially depends on the value of the two integer numbers $r$ and $q$ defining the parameter $\beta=r/q$. It is instructive to begin with the two simplest cases: $(r,q)=(0,1)$ and  $(r,q)=(1,1)$.

When $(r,q)=(0,1)$, Eq.~(\ref{f8}) simplifies to
%***************************************************
\begin{equation}
   \label{f9}
-\frac{J_y}{2}(b_{p+1}+b_{p-1}) - J_x\cos(2\pi\alpha p-\kappa)b_p +Fpb_p =E b_p \;.
\end{equation}
In the limit of large $F$ the spectrum of this equation is approximated by the ladder of energy bands
%***************************************************
\begin{equation}
\label{f10}
E_\nu(\kappa)=F\nu - J_x\cos(\kappa-2\pi\alpha \nu) \;,\quad  \nu=0,\pm1, \ldots \;.
\end{equation}
It is instructive to compare this equation with Eq.~(\ref{c5}), which gives energy spectrum of the system for zero magnetic field.

If $(r,q)=(1,1)$, Eq.~(\ref{f8}) takes the form
%****************************************************
\begin {equation}
   \label{f11}
 -[V(p;\kappa) b_{p+1} + V^*(p-1;\kappa) b_{p-1}] + dFp b_p=E b_p \;,
\end{equation}
where $d=1/\sqrt{2}$ and $V(p;\kappa)=\left( J_x e^{-i\pi\alpha p} e^{id\kappa}+ J_ye^{i\pi\alpha p} e^{-id\kappa}\right)/2$. Similar to the case $(r,q)=(0,1)$, the unperturbed spectrum of the system consists of flat bands separated by the Stark energy, {\em i.e.}, $E^0_\nu(\kappa)=dF\nu$. However, unlike the case $(r,q)=(0,1)$, the first order perturbative correction to this spectrum vanishes. The second order correction is given by
%****************************************************
\begin {equation}
   \label{f12}
\Delta E_\nu(\kappa)=\frac{|V(\nu-1;\kappa)|^2}{dF} - \frac{|V(\nu;\kappa)|^2}{dF}
=\frac{J_xJ_y}{2dF}[\cos(2\pi\alpha(\nu-1)-2d\kappa) -\cos(2\pi\alpha\nu-2d\kappa)] \;.
\end{equation}
According to the last equation, bandwidth decreases as $1/F$ when $F$ increases. This is actually a special case of a general perturbation theory result. It was proven in \cite{91} that for non-zero $r$ and $q$ the first not-vanishing corrections are of order $r+q-1$. Therefore, the bandwidth $\Delta E= \max E_\nu(\kappa)- \min E_\nu(\kappa)$ scales asymptotically as
%***************************************************
\begin{equation}
\label{f13}
\Delta E \sim \frac{1}{F^{r+q-1}} \;.
\end{equation}

%%%%%%%%%%%%%%%%%%%%%%%%%%%%%%%%%%%%%%%%%
\section{The case $\alpha=1/2$}
\label{secB}
%%%%%%%%%%%%%%%%%%%%%%%%%%%%%%%%%%%%%%%%%

We now analyze limiting cases of LS-states. In this section we focus on the important case $\alpha=1/2$, where the tight-binding Hamiltonian (\ref{b3}) resembles that of a lattice with two sublattices. We begin the analysis by considering the energy bands that exist for rational $\beta=F_x/F_y$.
%#############################################################
\begin{figure}[b]
\includegraphics[width=8.0cm]{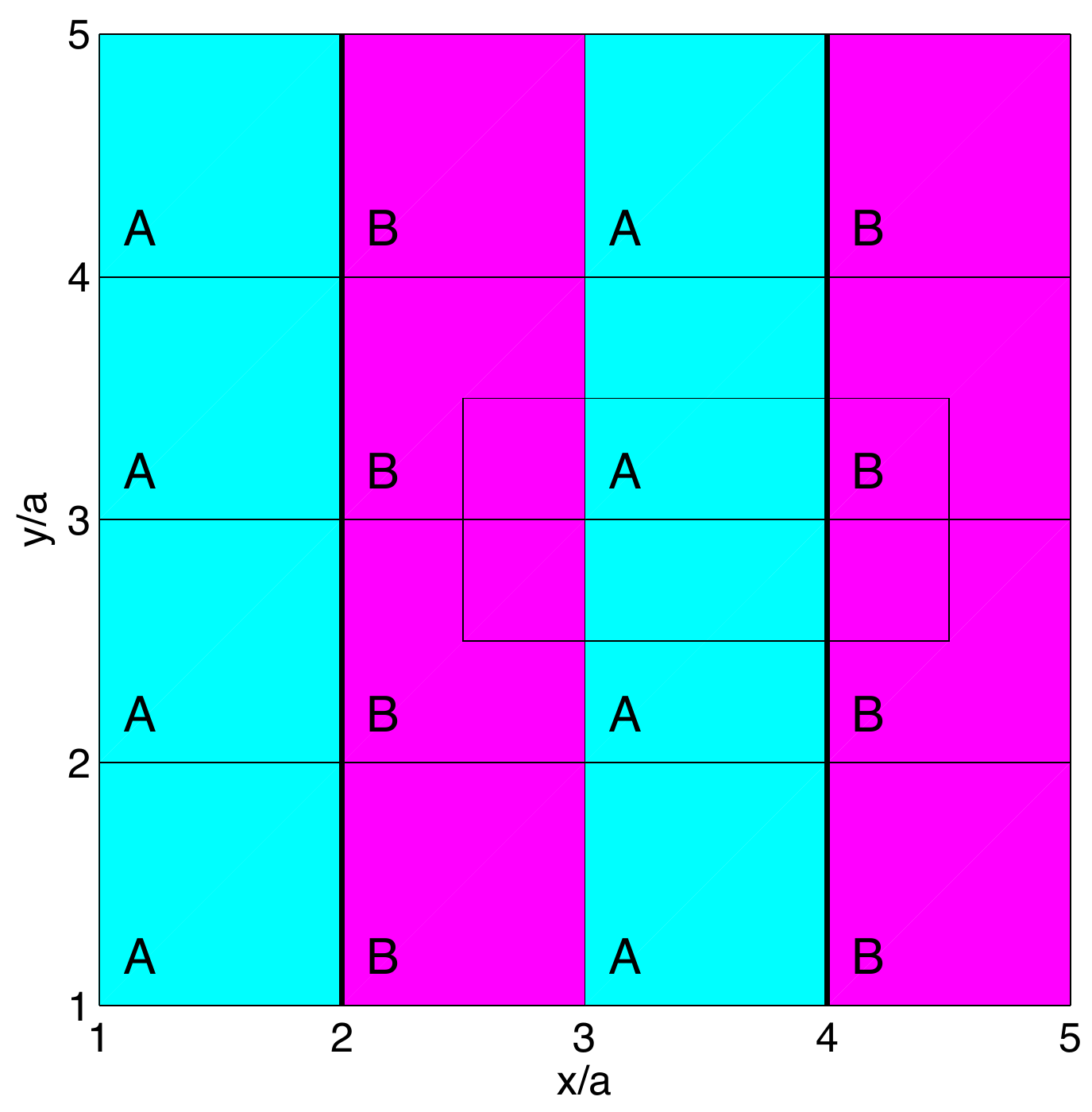}
\caption{A 2D square lattice with a uniaxial staggered magnetic field, where the tunneling between $B$ sites involves an additional phase $\phi$. If $\phi=\pi$ this also corresponds to a uniform magnetic field with $\alpha=1/2$. The elementary cell of the lattice is marked by a rectangle.}
\label{figB0}
\end{figure}

%%%%%%%%%%%%%%%%%%%%%%%%%%%%%%%%%%%%%%%%%
\subsection{Energy bands}
\label{secB1}

\begin{figure}[ht]
\includegraphics[width=8.2cm]{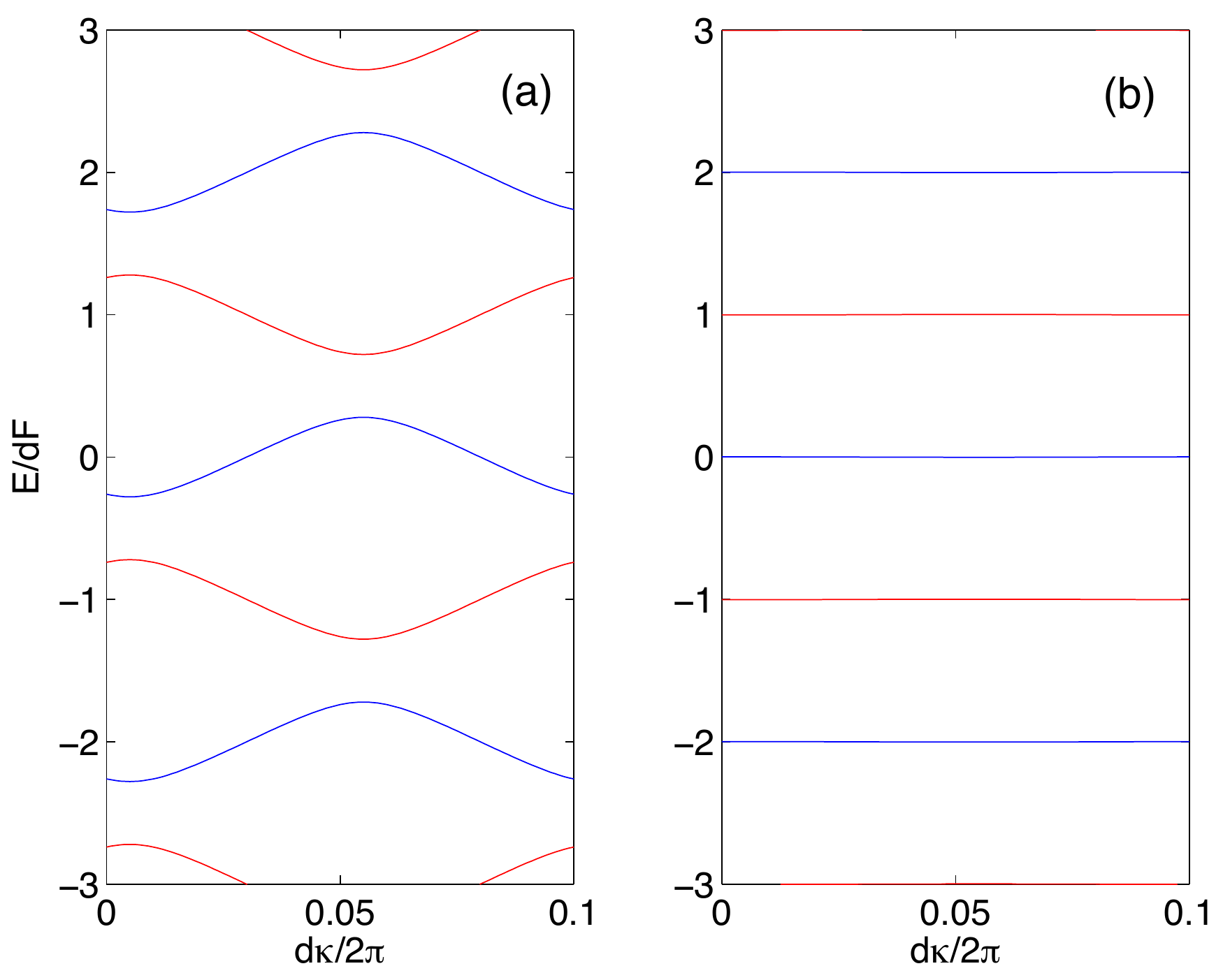}
%\hspace*{0.2cm}
\includegraphics[width=8.1cm]{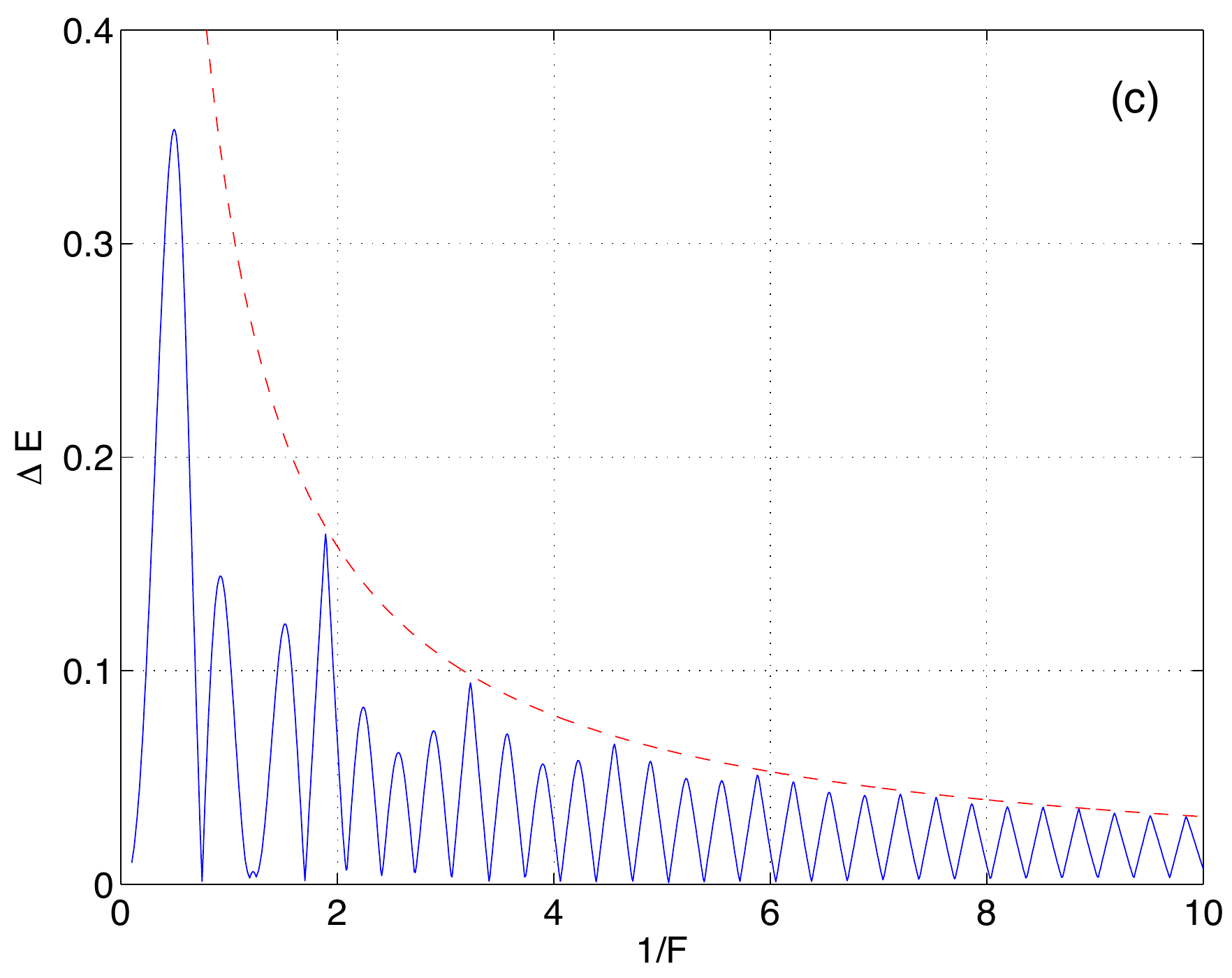}
\caption{Energy bands of the system (\ref{g1}) for $\beta=F_x/F_y=1/3$. The spectrum displayed in panel (a) is obtained for the value of $F$ corresponding to the maximum bandwidth, displayed in panel (c) versus $1/F$. Panel (b) shows the band collapse occurring at the first minimum of bandwidth. In panels (a-b) energy is measured in units of the ladder spacing $dF=F/\sqrt{10}$. This function is plotted in panel (c) as a dashed line. }
\label{figB1}
\end{figure}

Using the fact that for $\alpha=1/2$ the phase factor is $\exp(i2\pi\alpha l)=\pm 1$, we can rewrite the stationary Schr\"odinger equation, with Hamiltonian (\ref{b3}), as a system of two coupled equations, for  $A$- and $B$-lattice sites, as illustrated in Fig.~\ref{figB0}:
%***************************************************
\begin{eqnarray}
\nonumber
-\frac{J_x}{2}(\psi^B_{l',m'} + \psi^B_{l'-1,m'}) -\frac{J_y}{2}(\psi^A_{l',m'+1} + \psi^A_{l',m'-1})
+ (2F_x l' +F_y m')\psi^A_{l',m'}=E\psi^A_{l',m'} \;,\\
\label{g1}
-\frac{J_x}{2}(\psi^A_{l',m'} + \psi^A_{l'+1,m'}) +\frac{J_y}{2}(\psi^B_{l',m'+1} + \psi^B_{l',m'-1})
+ (2F_x l' +F_y m'+F_x)\psi^B_{l',m'}=E\psi^B_{l',m'} \;.
\end{eqnarray}
Notice that in Eq.~(\ref{g1}) the indices $l'$ and $m'$ denote different elementary cells, not lattice sites. To focus ideas, let us consider as a generic example the case $\beta=1/3$. The results of the numerical analysis of the system (\ref{g1}) for this value of $\beta$ are presented in the composite Fig.~\ref{figB1}, which shows the energy bands of LS-states for two particular values of the electric field $F$, in panels (a) and (b), and the width of these energy bands  as a function of $1/F$, in panel (c). Analyzing this figure leads to the following conclusions.

First of all, coherently with Eq.~(\ref{f13}), bandwidth is seen to decrease as $(1/F)^3$, when $F\rightarrow\infty$, in the leftmost part of panel (c). Next, moving on the curve to the right, as $F$ diminishes, the bandwidth takes its maximal value at a value $F^*\sim J$ of the electric field (the spectrum at this particular value of $F$ is shown in panel (a)), and then it displays an oscillatory behavior. This phenomenon is a 2D analogue of the phenomenon of band collapse, evident in panel (b), that is found in driven tilted 1D lattices \cite{Zhao91,Sias08,Ivan08}. In the  latter system, an external driving at frequency $\omega$, in multiphoton resonance with the Bloch frequency, $q\omega=\omega_B$, $q$ integer, has the effect of renormalizing the Bloch dispersion relation $E(\kappa)=-J\cos(a\kappa)$ into
%********************************************************
\begin{equation}
\label{g2}
E(\kappa)=-J_{\mbox{\tiny eff}} \cos(a\kappa) \;, \quad J_{\mbox{\tiny eff}}= {\cal J}_q\left(\frac{aF_\omega}{\hbar\omega_B}\right) \;,
\end{equation}
where ${\cal J}_q(z)$ is the $q$th order Bessel function and $F_\omega$ the driving amplitude. Rigorously, Eq.~(\ref{g2}) defines a quasi energy and not an energy.
In the 2D lattice under our consideration, the role of multi-photon resonance is played by the resonance between harmonics of Bloch frequencies associated with the two degrees of freedoms, so that a similar theory can be developed and the observed oscillations justified. In the next section we derive an analogue of Eq.~(\ref{g2}) by employing a perturbation approach borrowed from the theory of classical dynamical system.
%#############################################################

%%%%%%%%%%%%%%%%%%%%%%%%%%%%%%%%%%%%%%%%%%
\subsection{Staggered magnetic field}
\label{secB2}

To explain the results of the previous section it is convenient to consider a more general problem: a uniaxial staggered magnetic field, where hopping between $B$ sites in Fig.~\ref{figB1} and Eq. (\ref{g1}) involves a phase $\phi$, {\em i.e.}, $J_y \rightarrow J_y\exp(i\phi)$ \footnote{A uniaxial staggered field, with $\phi=\pi/4$, has been recently realized in cold atoms experiments \cite{Aide11}.}. The eigenvalue problem (\ref{g1}) can be seen as a particular case of this problem, when $\phi=\pi$, whence the name of $\pi$-flux.

We can analyze this problem following closely the steps already described in Sec.~\ref{secA2a} and  Sec.~\ref{secA4b}. First, we rotate coordinates to align the electric field with the vertical axis $\xi$ of a new coordinate frame $(\eta,\xi)$ and write the wave function as a plane wave in the $\eta$-direction. This results in a system of algebraic equations, parameterized by the transverse quasimomentum $\kappa$:
%***************************************************
\begin{eqnarray}
\nonumber
-\frac{J_x}{2}\left(b^B_{p-q}e^{-ir\tilde{d}\kappa} + b^B_{p}\right)
- \frac{J_y}{2}\left(b^A_{p-r}e^{iq\tilde{d}\kappa} + b^A_{p+r}e^{-iq\tilde{d}\kappa}\right)  + \tilde{d}\tilde{F}p b^A_{p}=Eb^A_{p} \;,\\
\label{g1a}
-\frac{J_x}{2}\left(b^A_{p+q}e^{ir\tilde{d}\kappa} + b^A_{p}\right)
- \frac{J_y}{2}\left(e^{i\phi}b^B_{p-r}e^{iq\tilde{d}\kappa} + e^{-i\phi}b^B_{p+r}e^{-iq\tilde{d}d\kappa}\right)  + \tilde{d}\tilde{F}\left(p+\frac{q}{2}\right) b^B_{p}=Eb^B_{p} \;,
\end{eqnarray}
where
%********************************************************
\begin{equation}
\tilde{d}=\frac{1}{\sqrt{r^2+q^2}} \;, \quad \tilde{F}=F\sqrt{\frac{r^2+q^2}{r^2+(q/2)^2}} \;,
\end{equation}
and $q$ is an even number. Notice that in  Eqs.~(\ref{g1a}) the orientation of the electric field is defined with respect to the primary axis of the double periodic lattice. Thus, for example, a direction of 45 degrees in the $(x,y)$ coordinate system corresponds to $(r,q)=(1,2)$. Albeit similar to Eq.~(\ref{f8}) in Sec.~\ref{secA4b}, the system of equations (\ref{g1a}) is actually simpler, because it lacks the  $p$-dependent phase factor in front of the amplitudes $b_p$. The price to pay for this simplification is the appearance of the sublattice index.

Next, in full analogy with Eq.~(\ref{c2a}) in Sec.~\ref{secA2a}, we introduce the generating functions
%********************************************************
\begin{equation}
   \label{g1c}
Y^{A,B}(\vartheta)=\frac{1}{2\pi}\sum_p b_p^{A,B}\exp(i\tilde{d}p\vartheta) \;,
\end{equation}
that transform the algebraic equations (\ref{g1a}) into the system of ordinary differential equations
%********************************************************
\begin{equation}
\label{g1d}
i\tilde{F}\frac{{\rm d}}{{\rm d}\vartheta} \left(
\begin{array}{c}
Y^A\\Y^B
\end{array}
\right)=\left(
\begin{array}{cc}
E+J_y\cos[\tilde{d}(q\kappa-r\vartheta)+\phi] & J_x(1+\exp[i\tilde{d}(r\kappa+q\vartheta)])/2 \\
J_x(1+\exp[-i\tilde{d}(r\kappa+q\vartheta)])/2 & E-\tilde{d}\tilde{F}q/2 + J_y\cos[\tilde{d}(q\kappa-r\vartheta)] 
\end{array}
\right)\left(
\begin{array}{c}
Y^A\\Y^B
\end{array}
\right)\;.
\end{equation}
Observe that, for any given $\kappa$, the functions $Y^{A,B}$ must be periodic functions of $\vartheta$, {\em i.e.} we require that $Y^{A,B}(\vartheta;\kappa)=Y^{A,B}(\vartheta+2\pi/\tilde{d};\kappa)$. This defines two equidistant spectra with level spacing  $\tilde{d}\tilde{F}$. When $\kappa$ is varied, these energy levels draw the energy bands. If $\phi=0$ it can be proved that these bands are flat, see Eq.~(\ref{g1h}) below. This fact can be also deduced without explicit calculations by noting that for $\phi=0$ the system (\ref{g1a}) corresponds to a tilted square lattice, whose energy spectrum is given  by Eq.~(\ref{c1}), that in turn implies degeneracy of every energy level, as soon as $F_x/F_y$ is a rational number.

%%%%%%%%%%%%%%%%%%%%%%%%%%%%%%%%%%%%%%%%%%
\subsubsection{Krilov-Bogoliubov-Mitropolskii technique}
\label{secB2a}

To find the energy bands analytically we have to solve Eq.~(\ref{g1d}) for $\phi\ne0$. In this subsection we discuss the approach of \cite{96}, which is based on the observation that the system (\ref{g1d}) can be formally viewed as a classical dynamical system, in which the variable $\vartheta$ plays the role of time. This enables us to employ the Krilov-Bogoliubov-Mitropolskii (KBM) technique \cite{Mitr71} from the theory of dynamical systems. In essence, this method consists of an averaging technique, together with an expansion in powers of a small parameter.

To simplify Eq.~(\ref{g1d}), we apply a sequence of substitutions:
$Y^A\rightarrow Y^A \exp(-iE\vartheta/\tilde{F})$, $Y^B\rightarrow Y^B \exp(-iE\vartheta/\tilde{F}) \exp(i\tilde{d}(q\vartheta+r\kappa)/2)$;
$y^A=Y^A+Y^B$, $y^B=Y^A - Y^B$;  and
$y^A\rightarrow y^A\exp(-i\tilde{F}^{-1} \int [J_y\cos(\phi/2)\cos(\Lambda_1+\phi/2)+J_x\cos \Lambda_2]{\rm d}\vartheta$,
$y^B\rightarrow y^B\exp(-i\tilde{F}^{-1} \int [J_y\cos(\phi/2)\cos(\Lambda_1+\phi/2)-J_x\cos \Lambda_2]{\rm d}\vartheta$,
where $\Lambda_1$ and $\Lambda_2$ are defined below, in Eq.~(\ref{g1f}). In terms of the new functions $y^{A,B}(\theta)$ the system of differential equations (\ref{g1d}) takes the form:
%********************************************************
\begin{equation}
\label{g1e}
i\frac{{\rm d}}{{\rm d}\vartheta} \left(
\begin{array}{c}
y^A\\y^B
\end{array}
\right)=\frac{1}{\tilde{F}}\left(
\begin{array}{cc}
0 & G \\ G^* & 0
\end{array}
\right)\left(
\begin{array}{c}
y^A\\y^B
\end{array}
\right) \;,
\end{equation}
where
%***************************************************
\begin{equation}
\label{g1f}
G=-J_y\sin\left(\frac{\phi}{2}\right) \sin\left(\Lambda_1+\frac{\phi}{2}\right)
\exp\left(i\frac{4J_x}{F} \frac{\sqrt{r^2+(q/2)^2}}{q} \sin\Lambda_2 \right) \;, \quad
\Lambda_1=\tilde{d}(q\kappa-r\vartheta) \;,\quad \Lambda_2=\frac{\tilde{d}}{2}(r\kappa+q\vartheta) \;.
\end{equation}
Notice that for $\phi=0$ Eq.~(\ref{g1e}) has the trivial, constant solution $y^{A,B}(\vartheta)=y_0^{A,B}$, where $y_0^{A,B}$ are the initial data.

Let us now  apply the KBM method. Restricting ourselves to second order in the parameter $\epsilon=1/\tilde{F}$, the KBM equation for the column function $(y^A,y^B)^T$ reads
%*******************************************************
\begin{equation}
\label{gg}
i\frac{{\rm d}}{{\rm d} \vartheta}\left(
\begin{array}{c}
  y^A\\ y^B
\end{array}\right)
=\epsilon\begin{pmatrix}
  0 & \langle G\rangle \\
  \langle G^*\rangle & 0
\end{pmatrix} \begin{pmatrix}
  y^A \\ y^B
\end{pmatrix}
+\epsilon^2\begin{pmatrix}
  \langle-iG'^{*}G\rangle & 0 \\
  0 &  \langle-iG'G^{*}\rangle
\end{pmatrix}\begin{pmatrix}
  y^A\\ y^B
\end{pmatrix}  \;,
\end{equation}
where angular brackets denote the average over the period $2\pi/\tilde{d}$ and the prime sign is a shorthand notation for the integral of the non--constant component of $G$. Namely, expressing $G$ via its Fourier series coefficients $G_\nu$, we define $G'(\vartheta):=\sum_{\nu\ne0}\frac{\exp(i\nu\vartheta)}{i\nu} G_\nu$. The solution of Eq.~(\ref{gg}) is $(y^A,y^B)^T=\exp(-i\epsilon\lambda \vartheta) (y^A_0,y^B_0)^T$, where
%****************************************************
\begin{equation}
\label{g1g}
\lambda=\pm \sqrt{|\langle G\rangle|^2 + \epsilon^2|\langle G'G^*\rangle |^2 } \;.
\end{equation}
The quantity $\lambda=\lambda(\kappa)$ is the correction to the flat energy bands of the case $\phi=0$. Using the explicit form of the periodic function $G=G(\vartheta;\kappa)$ this correction can be expressed in terms of Bessel functions.
%#############################################################
\begin{figure}[t]
\includegraphics[width=8.5cm]{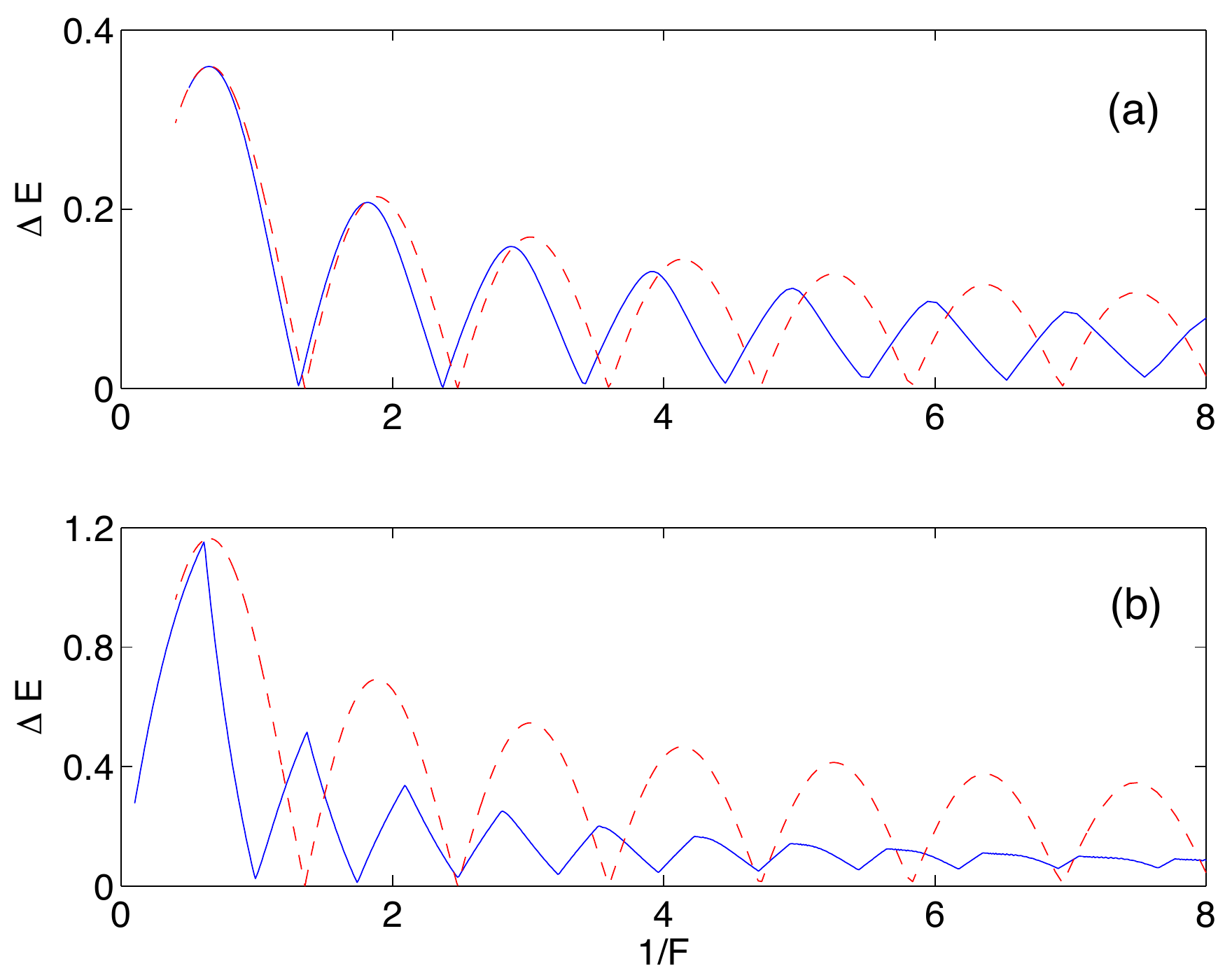}
\caption{Staggered magnetic field with $\phi=\pi/5$ (a) and $\phi=\pi$ (b). The width of the energy bands is shown versus $1/F$ for $\beta=F_x/F_y=1$. The dashed lines are the analytical result (\ref{g1h}).}
\label{figB7}
\end{figure}

For instance, let us consider the KBM corrections for  $(r,q)=(1,2)$, a direction that, as mentioned above, corresponds to the diagonal orientation of the electric field in Fig.~\ref{figB0}. Retaining only the first order in $\epsilon$, where according to Eq.~(\ref{g1g}) $\lambda=\langle G \rangle$, and using Eq.~(\ref{g1f}) we obtain the following expression for the bandwidth:
%****************************************************
\begin{equation}
\label{g1h}
\Delta E=\left| J_y\sin\left(\frac{\phi}{2}\right) {\cal J}_1\left(\frac{2\sqrt{2}J_x}{F}\right) \right| \;.
\end{equation}
Eq.~(\ref{g1h}) predicts a non-monotonic behavior of the bandwidth, just as Eq.~(\ref{g2}) does the same for a driven 1D lattice. However, while the latter is an exact result, Eq.~(\ref{g1h}) is a perturbative estimate valid for large $F$ and small $\phi$. Comparison of the analytical expression (\ref{g1h}) with the numerical solution of Eq.~(\ref{g1a}) is shown in Fig. \ref{figB7}, for $\phi=\pi/5$. It reveals a satisfactory agreement in the interval  $0\le 1/F\le2$, which includes two bandwidth maxima and one collapse point at $1/F\approx1.3$. In the same figure, when $\phi=\pi$, quantitative agreement takes place only till $1/F \simeq 1/2$, which corresponds to the first maximum.  Of course, these validity intervals can be further extended into the region of small $F$ by including second and higher order corrections.

%%%%%%%%%%%%%%%%%%%%%%%%%%%%%%%%%%%%%%%%%%%%%%%%%%%%%%%%%%%%
\subsection{Wave packet dynamics: rational {vs.} irrational $\beta$}
\label{secB3}

Let us now consider the dynamics of a wave--packet in real space. It is helpful to think of a Gaussian shaped packet, although the analysis is not restricted to this case. To characterize such wave-packet we use the moments of the position operator over the discrete lattice. Define indeed
$M_{1,x}:=\sum_{l,m} l|\psi_{l,m}(t)|^2$,  $M_{1,y}:=\sum_{l,m} m|\psi_{l,m}(t)|^2$ and $M_2:=\sum_{l,m} (l^2+m^2)|\psi_{l,m}(t)|^2$. Clearly, $M_{1,x}$ and $M_{1,y}$ provide the position of the center of mass of the packet, while  $M_2$ is a measure of its variance. Using these values we also define the dispersion
%****************************************************
\begin{equation}
\label{g3}
\sigma=\sqrt{M_2-M_{1,x}^2 -M_{1,y}^2} \;.
\end{equation}

\begin{figure}[t]
\includegraphics[width=8.5cm]{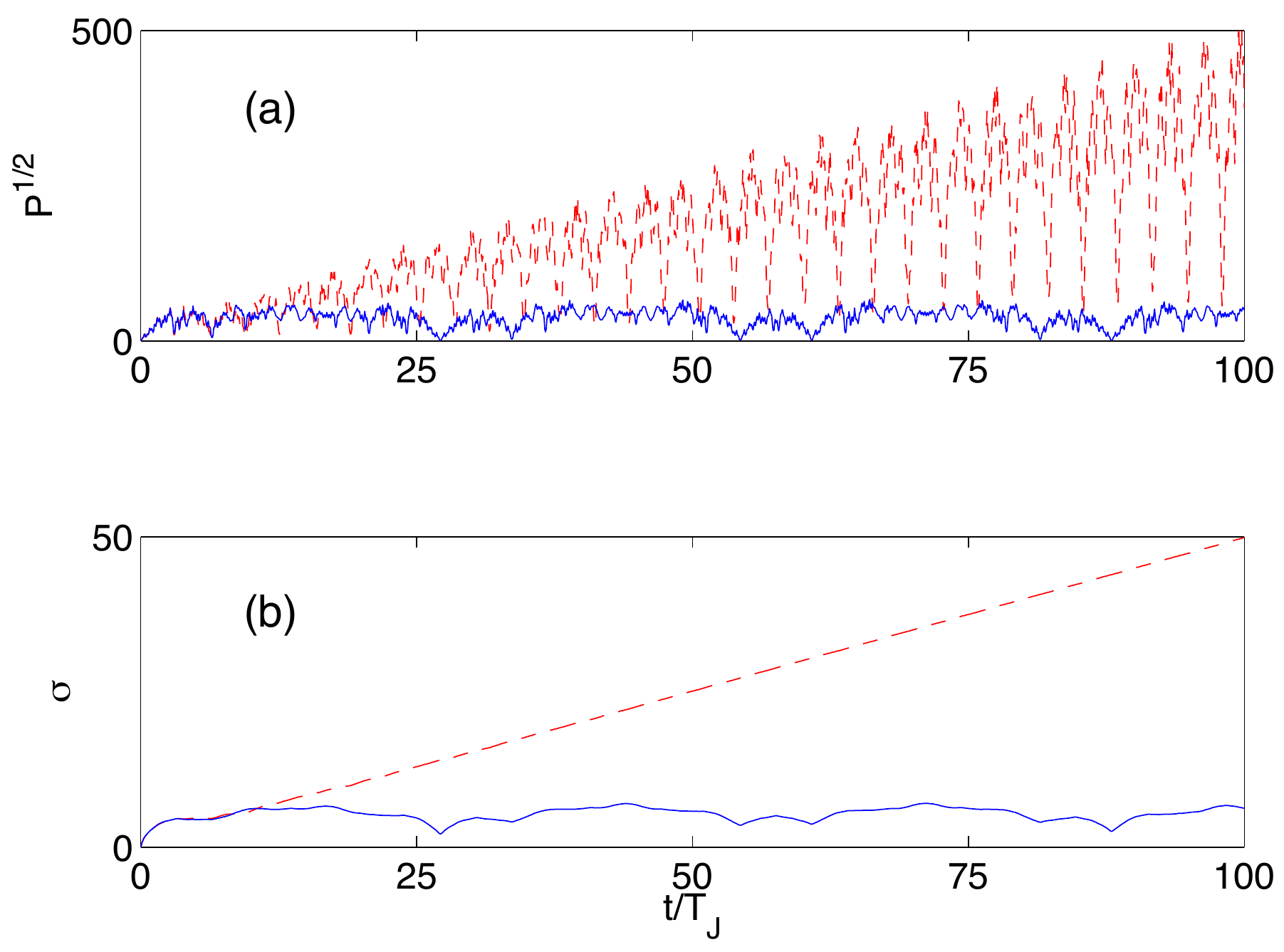}
\includegraphics[width=8.5cm]{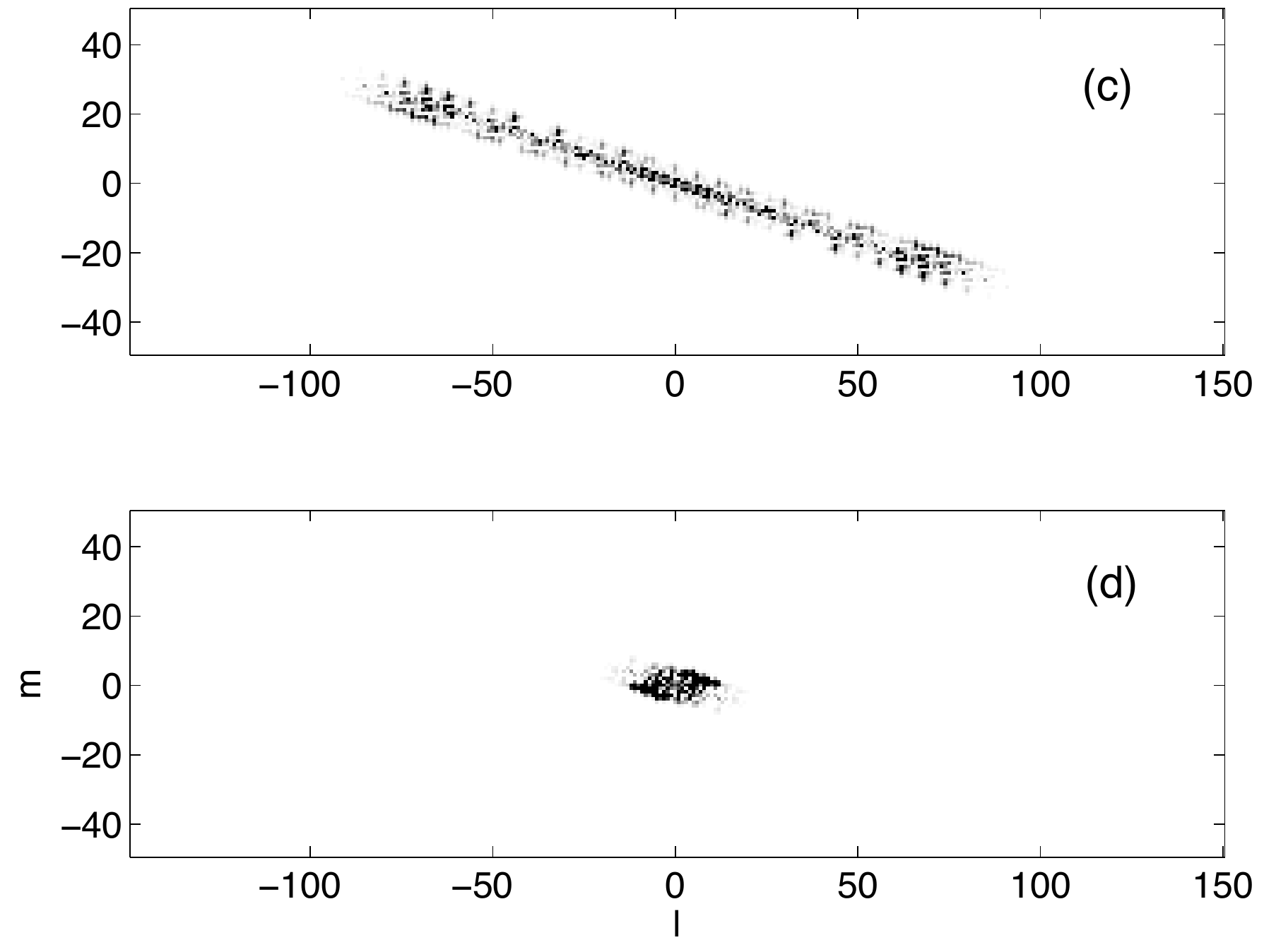}
\caption{Panels (a) and (b): Square root of the participation ratio (\ref{g4}), panel (a), and dispersion (\ref{g3}), panel (b), for $\beta=1/3$, red dashed lines, and $\beta=(\sqrt{5}-1)/4$, blue solid lines. Time is measured in units of the tunneling period $T_J=2\pi/J$. The intensity of the electric field is $F=0.5$. Initially only one site at the lattice origin has been populated. Panels (c) and (d): Occupation probabilities (in grey tones)  $|\psi_{l,m}(t)|^2$ of lattice sites at $t=100T_J$, for rational $\beta=1/3$ (panel c) and irrational $\beta=(\sqrt{5}-1)/4$ (panel d).}
\label{figB3}
\end{figure}

Let us first consider these dynamics when the parameter $\beta$ takes on rational values. Since in this case the energy spectrum is continuous, the dynamics of any localized wave packet is a ballistic spreading, in which asymptotically the dispersion $\sigma(t)$ grows linearly, $\sigma(t)\simeq At$, for large values of time $t$. This quantity is plotted as a red dashed line in Fig.~\ref{figB3}(b), for $\beta=1/3$, while Fig.~\ref{figB3}(c) shows the population of lattice sites at $t=100T_J$, for the same value of $\beta$. The initial wave-function in these simulations is localized at the lattice origin: $\psi_{l,m}(t=0)=\delta_{l,0}\delta_{m,0}$. The coefficient $A$ in the asymptotic linear growth of the dispersion is obviously proportional to the bandwidth $\Delta E$ shown in Fig.~\ref{figB1}(c). For example,  by choosing the value $F=1/1.85$,  where bandwidth is twice larger than for $F=0.5$,  the final value of dispersion is $\sigma(t=100T_J)=108$.
%#############################################################

In addition to the wave-packet dispersion, we also plot in Fig.~\ref{figB3}, panel (a), the time evolution of the participation ratio, defined as
%***************************************************
\begin{equation}
\label{g4}
P[\psi(t)]=\left(\sum_{l,m} |\psi_{l,m}(t)|^4 \right)^{-1} \;.
\end{equation}
This quantity is a measure of the number of sites which are significantly populated: for this reason, it has been frequently employed in theoretical and numerical studies. Unlike the dispersion, it possesses fine features that are sensitive to the initial condition. However, on average $P[\psi(t)]$ grows quadratically in time, when $\beta$ is rational (red dashed line in the figure), in the same way as the second moment.

The situation drastically changes when considering an irrational value of $\beta$. The blue solid curves in Fig.~\ref{figB3}(a-b) refer to the case $\beta=(\sqrt{5}-1)/4\approx0.3090$, which is close to $\beta=1/3\approx0.3333$ considered above. Fig.~\ref{figB3}(d) shows the wave packet at $t=100T_J$. For irrational values of $\beta$, LS-states are localized, and the spectrum is discrete. Thus the dispersion $\sigma(t)$ and the participation ratio $P[\psi(t)]$ must show saturation, that is, they cannot grow indefinitely, a fact that is clearly observed in the numerical data. The saturation level of both quantities is obviously determined by the  localization length of LS-states. Therefore, the main theoretical question to be answered in the case of irrational $\beta$ concerns the scaling properties of the localization length of LS-states, to which we now turn.

%%%%%%%%%%%%%%%%%%%%%%%%%%%%%%%%%%%%%%%%%%
\subsubsection{Localization length}
\label{secB3a}

\begin{figure}[t]
\includegraphics[width=8.5cm]{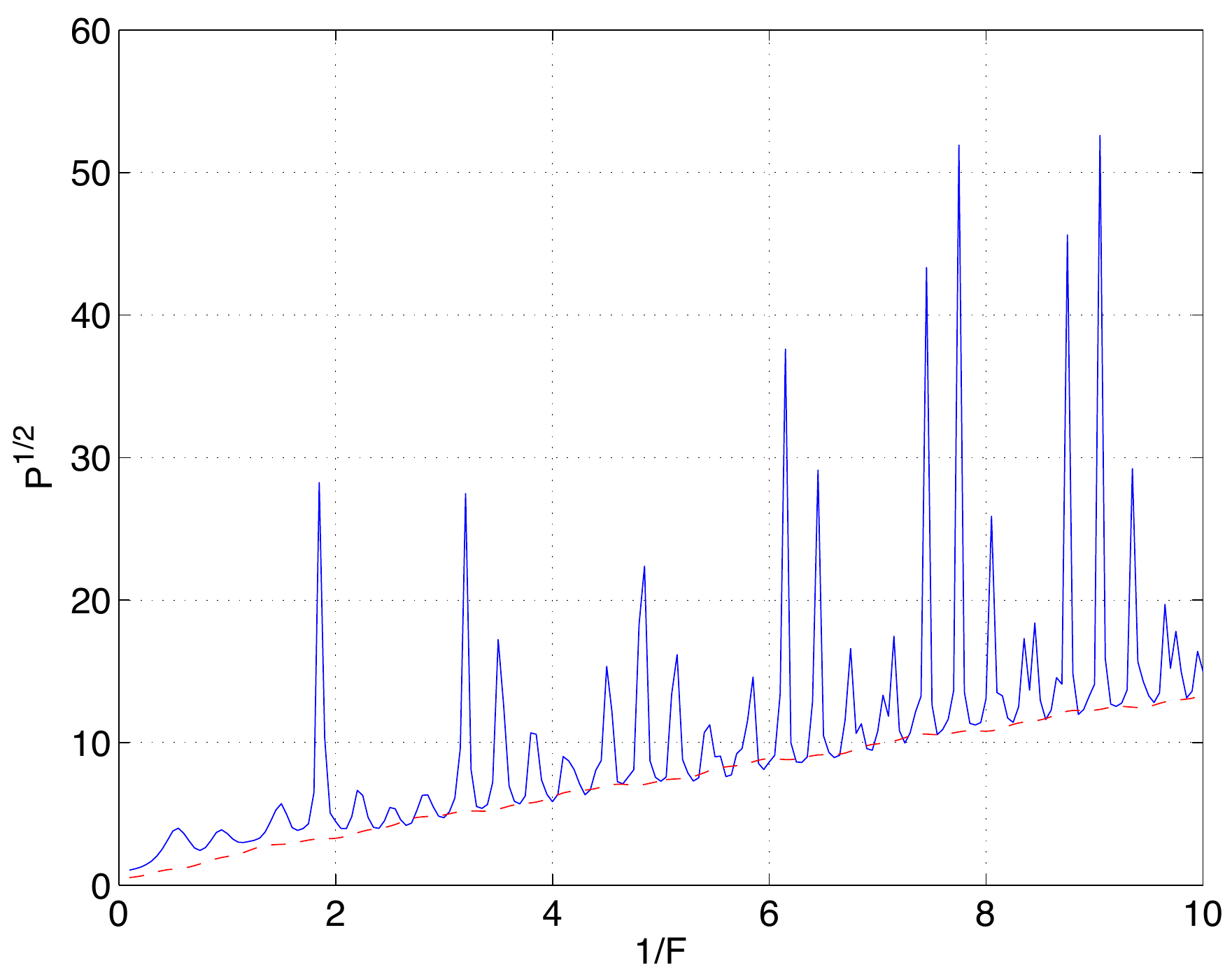}
\caption{Square root of the participation ratio $P[\Psi]$ of localized LS-states ($\alpha=1/2$), solid line, and WS-states ($\alpha=0$), dashed line, versus $1/F$, for $\beta=(\sqrt{5}-1)/4$. The participation ratio of WS-states is scaled by a factor four.}
\label{figB5}
\end{figure}

To characterize the localization property of LS-states we can use the  participation ratio introduced above: we compute $P[\Psi^{(n,k)}]$, following Eq. (\ref{g4}).
%***************************************************
%\begin{equation}
%\label{g5}
%P=\left(\sum_{l,m} |\Psi^{(n,k)}_{l,m}|^4 \right)^{-1} \;.
%\end{equation}
%
Notice that according to Eq.~(\ref{e2}) all states $\Psi^{(n,k)}$ are characterized by the same participation ratio, so that the quantum numbers $n$ and $k$ can be omitted from the notation. A reference frame for the quantity $P[\Psi]$ is provided by the participation ratio of WS-states ($\alpha=0$). It follows from Eq.~(\ref{c3}) and from the properties of Bessel functions that these states are  essentially null outside a rectangular region of sides $J_x/F_x$ and $J_y/F_y$; therefore, their participation ratio is proportional to the product of these numbers and scales as $1/F^2$.  The solid and dashed lines in Fig.~\ref{figB5} report this quantity for LS-states and WS-states, respectively. A remarkable feature of LS-states is the presence of a resonance-like structure, super-imposed to the general trend given by the law
%***************************************************
\begin{equation}
\label{g5b}
P\simeq 1/F^2 \;.
\end{equation}
Comparing Fig.~\ref{figB5} with Fig.~\ref{figB1}(c) we notice that resonance peaks are positioned at the values of $F$ where the bandwidth of extended LS-states, for rational $\beta=1/3$, are locally maximal. This fits physical intuition: in fact, band collapses and revivals are originated by destructive and constructive interference. Now, for irrational $\beta$, interference is always destructive. However, the overall destructive interference does not prevent constructive interference to build up in finite regions of configuration space.  We believe that this is indeed the case, when an irrational value $\beta$ is well approximated by a simple rational number $\beta'$, and the value of $F$ is adjusted in such a way to provide constructive interference for $\beta'$.
%#############################################################

Concluding this section we stress that the scaling law (\ref{g5b}) is not universal and is valid only for the currently considered Peierls phase, $\alpha=1/2$. It will be shown later, in Sec.~\ref{secC4}, that for $\alpha\ll 1$ a completely different scaling law describes the localization length.

%%%%%%%%%%%%%%%%%%%%%%%%%%%%%%%%%%%%%%%%%%%
\subsection{Bloch oscillations of a delocalized packet}
\label{secB4}

To study BO of a delocalized wave packet we first need to know the energy dispersion relation. This can be found by setting  $F=0$ in Eq.~(\ref{g1}) and  using the substitution
%***************************************************
\begin{equation}
   \label{g6}
\Phi_{l',m'}=\left(
\begin{array}{c}
\psi^A \\ \psi^B
\end{array} \right)
e^{i2\kappa_x l'}e^{i\kappa_y m'} \;,
\end{equation}
where the factor two in the exponent takes into account the doubling of the lattice period. After elementary calculations one arrives at the relation
%***************************************************
\begin{equation}
\label{g7}
E_{1,2}(\kappa)=\mp\sqrt{J_x^2\cos^2(\kappa_x) + J_y^2\cos^2(\kappa_y)} \;.
\end{equation}
The dispersion relation (\ref{g7}) is plotted  in Fig.~\ref{figA1}(b) in the extended Brillouin zone $-\pi<\kappa_x,\kappa_y\le \pi$.  A remarkable feature of this dispersion relation is the presence of Dirac's cones, which unveil the links of this problem to the phenomena of BO in honeycomb and honeycomb-like lattices \cite{92,Krue12,Tarr12} and BO in simple square lattices in the presence of spin-orbit coupling \cite{Lars10,Zhan12}.

%#############################################################
\begin{figure}[t]
\includegraphics[width=8.5cm]{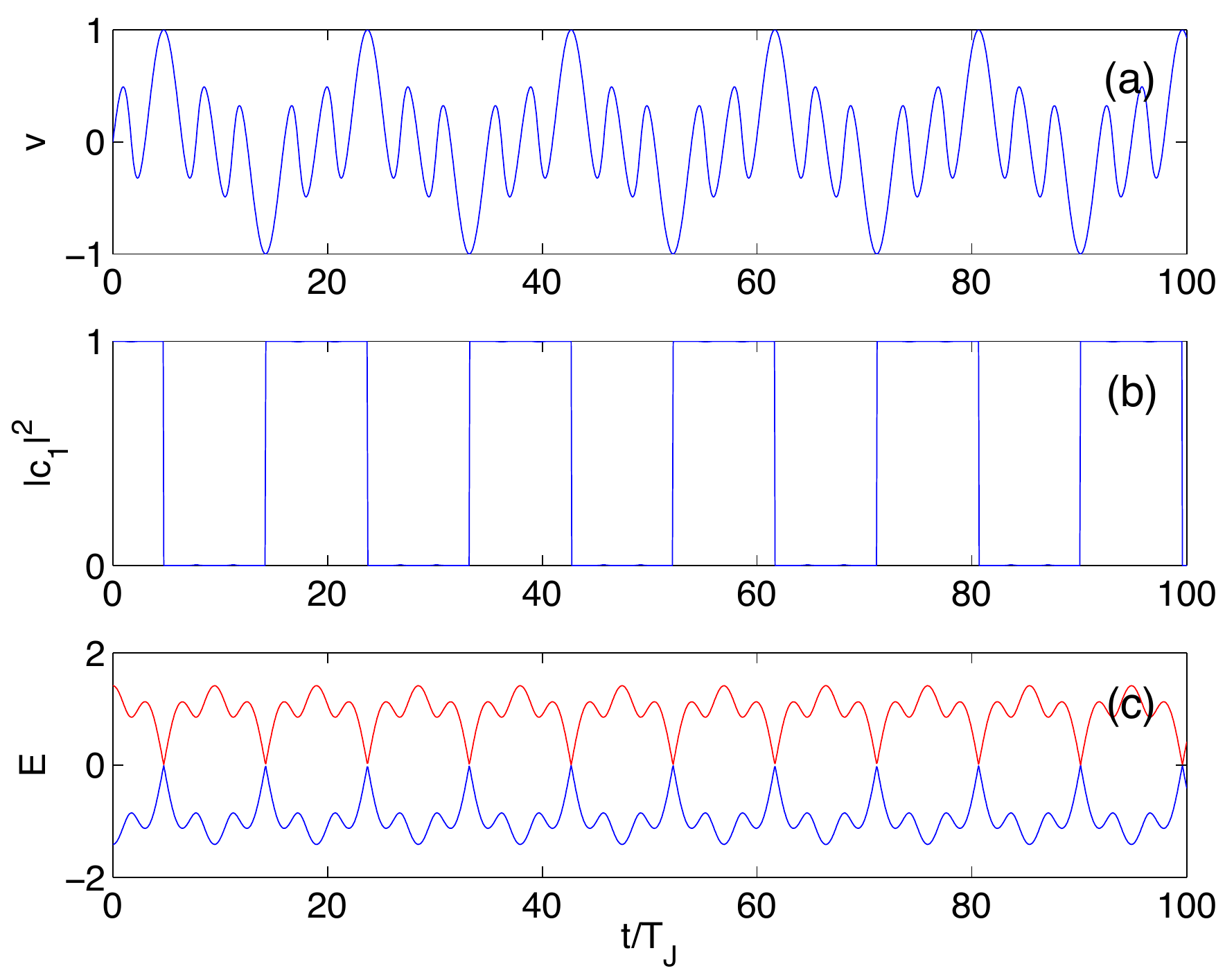}
\includegraphics[width=8.5cm]{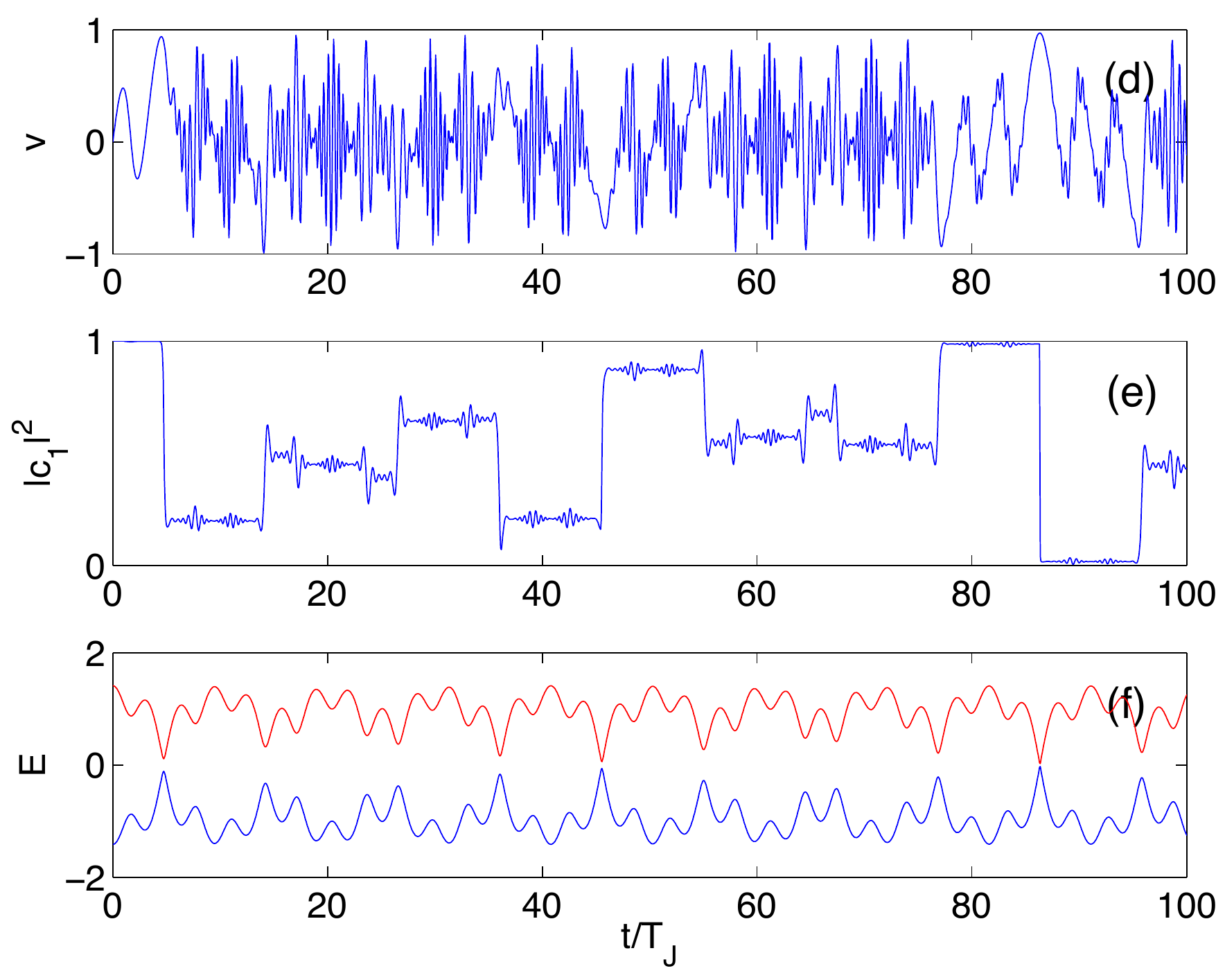}
\caption{Projection of the mean velocity $v(t)$ on the direction of the electric field ${\bf F}$, panels (a) and (d), population of the lower band $|c_1(t)|^2$, (b) and (e), and the dispersion relation (\ref{g7}) along the line $\kappa={\bf F}t$, (c) and (f). Parameters are $F=1/6$ and $\beta=1/3$, panels (a-c), and $\beta=(\sqrt{5}-1)/4$, panels (d-f).}
\label{figB6a}
\end{figure}

The next step of the analysis is to extend the Brillouin zone by periodicity to the whole plane and to draw the line $\kappa(t)=\kappa_0-{\bf F}t$. If $\beta=F_x/F_y$ is a rational number, this path either goes in between Dirac's points, or it exactly encounters them. In the limit of small $F$, we recover the familiar Bloch oscillations (\ref{c7}), see Fig.~\ref{figB6a}(a-c).  The case of irrational $\beta$ is more complex. Here, the line $\kappa(t)$ passes arbitrarily close to Dirac points: thus, independently of the smallness of $F$, interband LZ-transitions occur. This is illustrated in the panel (e) in Fig.~\ref{figB6a}, which shows the population of the lower band, see also Eq.~(\ref{c9}) in Sec.~\ref{secA2b}. As a result, dynamics of the mean velocity does not obey Eq.~(\ref{c7}) but is a complex quasi-periodic process resembling a random process Ref.~\cite{92}.

\begin{figure}[t]
\includegraphics[width=8.5cm]{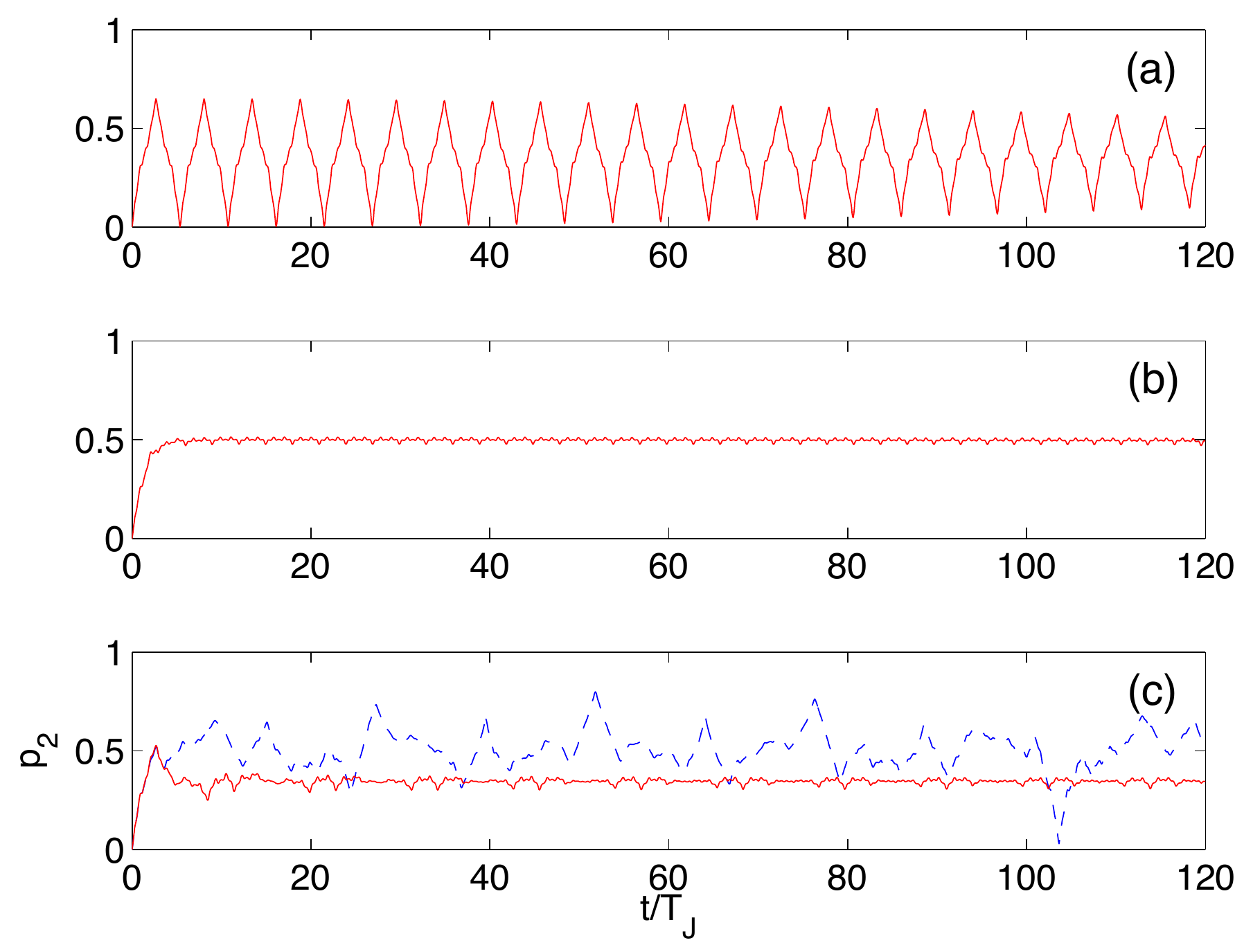}
\includegraphics[width=8cm]{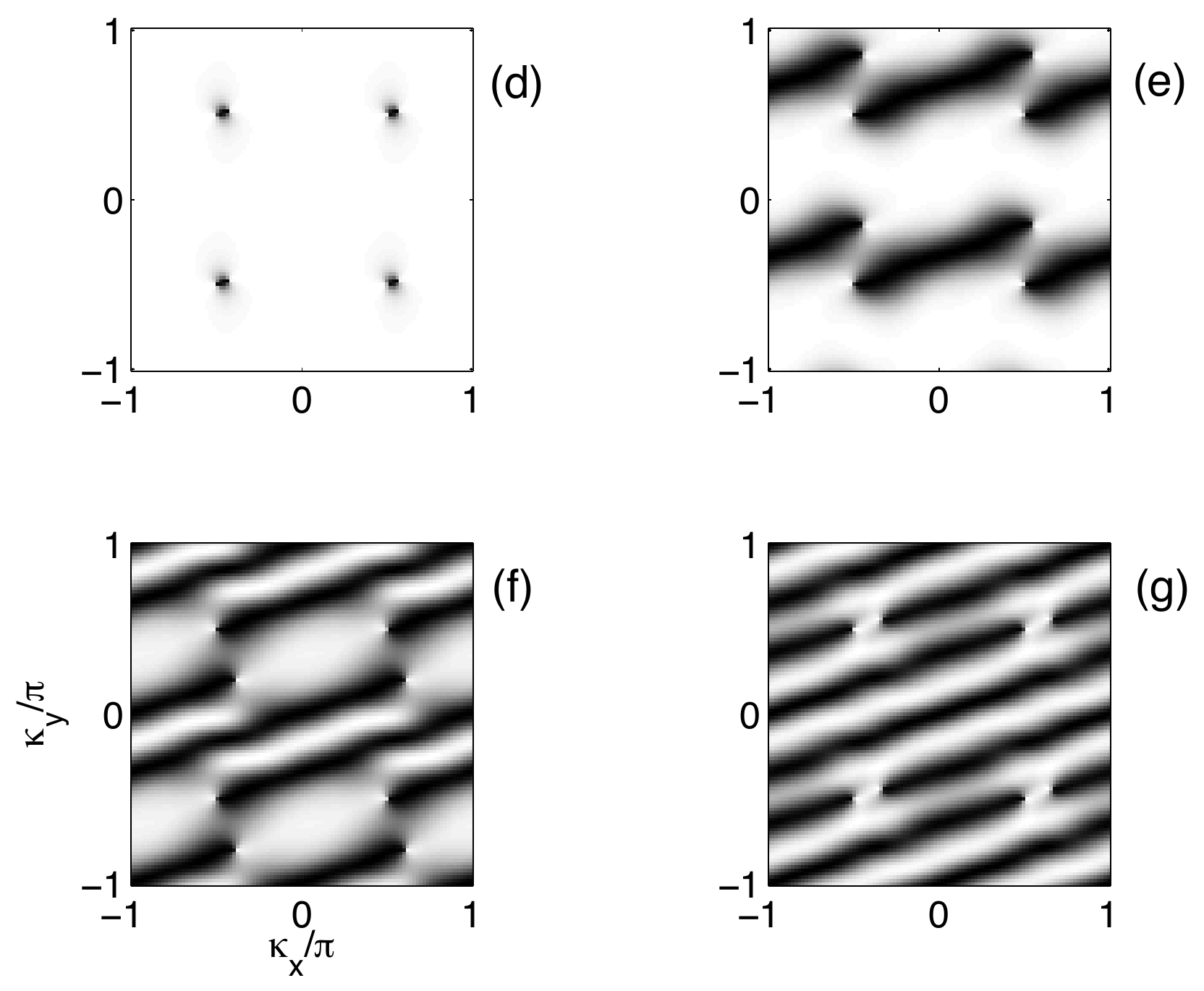}
\caption{Total population of the upper band as a function of time for $\beta=1/3$ and $F=1/1.7$ (a), $F=1/8.9$ (b) and  $F=1/8$ (c). Dashed line in the panel (c) refer to the irrational $\beta=(\sqrt{5}-1)/4$. Panels (d-g) show the short time dynamics (which is essentially the same for all considered cases) of the function $p_2(\kappa_x,\kappa_y; t)$ for $t/T_J=1/16, 1, 2, 3$.}
\label{figB9}
\end{figure}

It is important at this point to briefly discuss an application to ongoing laboratory experiments with cold atoms in optical lattices with Dirac's cones \cite{Tarr12}. To apply the theoretical results of the previous paragraph to real experiments, we need to perform an average over the quasimomentum distribution of the initial state of a laboratory quantum system. As an example, let us consider an ensemble of spin-polarized Fermi atoms, with Fermi energy $E_F=0$. For this Fermi energy the lower band in Eq.~(\ref{g7}) is completely filled, while the upper band is empty. As stated above, a static force induces interband transitions that predominantly take place in the vicinity of  Dirac points, see Fig.~\ref{figB9}(d-g). We are interested in the total population of the upper band,
%***********************************************************
\begin{equation}
p_2(t)=\int\int p_2(\kappa_x,\kappa_y; t){\rm d} \kappa_x {\rm d} \kappa_y \;,
\end{equation}
which is the quantity measured in the experiment \cite{Tarr12}. Solid lines in Fig.~ \ref{figB9} show $p_2(t)$  for rational $\beta=1/3$ and three different values of $F$:  $F=1/1.7$ (a), $F=1/8.9$ (b) and  $F=1/8$ (c). Notice that according to Fig.~\ref{figB1}(c) these values correspond to two particular cases,  where the energy bands of LS-states are almost flat and almost touch each other ({\em i.e.}, they take the maximal possible width), and one generic case where $0<\Delta E<F$. It is seen that, after averaging over the quasimomentum, we observe periodic oscillations of the band population only in the case of flat bands while for non-flat bands $p_2(t)$ rapidly approaches a constant value. The dashed line in panel (c) refers to irrational $\beta=(\sqrt{5}-1)/4\approx 1/3$. As mentioned above, here $p_2(t)$ is a quasi-periodic process where the population of the upper band periodically drops to almost zero.
%#############################################################

%This example also serves to illustrate the power of the theory of LS-states. In fact, without knowledge the spectrum of LS-states the observed qualitative changes in population dynamics, which happen at rather small variation of the system parameters, might come as a big surprise.

%%%%%%%%%%%%%%%%%%%%%%%%%%%%%%%%%%%%%%%%%
\section{The case $\alpha\ll 1$}
\label{secC}
%%%%%%%%%%%%%%%%%%%%%%%%%%%%%%%%%%%%%%%%%

In the previous section we have analyzed the system (\ref{b3}) for the Peierls phase $\alpha=1/q$ with $q=2$. In principle, the methods of Sec.~\ref{secB} can be generalized to cover cases with larger values of $q$, such as $q=3$ or $q=4$, to begin with. However, the analytical complexity of the methods increases tremendously with $q$, and the case $q\gg 1$ requires different techniques.  Among these, the semiclassical approach proves to be very useful.

%%%%%%%%%%%%%%%%%%%%%%%%%%%%%%%%%%%%%%%%%%
\subsection{The Driven Harper Model}
\label{secC1}

The tight-binding Hamiltonian (\ref{b3}) has a classical counterpart that can be derived as follows. The shift operators $\widehat{T}_x=\exp(\partial/\partial \tilde{x})$ and $\widehat{T}_y=\exp(\partial/\partial \tilde{y})$ permit to re-write this Hamiltonian in the form
%***************************************************
\begin{equation}
   \label{j1}
\widehat{H}=-J_x\cos\left(-i\frac{\partial}{\partial \tilde{x}}\right)
-J_y\cos\left(-i\frac{\partial}{\partial \tilde{y}}+2 \pi \alpha \tilde{x}\right)+F_x\tilde{x}+F_y\tilde{y} \;,
\end{equation}
where the fictitious continuous variables $\tilde{x}$ and $\tilde{y}$ are associated with the indexes $l$ and $m$, respectively. Next, we introduce the coordinate operators $\hat{x}=2\pi\alpha \tilde{x}$ and $\hat{y}=2\pi\alpha \tilde{y}$ and the momentum operators $\hat{p}_x=-i2\pi\alpha\partial/\partial x$ and $\hat{p}_y=-i2\pi\alpha\partial/\partial y$. The commutator of these operators is $i2\pi\alpha$ and, hence, the Peierls phase plays the role of an effective Planck constant in the semiclassical analysis:
%***************************************************
\begin{equation}
   \label{j2}
[\hat{x},\hat{p_x}]=i\hbar_{\rm eff} \;,\quad  [\hat{y},\hat{p_y}]=i\hbar_{\rm eff} \;,\quad       \hbar_{\rm eff}=2\pi\alpha \;.
\end{equation}
Finally, substituting quantum operators with classical canonical variables yields the classical Hamiltonian
%***************************************************
\begin{equation}
H_{cl}=-J_x\cos(p_x)-J_y\cos(p_y+x)+F'_x  x +F'_y y \;,
\end{equation}
where $F'_{x,y}=F_{x,y} /2\pi\alpha$. Instead of rescaling the electric field we can scale the hopping matrix elements:
%***************************************************
\begin{equation}
\label{j3}
H_{cl}=-J'_x\cos(p_x)-J'_y\cos(p_y+x)+F_x  x +F_y y \;, \quad J'_{x,y}=2\pi\alpha J_{x,y} \;.
\end{equation}
To be certain, we shall stay with the second option.

The 2D classical system (\ref{j3}) can be reduced to a 1D, time--dependent system by solving Hamilton's equation of motion for the momentum $p_y$: this is trivially done, yielding $p_y(t)=p_y(0)-F_y t$. We then have
%***************************************************
\begin{equation}
H_{cl}(t)=-J'_x\cos(p_x)-J'_y\cos(x-F_y t)+F_x  x  \;,
\end{equation}
where we dropped the irrelevant `initial phase' $p_y(0)$. Furthermore, using the canonical substitution $p_x \rightarrow p_x+F_xt$ the 1D Hamiltonian can be presented in symmetric form,
%***************************************************
\begin{equation}
\label{j4}
H_{cl}(t)=-J'_x\cos (p+F_x t) - J'_y\cos(x-F_y t) \;, \quad  p\equiv p_x \;,
\end{equation}
which reveals the gauge invariance of the original problem\footnote{It is easily proven that changing the Landau gauge from ${\bf A}\sim (0,x)$ to ${\bf A}\sim (-y,0)$ corresponds to exchanging coordinate and momentum in the Hamiltonian (\ref{j4}).}. In what follows we shall call the system (\ref{j4}) the {\em classical driven Harper model}. The {\em quantum} driven Harper model is obtained by quantizing the above Hamiltonian. The corresponding time dependent  Schr\"odinger equation is
%****************************************************************
\begin{equation}
\label{e9}
i\dot{b}_{l}=-\frac{J_x}{2}\left(b_{l+1}e^{iF_x t}+b_{l-1}e^{-iF_x t}\right) -J_y \cos(2\pi\alpha l +\kappa -F_y t) b_{l}  \;,
\end{equation}
which is directly related to the 1D Schr\"odinger equation (\ref{e8}) considered in Sec.~\ref{secA4}. In fact, one obtains these equation from each other by using the substitution $b_l(t)\leftrightarrow b_l(t)\exp(-iF_x lt)$.

A remark on the topology of the classical phase space is now in order. Our prime interest is the cylinder, $-\pi<p\le\pi$ and $-\infty<x<\infty$, where the dynamics of the classical system (\ref{j4}) can be compared with the dynamics of the quantum system (\ref{e9}). However, the system (\ref{j4}) can be formally studied also on the torus,  $-\pi<p,x\le\pi$, and on the plane,  $-\infty<p,x<\infty$. In the last case, it is easy to prove that the driven Harper model is completely integrable. Indeed, using the canonical substitution $p'=p+F_x t$ and $x'=x-F_y t$ the new Hamiltonian appears to be time-independent,
%*****************************************************
\begin{equation}
\label{j5}
H'_{cl}=-J'_x\cos(p') - J'_y\cos(x') +F_x x'+F_y p'  \;,
\end{equation}
and, hence, the right hand side of (\ref{j5}) is the global integral of the motion.
It is also instructive to note that, in terms of the Hamiltonian (\ref{j4}), the cyclotron frequency $\omega_c$ (\ref{b6}) is the frequency of small oscillations of the standard (not driven) classical Harper model,
%***************************************************
\begin{equation}
   \label{j4a}
H_{cl}=-J'_x\cos (p) - J'_y\cos(x) \;.
\end{equation}

Although the driven Harper is completely integrable for any set of parameters, it has qualitatively different dynamical regimes, depending on both the rationality of the parameter $\beta$ and the relative size of cyclotron frequency $\omega_c$ (\ref{b6}) and Bloch frequency $\omega_B$ (\ref{b7}). We now discuss these cases separately.
%#############################################################
\begin{figure}[t]
\includegraphics[width=10.5cm]{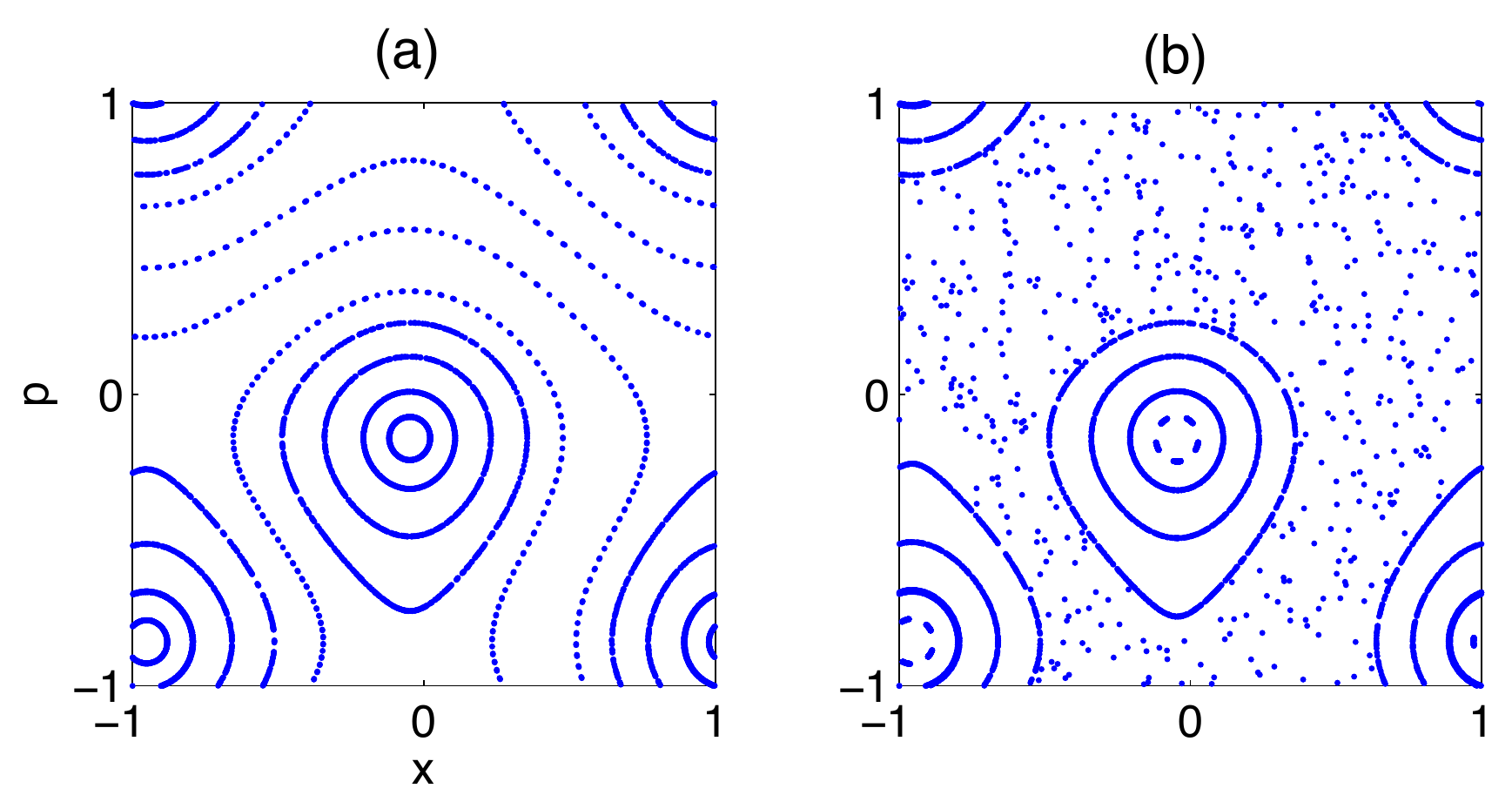}
\caption{A portion of phase space (stroboscopic map over $T_y=2\pi/F_y$) of the classical driven Harper model  for rational $\beta=1/3$ (a) and irrational $\beta=(\sqrt{5}-1)/4$ (b). The other parameters are $J'_{x,y}=2\pi \cdot 0.1545$ and $F=0.3$. Transporting islands are seen as stability islands surrounding elliptic points at $(x,p)\approx(0,0)$ and $(x,p)\approx(-\pi,-\pi)$. In the case of rational $\beta$ phase trajectories are closed on the torus and the stroboscopic map reproduces these trajectories.  For irrational $\beta$ any trajectory, which does not belong a stability island, never repeats itself on the torus and appears as a scattered array of points resembling -- rather, faking -- a chaotic trajectory.}
\label{figC1}
\end{figure}

%%%%%%%%%%%%%%%%%%%%%%%%%%%%%%%%%%%%%%%
\subsubsection{High-frequency regime}
\label{secC1a}

In the high-frequency regime, $\omega_B>\omega_c$, and for irrational $\beta$, any phase trajectory of the system (\ref{j4}) is bounded, implying that the mean velocity $\bar{v}$,
%***************************************************
\begin{equation}
   \label{j6}
\bar{v}=\lim_{t\rightarrow\infty} \frac{x(t)}{t} \;,
\end{equation}
is null. The particle can have nonzero mean velocity only if $\beta$ is a rational number. This can be proven using adiabatic perturbation theory, where one distinguishes between fast variables $p,x$ and slow variables $p',x'$. We demonstrate this for two particular cases: $\beta=0$ and $\beta=1$.

If $\beta=0$ the slow variable is $p'(t)\approx p_0$, where $p_0$ is the initial momentum. Then $x\approx x_0+J'_x\sin(p_0) t$, and $\bar{v}=J'_x\sin(p_0)$. If a classical ensemble of particles is uniformly distributed over the `elementary cell' $-\pi\le p,x <\pi$, we observe ballistic spreading, where the mean-squared displacement  $\sigma=\sqrt{\langle x^2 \rangle -\langle x \rangle^2}$ asymptotically follows a linear law, $\sigma(t)=At$, with $A=J'_x/\sqrt{2}$.

Next, consider the case $\beta=1$. As in the former, at zero order we have $p'(t)=p_0$. However, at first  order, the momentum $p'(t)=p_0+(J'_y/F_y)\cos(x+F_y t)$ is a periodic function of time.  Substituting this solution into Hamilton's equation for the conjugate variable $x$, we find $x(t)=J'_x\int_0^t \sin[p(t)-F_x t] {\rm d}t \sim t$, where the proportionality coefficient can be expressed  through the Bessel function ${\cal J}_1(J'_y/F_y)$.  This implies that the ensemble of particles defined above will possess a dispersion $\sigma(t)=At$ with $A\sim J'_x J'_y/F$. These rates of ballistic spreading are two particular cases of a general result,
%******************************************************
\begin{equation}
\label{j7}
A\sim F^{-(r+q-1)} \;,\quad F\gg \omega_c \;,
\end{equation}
that coincides with the rate of wave-packet spreading derived in Sec.~\ref{secA4d} using quantum perturbation theory.
%#############################################################
\begin{figure}[t]
\includegraphics[width=8.5cm]{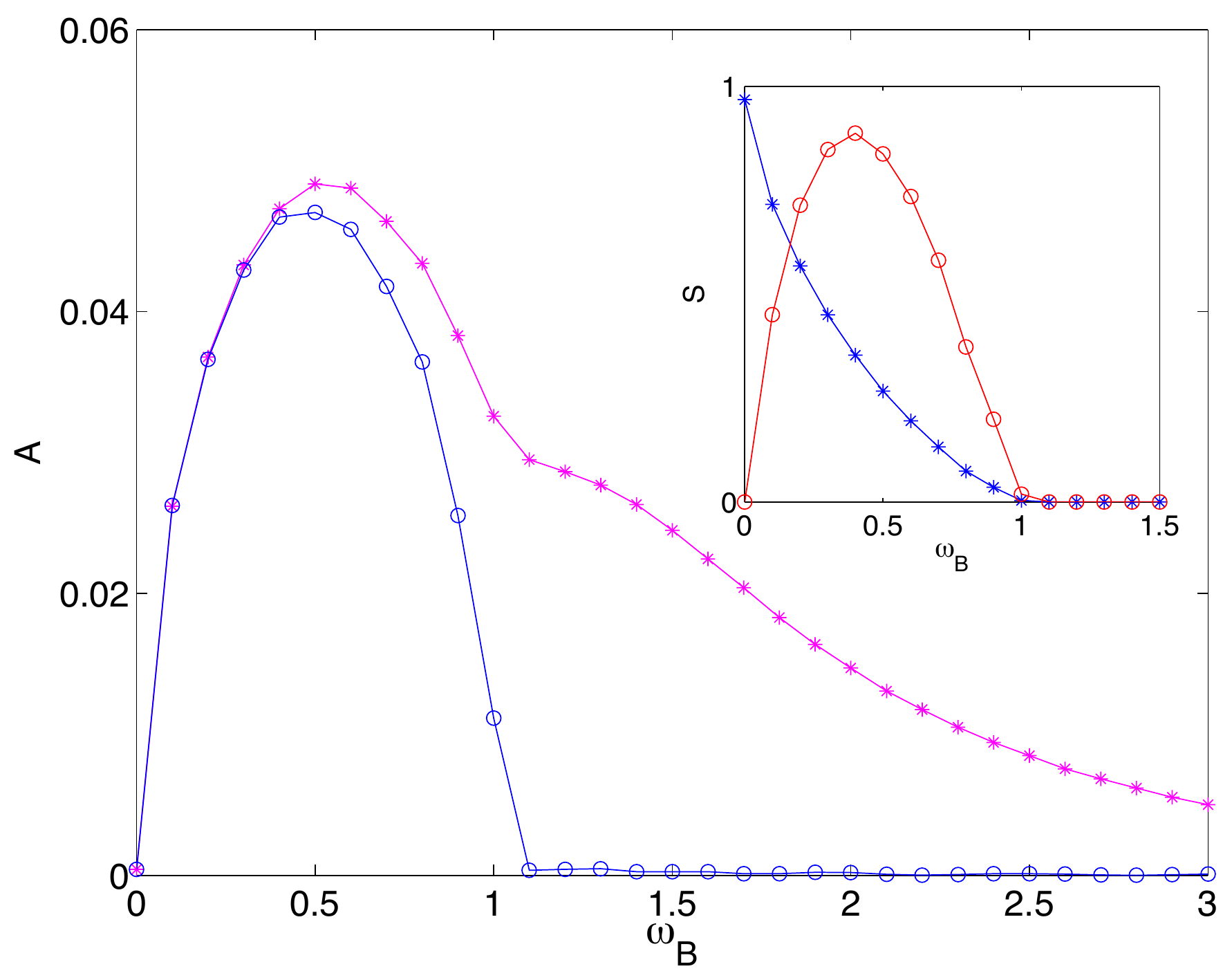}
\caption{Rate of ballistic spreading for an ensemble of classical particles, versus the Bloch frequency $\omega_B=F$, for rational $\beta=1/3$ (upper curve, stars) and irrational $\beta=(\sqrt{5}-1)/4$ (lower curve, open circles). The inset shows the relative size, $S$, of transporting islands (stars) and the function (\ref{j8}) (open circles), where we arbitrary set the proportionality coefficient to 4.}
\label{figC2}
\end{figure}

%%%%%%%%%%%%%%%%%%%%%%%%%%%%%%%%%%%%%%%
\subsubsection{Low-frequency regime}
\label{secC1b}

The regime of low-frequency driving, $\omega_B<\omega_c$, is more subtle than the previous, because here the phase space of (\ref{j4}) contains two chains of transporting islands, see Fig.~\ref{figC1}. Remark that these chains exist for both rational and irrational values of $\beta$. In a classical ensemble of particles, those with initial conditions lying in the transporting islands move in the negative direction, with velocity $\bar{v}=F_y$, while the others travel in the positive direction, and therefore $\sigma(t)=At$. At the same time, for the statistical ensemble under consideration, the mean current and, hence, the mean displacement vanish. The value of the proportionality coefficient $A$ obtained numerically is depicted in Fig.~\ref{figC2}, for the two values of $\beta$ used in Fig.~\ref{figC1}.  As expected, for small values of $F$ we find
%******************************************************
\begin{equation}
\label{j8}
A\sim F S(F) \;,
\end{equation}
where $S(F)$ is the relative size of transporting islands, see the inset in Fig.~\ref{figC2}.  If $\beta$ is a rational number this dependence leaves place, for large $F$, to the asymptotic dependence in Eq. (\ref{j7}). For irrational $\beta$  Eq.~(\ref{j8}) gives $A=0$ as soon as the transporting islands disappear, which is consistent with Eq. (\ref{j7}) as well.

%%%%%%%%%%%%%%%%%%%%%%%%%%%%%%%%%%%%%%%%%%%%%%
\subsection{Transporting states}
\label{secC2}

\begin{figure}[t]
\includegraphics[width=8.5cm]{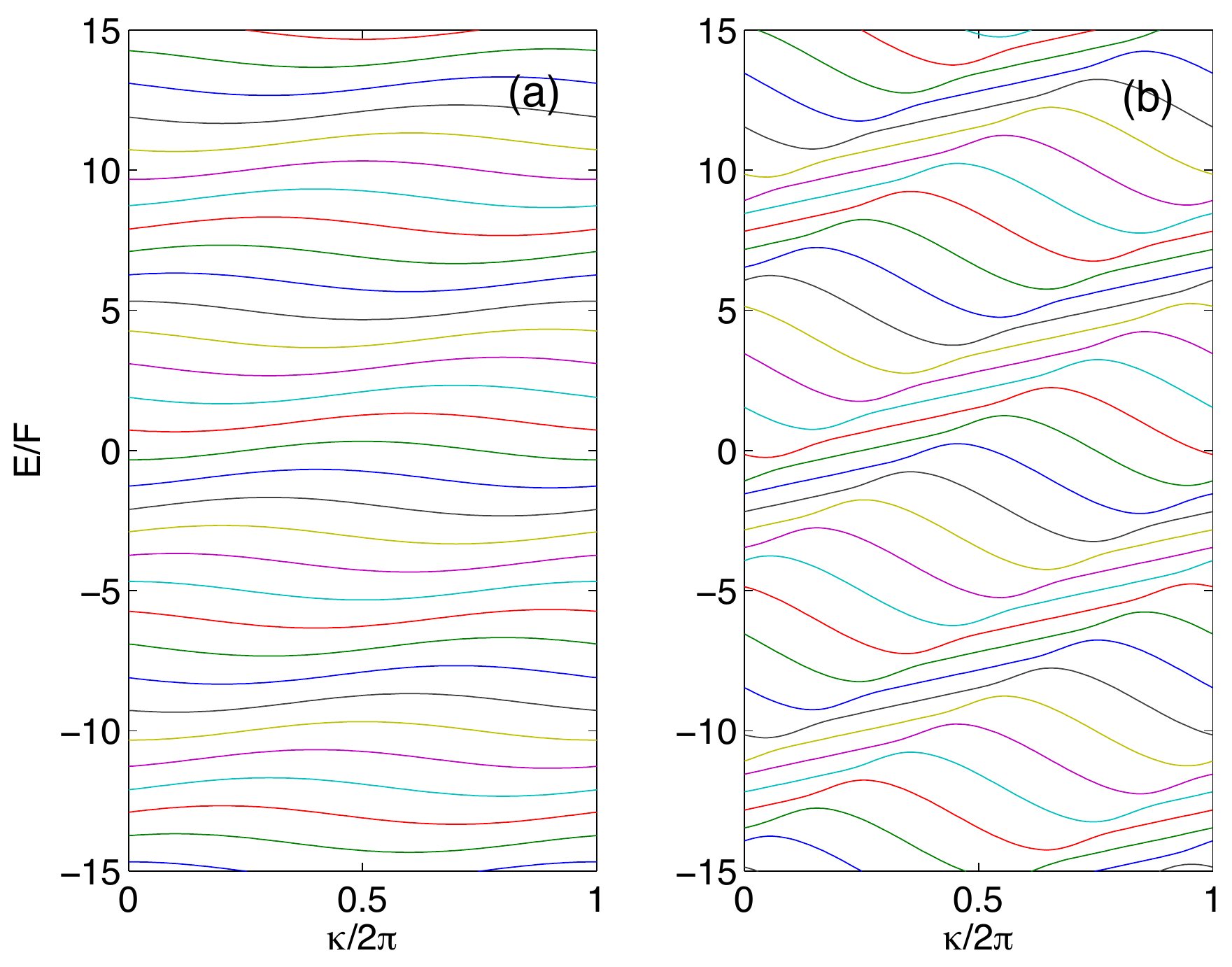}
\includegraphics[width=8.5cm]{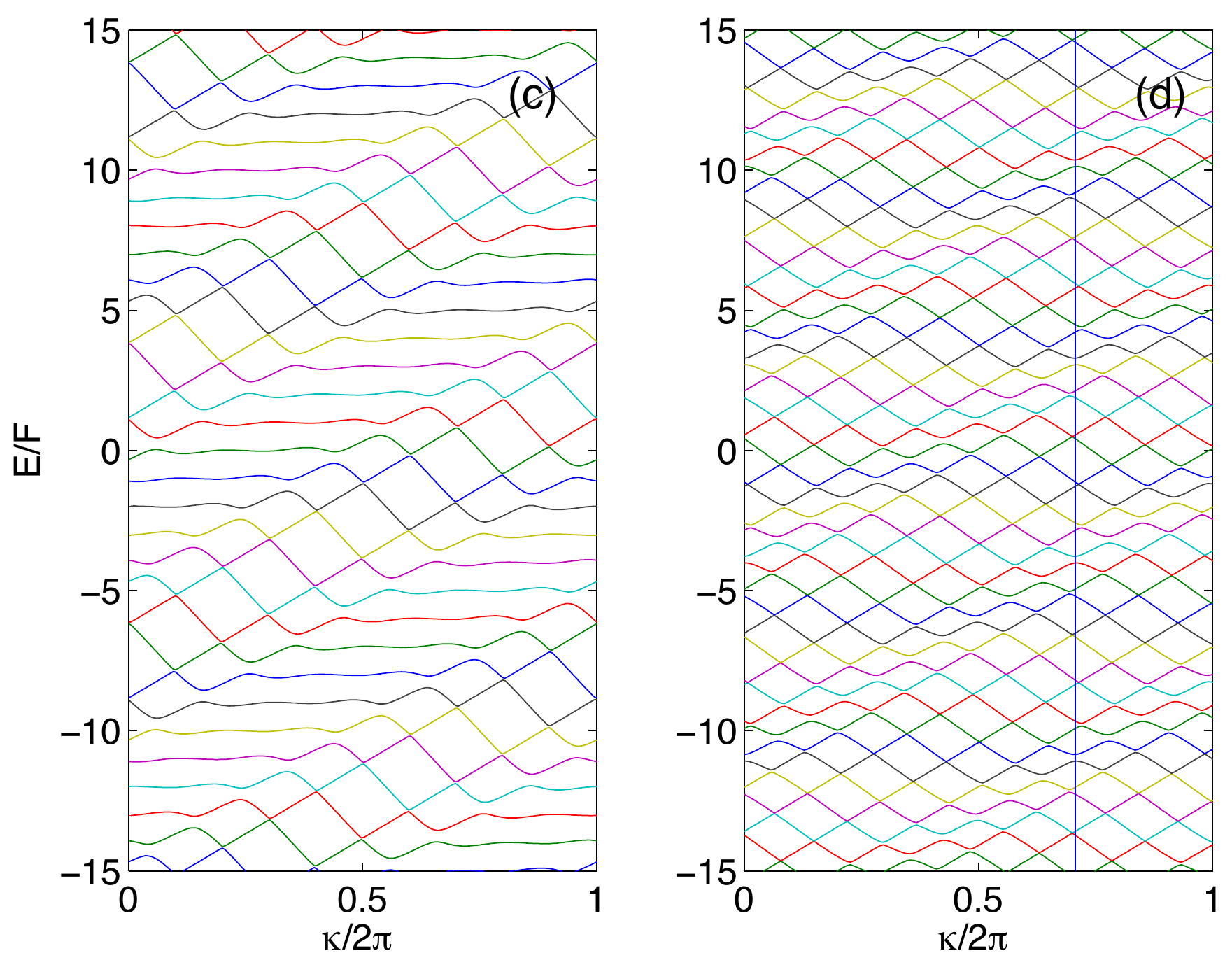}
\caption{The energy spectrum $E=E_\nu(\kappa)$ for $F=3$ (a), $F=2\pi\alpha$ (b) and $F=0.3$ (c-d). The Peierls phase is $\alpha=1/10$, while $\beta=0$ in panels (a-c) and $\beta=1$ in panel (d). Notice that the energy axis is given in units of $F$. The vertical line in the panel (d) marks the first Brillouin zone, the size of which scales as $1/\sqrt{r^2+q^2}$.}
\label{figC3}
\end{figure}

The main physical consequence that follows from the classical analysis developed in the previous section is the existence of a bifurcation at $F=F_{cr}$,
%******************************************************
\begin{equation}
\label{j9}
F_{cr}=2\pi\alpha\sqrt{J_xJ_y}=\omega_c \; ,
\end{equation}
which leads to the appearance of transporting islands. One immediately finds signatures of these islands in the energy spectrum of extended LS-states. In fact, let us focus for the moment on the case $\beta=0$: Figure \ref{figC3}(a-c) can serve to appreciate the variation of the energy spectrum, as a function of the electric field intensity, when $F$ decreases and $\alpha=1/10$ is fixed. For $F=3$ the energy bands of LS-states are well approximated by  Eq.~(\ref{f10}). At $F=\omega_c$ we observe the formation of a specific pattern, which emerges more clearly as  $F$ gets smaller. This pattern is made of straight lines of slope $v^*$:
%******************************************************
\begin{equation}
\label{k1}
E(\kappa) = v^* \kappa, \;\;
v^*=F/2\pi\alpha  \;.
\end{equation}
If we now compute the expectation value of the operator $\hat{v}_x$ on any state $\Psi^{(\nu,\kappa)}$ associated with a straight line, we find that it is also given by $v^*$ in Eq.~(\ref{k1}): $\langle\Psi^{(\nu,\kappa)}| \hat{v}_x | \Psi^{(\nu,\kappa)}\rangle=v^*$. Thus a particle in this quantum state travels the distance of $1/\alpha$ lattice periods during a Bloch period $T_B=2\pi/F$. Therefore, we shall refer to $v^*$ in Eq.~(\ref{k1}) as the {\em drift velocity} \footnote{Rewriting Eq.~(\ref{k1}) in dimensional units gives $v^*=cF/B$. Note that this coincides with the drift velocity of a charged particle in free space subject to an electric, $F$, and a magnetic, $B$, field.}. Correspondingly, LS-states associated with straight lines in Fig.~\ref{figC3}(c) will be called {\em transporting states}.  We stress that transporting states exist for any rational value of $\beta$, as soon as $F<F_{cr}$ or, what is the same, $\omega_B<\omega_c$. This statement is illustrated in Fig.~\ref{figC3}(d), which shows the band structure of extended LS-states for $\beta=1$. A number of straight lines with slope $v^*$, Eq. (\ref{k1}), are clearly observable in the figure.
%#############################################################

Let us now introduce an interesting construction: a moving, non spreading wave--packet. In fact, using the extended transporting states  $\Psi^{(\nu,\kappa)}$, we can envision to build a localized wave-packet,
%******************************************************
\begin{equation}
\label{k2}
\psi_{l,m}=\int  g(\kappa) \Psi^{(\nu,\kappa)}_{l,m} {\rm d}\kappa \;, \int |g(\kappa)|^2{\rm d}\kappa <\infty \;.
\end{equation}
Because of the linear dispersion relation, Eq. (\ref{k1}), this packet would propagate at velocity $v^*$, without changing its shape. This program, however, is hampered by a hidden difficulty. In fact, when implementing Eq. (\ref{k2}), we must follow a given straight line, but this line is interrupted by avoided crossings. If $F\ll \omega_c$ these avoided crossings are exponentially small and one can substitute them by real crossings, thus analytically linking transporting states in different energy bands. On the other hand, the number of avoided crossings  proliferates exponentially if $\beta$ tends to an irrational number. Thus, in practice, one can construct localized transporting states only when $\beta$ is a simple rational number. %\footnote{This also gives an intuitive explanation of the fact that there are no exact transporting states for irrational values of $\beta$.}.
Observe that, in the semiclassical limit, the state (\ref{k2}) is supported in one or several transporting islands.
Further details on the construction and properties of localized transporting states can be found in \cite{91}.

As a verification of this picture, we have constructed the wave packet (\ref{k2}) with $\beta=0$ and $\beta=1$ and we have computed numerically its time evolution via Schr\"odinger equation, with the Hamiltonian (\ref{b3}).  We observed that indeed this packet travels extremely large distances without experiencing a noticeable dispersion.  However, when analyzing the packet in logarithmic scale -- a technique to display fine mathematical features, although seldom observable in a physical experiment -- one notices that the full wave-packet dynamics rather resembles that of a comet, with the comet head moving at the drift velocity $v^*$ and the tail extending in the opposite direction. We discuss this behavior in the next subsection, where we intentionally chose a parameter region where avoided crossings of transporting LS-states are {\em not} exponentially small. In the semiclassical theory, this corresponds to a situation where the size of transporting island in Fig.~\ref{figC1} is comparable with the value of the effective Planck constant  $\hbar_{\rm eff}=2\pi\alpha$.

%%%%%%%%%%%%%%%%%%%%%%%%%%%%%%%%%%%%%%%%%%%%%%%%%%%%%
\subsection{Localization length of LS-states}
\label{secC3}

\begin{figure}[t]
\includegraphics[width=8.5cm]{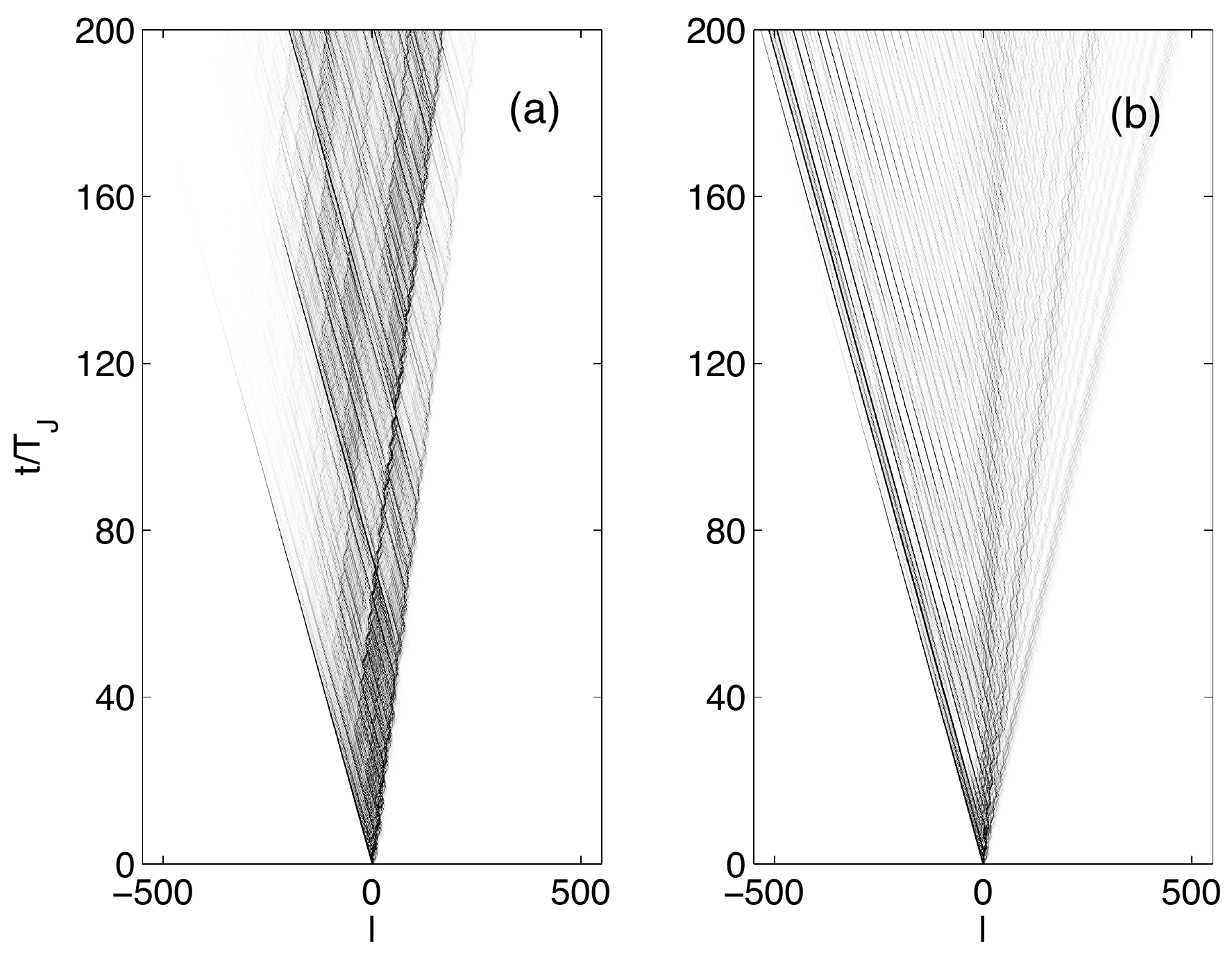}
\caption{Grey tone image of a quantum wave-packet (black maximum) versus space $l$ and time $t$.  The left panel is for irrational $\beta=(\sqrt{5}-1)/4$, the right panel has $\beta=1/3$. The other parameters are $J_x=J_y=1$, $\alpha=0.1545$ and $F=0.45$.}
\label{figC5}
\end{figure}

The comet-like behavior of the wave packet observed in the previous subsection can be analyzed in terms of the quantum driven Harper model (\ref{e9}) \footnote{We note in passing that the system (\ref{e9}) -- besides its relation to the system (\ref{b3}) -- is of independent relevance, since it can be realized with cold atoms in (quasi) 1D lattices. The proposed experiment \cite{87} is a minor modification of the experiment \cite{Roat08}, where the authors simulated the Aubry-Andr\'e model \cite{Aubr80}, which includes the standard (not driven) Harper model as a particular case.}. The fundamental difference between classical and quantum driven Harper dynamics is that in the former a particle may tunnel out of a stability island, while in the latter, once initially located inside it, it is forever captured therein. This is displayed in Fig.~\ref{figC5}, that shows the evolution of a localized wave-packet, which is initially supported by the central transporting island. Tunneling out of this island, as well as the opposite process of recapture into other islands in the chain, are clearly observed.  However, this in and out tunneling process results in different asymptotic behaviors, depending on rationality of the parameter $\beta$: for rational $\beta$ it generates a ballistic spreading of the wave-packet, while for irrational $\beta$ it leads to wave-packet localization.
%#############################################################

One can estimate the characteristic localization length as follows \footnote{We adapt to our case the technique of \cite{Hufn02,Back05}, originally developed for a model quantum system which, like ours, has a chain of transporting islands in the classical limit.}.  Consider an initially populated transporting island. The rate of tunneling out of this island is determined by the ratio between the size of  stability island, $S=S(F)$, and the effective Planck constant, $\hbar_{\rm eff}=2\pi\alpha$: the island is depleted after a time which is exponential in $S/\hbar_{\rm eff}$. During this time the quantum particle is transported at a distance $F_y/2\pi \alpha$ in units of the lattice period. Therefore, the localization length $L_{\rm loc}$ is
%******************************************************
\begin{equation}
\label{k3}
L_{\rm loc} \approx \frac{F}{\alpha}\exp\left(C\frac{S(F)}{\alpha}\right) \;,
\end{equation}
where $C$ is some constant.

\begin{figure}[t]
\includegraphics[width=8.0cm]{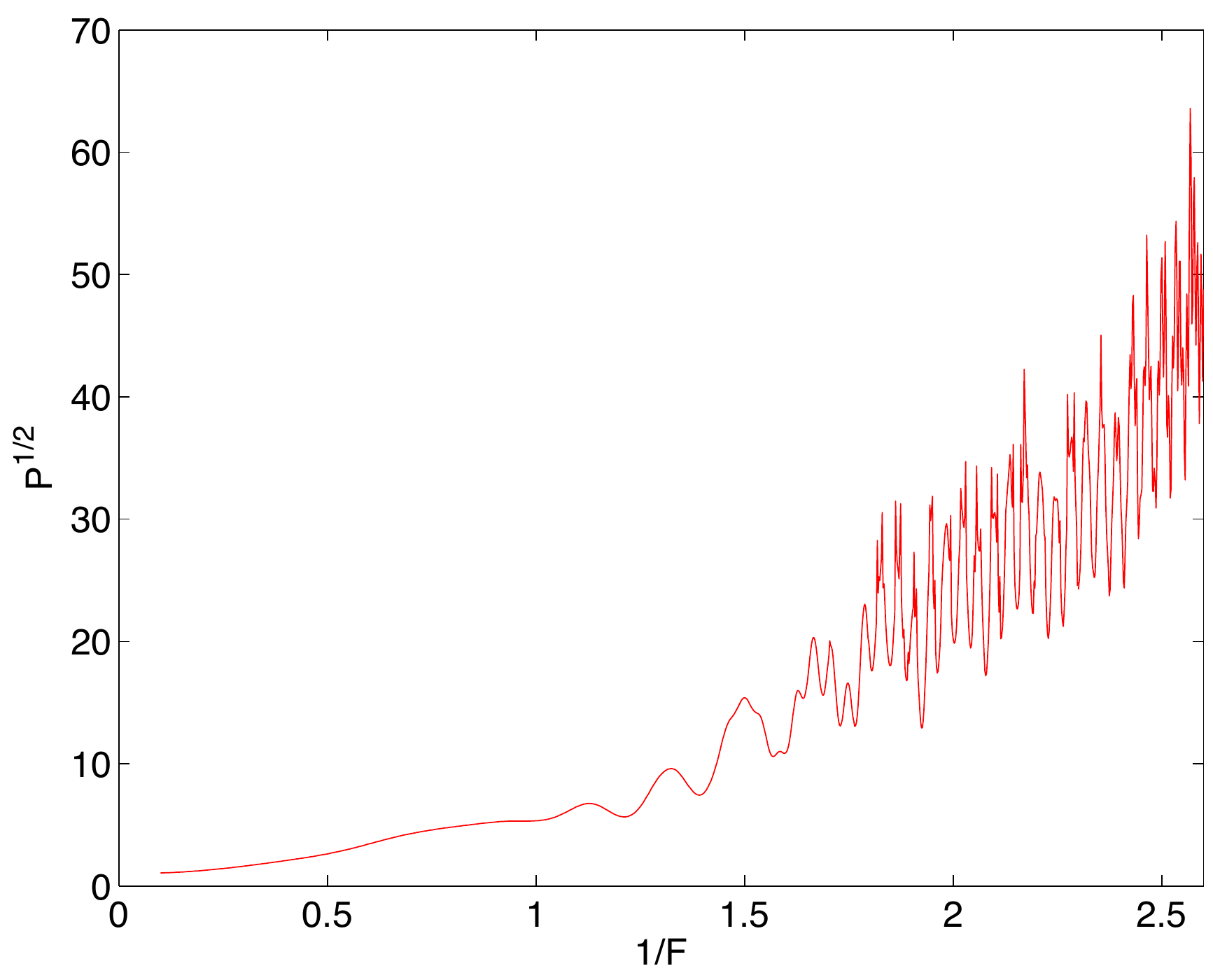}
\caption{Square root of the participation ratio $P[\Psi]$ of localized LS-states versus $1/F$, for $\alpha=1/10$ and  $\beta=(\sqrt{5}-1)/4$.}
\label{figC6}
\end{figure}

The estimate (\ref{k3}) equally applies to the localization length of eigenstates of the evolution operator (\ref{e10}), which describes the  stroboscopic dynamics of the quantum driven Harper model. Furthermore, as shown in Sec.~\ref{secA4}, the localization length of evolution operator eigenstates provides a reliable estimate of that of 2D LS-states. Figure \ref{figC6} shows the participation ratio (\ref{g5b}) of localized LS-states as a function of $1/F$, calculated by straightforward diagonalization  of the original 2D Hamiltonian, for irrational $\beta=(\sqrt{5}-1)/4$. The dramatic increase of participation ratio for small $F$ qualitatively confirms the exponential scaling law (\ref{k3}).
%%******************************************************
%\begin{equation}
%\label{k4}
%P\sim \exp\left(C\frac{S(F)}{\alpha}\right) \;,\quad F>\omega_c \;.
%\end{equation}
%%
In Fig.~\ref{figC6} one also observes wild oscillations of the participation ratio superimposed to the overall exponential increase. The origin of these oscillations remains an open problem.
%#############################################################

%%%%%%%%%%%%%%%%%%%%%%%%%%%%%%%%%%%%%
\subsection{Dynamics of an incoherent wave packet}
\label{secC4}

To conclude this section we discuss the dynamics of an incoherent wave--packet, that can be numerically rendered by a wide 2D Gaussian packet, with scrambled phases. The prominent feature of this dynamics is dispersion without directed transport. In fact, the wave-packet mainly spreads in both directions orthogonal to ${\bf F}$, as a consequence of the global structure of LS-states.

\begin{figure}[t]
\includegraphics[width=8.5cm]{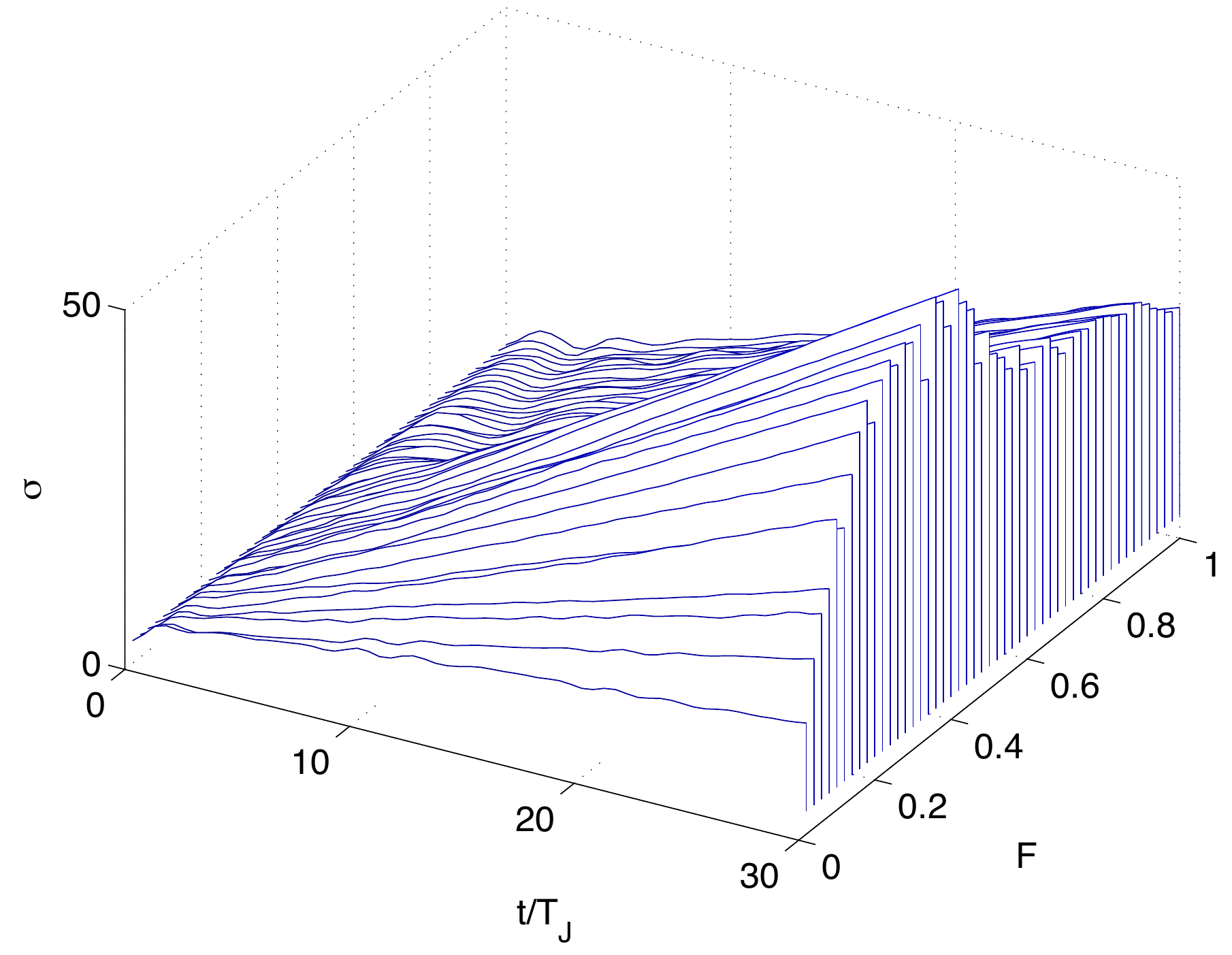}
\includegraphics[width=8.5cm]{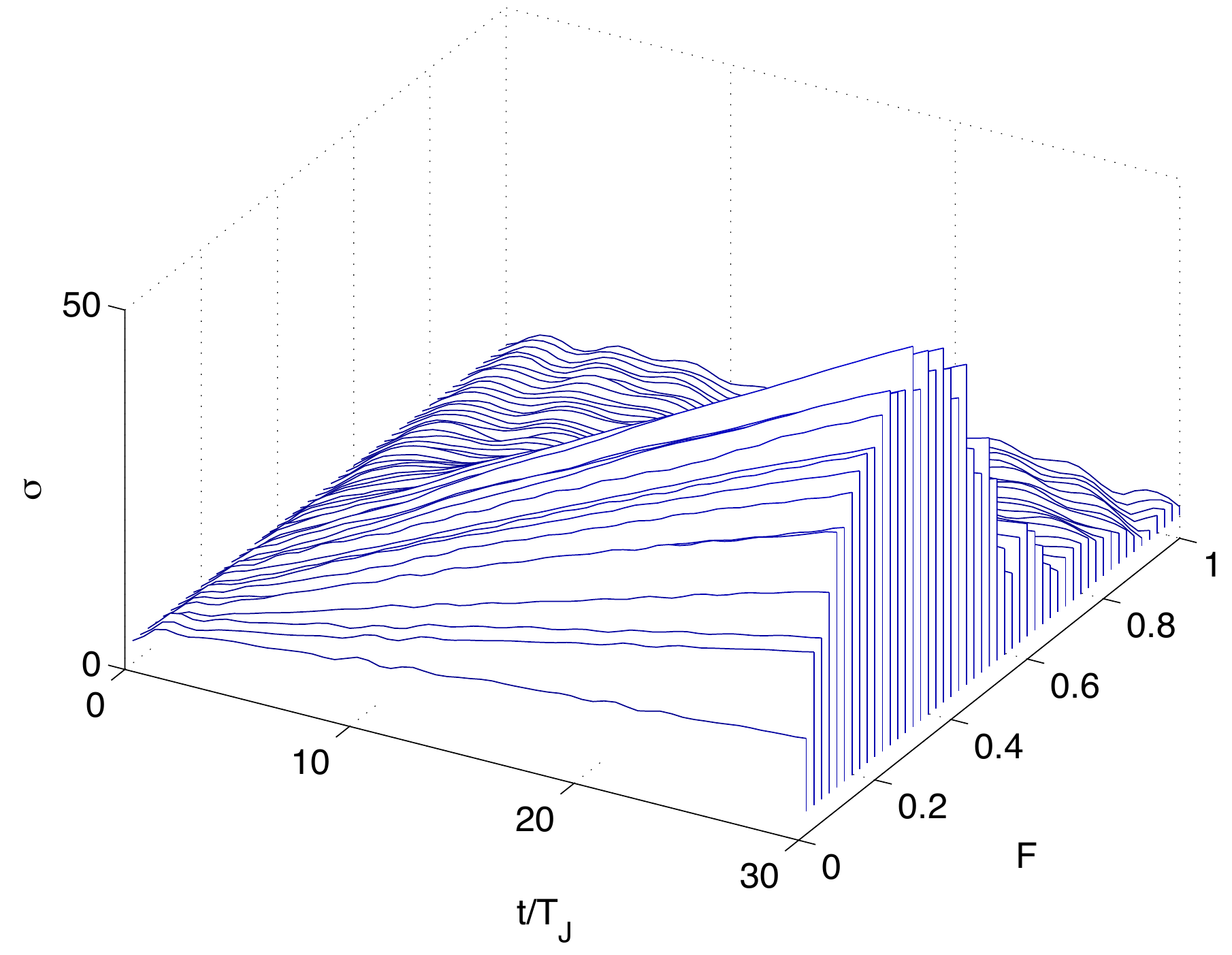}
\caption{The dispersion (\ref{g3}) as a function of time, for several values of $F$ in the interval $0<F\le 1$. The other parameters are $J_x=J_y=1$, $\alpha=1/10$, and $\beta=1/3$ (a), $\beta=(\sqrt{5}-1)/4$ (b). The initial wave function is a 2D, phase--scrambled Gaussian packet approximately 10 sites wide in each direction. The lattice size is $300\times100$ sites.}
\label{figC7}
\end{figure}

We have calculated numerically the dispersion $\sigma=\sigma(t)$ of the packet, as a function of time, until a maximum time $t_{max}=30T_J$. It is depicted in Fig.~\ref{figC7}, when $F$ lies in the interval $0<F\le1$, in two cases: rational $\beta=1/3$ and irrational $\beta=(\sqrt{5}-1)/4$. As expected, in the former case we observe a linear growth for any value of $F$, in accordance with the continuous spectrum of LS-states and in agreement with classical analysis. Yet, Fig.~\ref{figC7} presents a paradox: at first glance, the case of irrational $\beta$ also agrees with the classical results: $\sigma(t)$ grows linearly in time for $F<\omega_c\approx 0.6$ and does not grow for $F>\omega_c$.  However, an unbounded growth of the dispersion contradicts discreteness of the spectrum of LS-states -- that we know to hold for irrational $\beta$. Thus, the linear growth observed for $F<0.6$ must be a transient effect.  In fact, simulating the system dynamics for a longer time span, $t_{max}=60T_J$ (and, correspondingly, a system size twice as large) we are able to detect signatures of saturation of the dispersion, in the range $0.4<F<0.6$. For smaller $F$, due to the exponential increase of the localization length of LS-states, numerical detection of saturation requires even larger system sizes.
%#############################################################

%%%%%%%%%%%%%%%%%%%%%%%%%%%%%%%%%%%%%%%%%%%%%
%
%%%%%%%%%%%%%%%%%%%%%%%%%%%%%%%%%%%%%%%%%%%%%
\section{Finite lattices}
\label{secD}

We have so far considered only infinite lattices. However, in laboratory experiments one deals with finite lattices: this may introduce new, finite size effects. Clearly, we must first specify the boundary conditions. In this section we consider two qualitatively different types of boundary: harmonic confinement, which is the default boundary condition in experiments with cold atoms in optical lattices, and Dirichlet boundary conditions, that are typically realized in solid-state and photonics systems. To keep the paper within reasonable length we restrict ourselves to the case $\alpha\ll 1$.

%%%%%%%%%%%%%%%%%%%%%%%%%%%%%%%%%%%%%%%%%%%%%
\subsection{Parabolic lattices}
\label{secD1}

In the case of harmonic confinement, the tight-binding Hamiltonian of a quantum particle in the presence of (real or synthetic) electric and magnetic fields reads
%***********************************************
\begin{equation}
\label{m1}
\left(\widehat{H} \psi\right)_{l,m}= -\frac{J}{2}\left(\psi_{l+1,m}+\psi_{l-1,m}\right)
-\frac{J}{2}\left(e^{i2\pi\alpha l} \psi_{l,m+1}+e^{-i2\pi\alpha l} \psi_{l,m-1}\right)
+\frac{\gamma}{2}(l^2 + m^2)\psi_{l,m} \;,
\end{equation}
where $\gamma$ is the strength of confinement. The latter quantity can be expressed through the particle mass $M$ and the frequency of the harmonic potential $\omega_{\rm hc}$, as $\gamma=Ma^2\omega_{hc}^2$.  Note that electric field does not enter the Hamiltonian (\ref{m1}), because it simply shifts the origin of the parabolic lattice and we assume that this fact has already been taken into account. It is easy to prove that the spectrum of the Hamiltonian (\ref{m1}) is discrete and is bounded from below by the energy $-2J$ (here we set again $J_x=J_y=J$). Moreover,  there are no energy gaps for $\alpha\ne0$ and the density of states is a continuous function of the energy $E$, that approaches the constant density $2\pi/\gamma$ for $E>2J$ \cite{Gerb10}. This makes the problem very different  from that of a plane lattice and of lattices with stronger confinement \footnote{The case of stronger confinement, where one indeed finds similarities with the spectrum of a plane lattice, was analyzed in some details in \cite{Gerb10,Buch12,Gold12,Gold13b}.}.

The crucial observation which helps us to understand the properties of the system (\ref{m1}), is that locally the harmonic confinement can be substituted by the gradient force
%*********************************************
\begin{equation}
   \label{m2}
{\bf F}=\gamma {\bf r} \;,\quad {\bf r}=(x,y)  \;,
\end{equation}
which points to the lattice origin. Thus, we can reduce the problem under consideration to the previous problem of LS-states. Using this analogy, we can explain the local properties of the spectrum and the structure of eigenstates of the system (\ref{m1}).

%%%%%%%%%%%%%%%%%%%%%%%%%%%%%%%%%%%%%%%%%%%%%%%
\subsubsection{Classical approach}
\label{secD1a}

If $\alpha\ll1$, we can employ the the semiclassical approach of Sec.~\ref{secC1}. The classical counterpart of (\ref{m1}) reads
%*********************************************
\begin{equation}
   \label{m3}
H_{cl}=-J\cos(p_x) - J\cos(p_y + x) +\frac{\gamma'}{2}(x^2+y^2)  \;, \quad \gamma'=\gamma/(2\pi\alpha)^2 \;.
\end{equation}
Note that according to Sec.~\ref{secC1} the classical limit requires $\alpha\rightarrow 0$, while keeping the parameter $\gamma'$ constant. The latter condition ensures that the classical dynamics does not depend on the effective Planck constant $\hbar_{\rm eff}=2\pi\alpha$. In this section, however, to allow for a direct comparison of classical trajectories with quantum wave functions at finite $\alpha$, we use the scaled classical Hamiltonian ($x\rightarrow 2\pi\alpha x$, $y\rightarrow 2\pi\alpha y$),
%****************************************
 \begin{equation}
 \label{m4}
 H_{cl}=-J\cos(p_x) - J\cos(p_y+2\pi\alpha x) +\frac{\gamma}{2}(x^2+y^2)  \;,
 \end{equation}
which explicitly includes the parameter $\alpha$.

\begin{figure}[t]
\includegraphics[width=8.0cm]{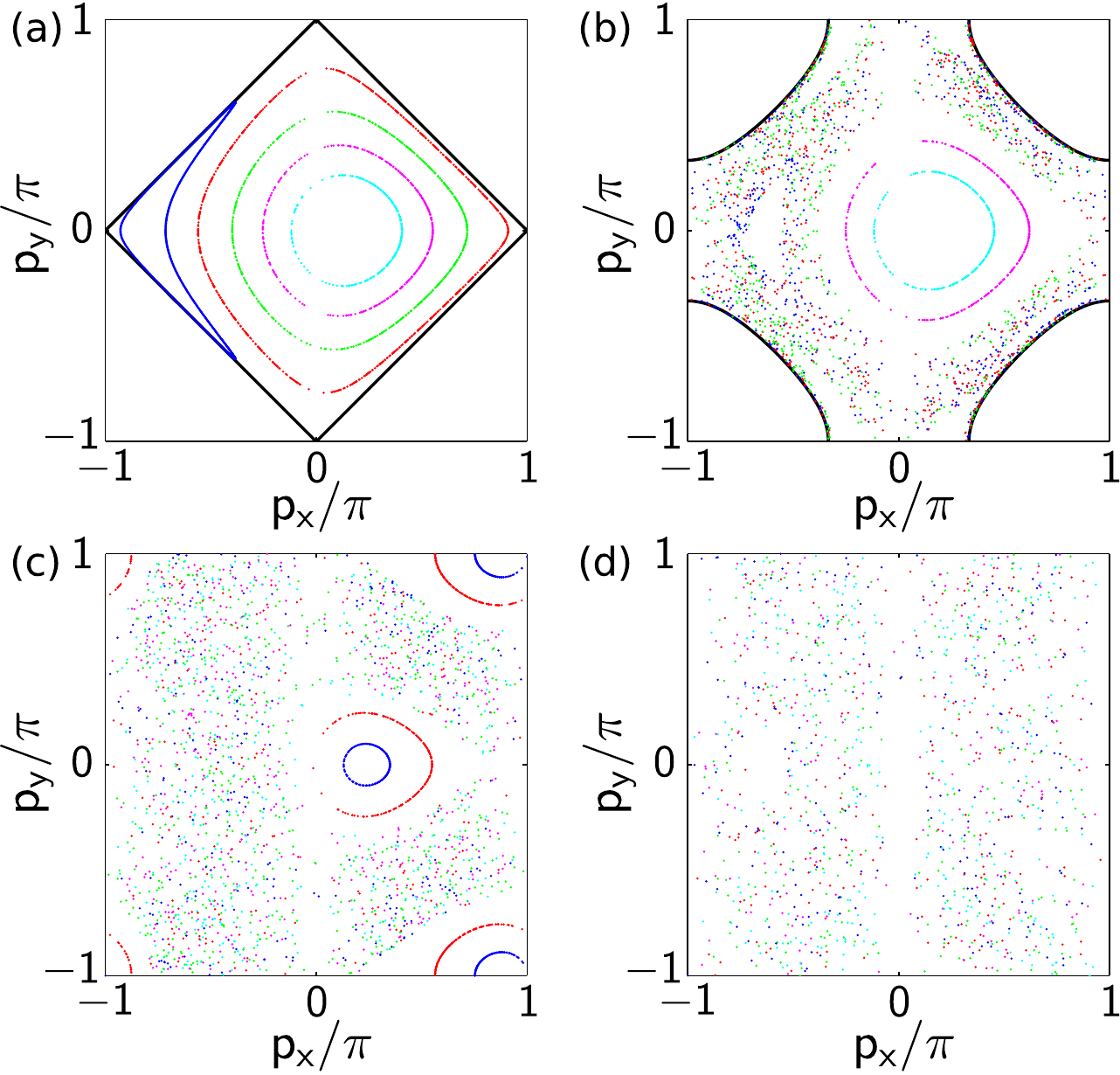}
\includegraphics[width=8.0cm]{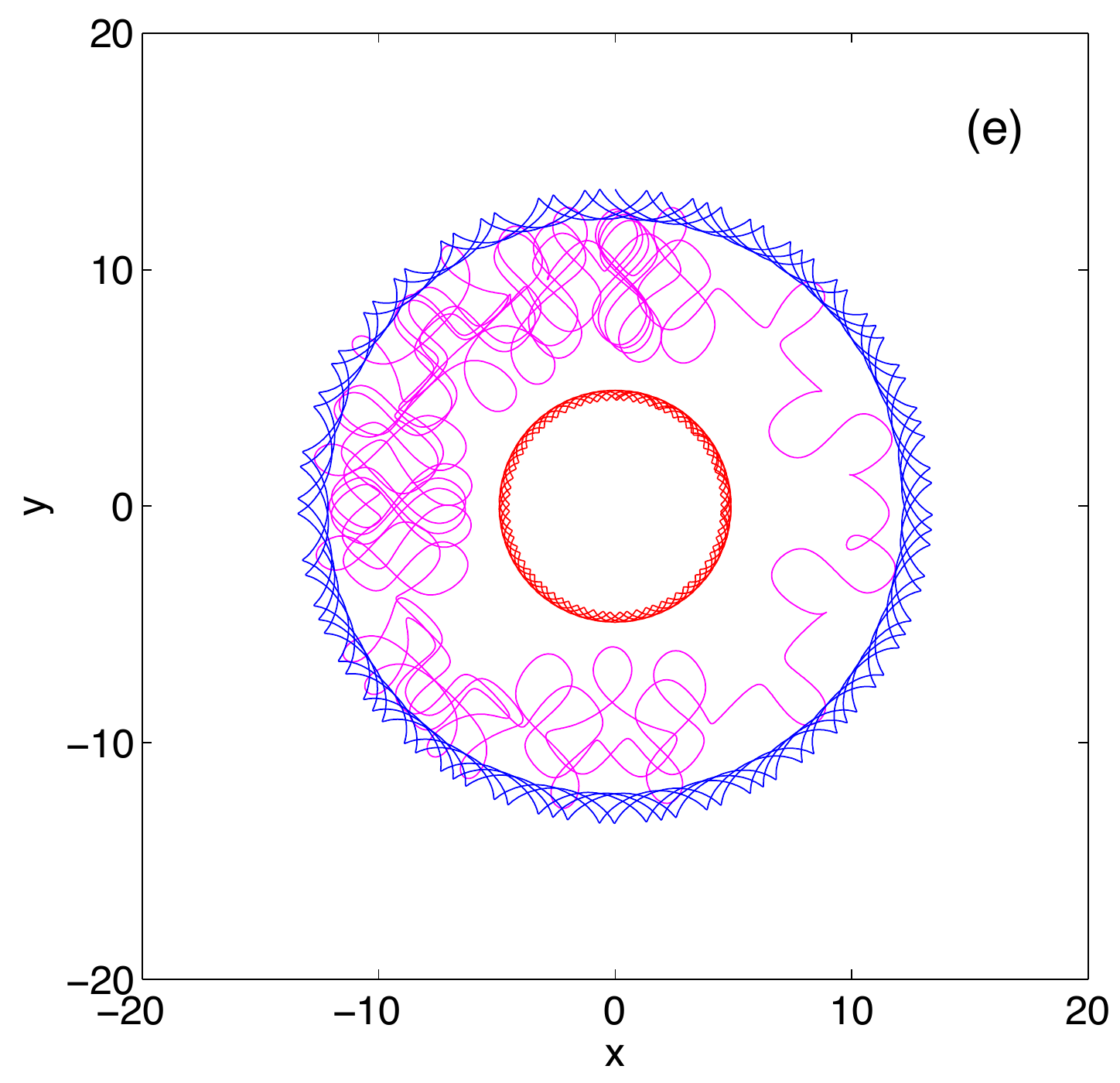}
\caption{Poincare cross-sections by the plane $\theta=0$ at energies $E = 0, 0.5, 3, 11$ (a-d). The system parameters are $J=1$, $\alpha=1/6$, and $\gamma=0.05$. The solid lines in the panels (a) and (b) restrict  the available phase space. The panel (e) shows examples of classical trajectories in the coordinate space $(x,y)$ for moderate $E=2.5$. Initial conditions for the depicted trajectories are chosen inside the middle island (blue line outer trajectory), inside the upper/lower island (red line inner trajectory), and in the chaotic sea (magenta line in between the inner and outer trajectories). The total evolution time corresponds to 5 periods at frequency $\Omega$ (\ref{m5}). }
\label{figD2}
\end{figure}

At a given energy $E$, any phase trajectory of (\ref{m4}) is uniquely described by the momenta $p_x$ and $p_y$ and by the angle $\theta=\arctan(x/y)$, that are cyclic variables: the energy shell of (\ref{m4}) lies within a three-dimensional torus, or it coincides with this torus when $E\ge 2J$. Fig. \ref{figD2}(a-d) shows the Poincare cross-sections of the energy shell with the plane $\theta=0$ for a few values of $E$. While for  $E<  0$ only regular trajectories are found, for $E>0$ the typical structure of a non-integrable system with mixed phase space appears.
%#############################################################

Let us first describe the case of moderate energy, where the phase space consists of two large stability islands surrounded by a chaotic sea. Here we have a one-to-one correspondence with the driven Harper model, see Fig.~\ref{figC1}. The classical particle captured in a stability island moves with the drift velocity (\ref{k1}) under the action of the gradient force $F=\gamma r$ and, hence, encircles the lattice origin with the frequency
%*********************************************
\begin{equation}
\label{m5}
\Omega=\gamma/2\pi\alpha   \;.
\end{equation}
At the same time, a trajectory located in the chaotic component encircles the lattice origin in the opposite direction, but of course no unique encircling frequency can be associated to this case. Examples of these trajectories are given in Fig.~\ref{figD2}(e).

Increasing the system energy, which implies an increase of the mean radius of  trajectories and, correspondingly, of the magnitude of the gradient force, transporting islands disappear. Transport is prohibited and the particle remains localized in a sector of the circle. This transition takes place at a critical energy $E_{cr} $: using the relations $E\approx \gamma r^2/2$ and $F_{cr}\approx\omega_c=2\pi\alpha J$  it can be estimated as
%*********************************************
\begin{equation}
\label{m6}
E_{cr} \approx \frac{(2\pi \alpha J)^2}{2\gamma}=E_R\left(\frac{\omega_c}{\omega_{\text{hc}}}\right)^2 \;,
\end{equation}
where $E_{R}=\hbar^2/Ma^2$ sets the relevant energy scale and is proportional to the recoil energy, in cold atom system.

Finally, as seen in Fig.~\ref{figD2}(a), the low-energy dynamics of the system is always regular, so that we only observe regular transporting trajectories.

%%%%%%%%%%%%%%%%%%%%%%%%%%%%%%%%%%%%%%%%%%%%%%%
\subsubsection{Quantum approach}
\label{secD1b}

\begin{figure}[t]
\includegraphics[width=7.5cm]{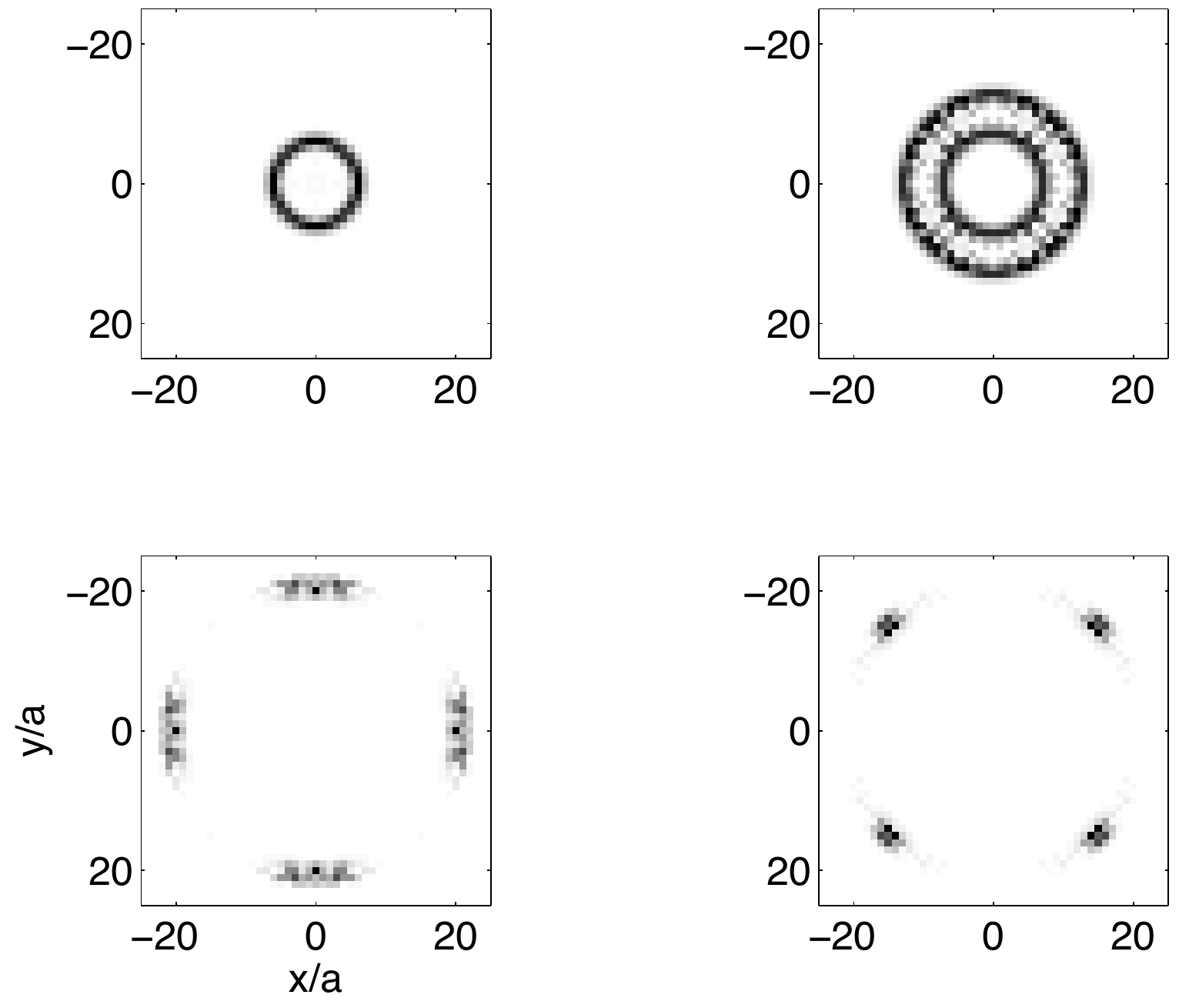}
\includegraphics[width=8.5cm]{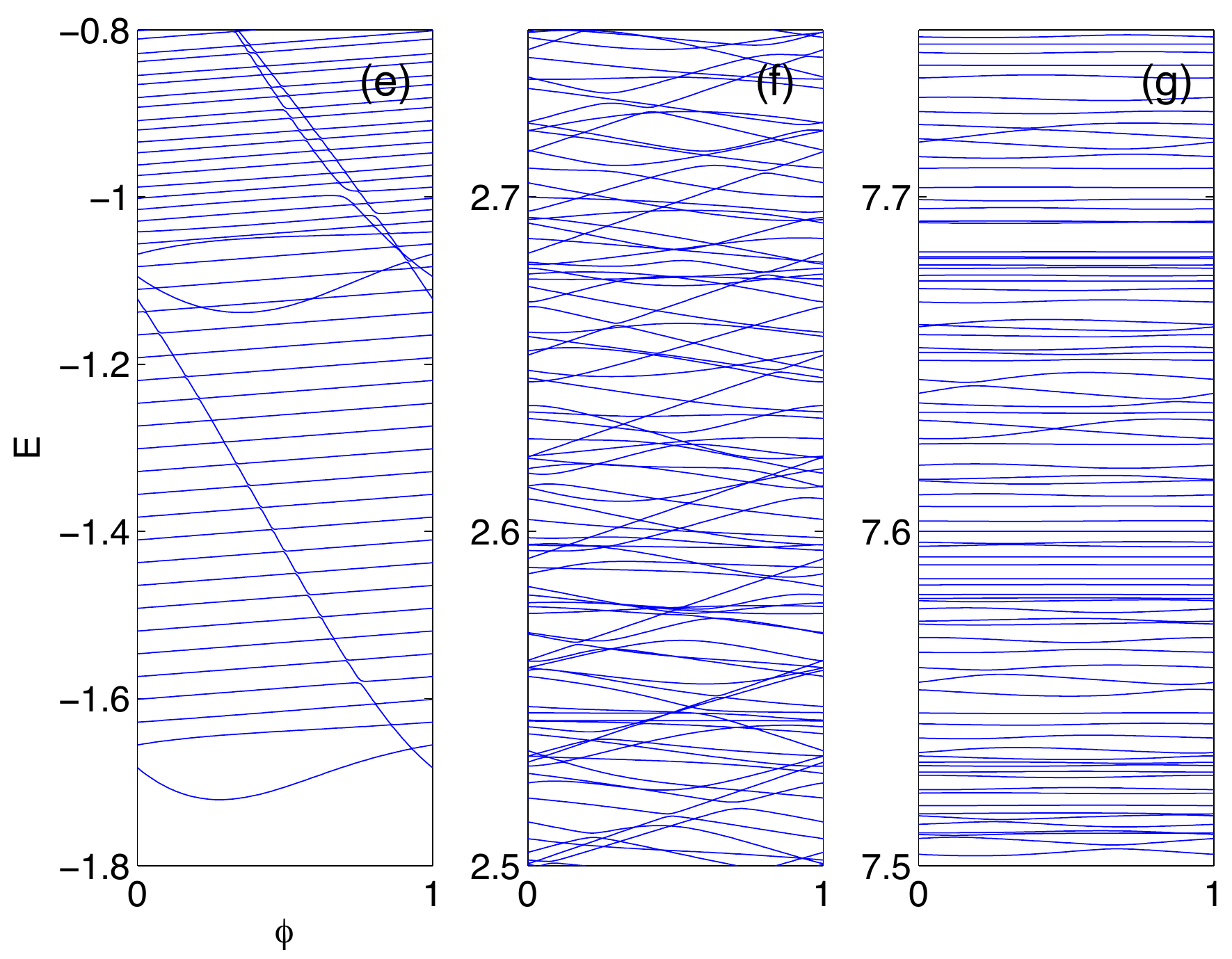}
\caption{The 25th (a),  349th (b), 1365nd (c), and 1370th (d) eigenstates of the Hamiltonian (\ref{m1}) with energies $E=-0.5511, 2.7671, 10.8544, 10.9069$, respectively. The squared absolute value of the wave-function $\Psi_{l,m}$ is shown as a gray-scale map. We used $J=1$, $\alpha=1/6$ and $\gamma=0.05$. Panels (e-g) show the energy spectrum of the Hamiltonian (\ref{m9}) versus the local flux $\phi$ inserted at the origin.  Notice the different energy ranges in the panels. Parameters are $J=1$, $\alpha=0.1$ and $\gamma=0.018$.}
\label{figD1}
\end{figure}

The classical results of the previous subsection suggest the following classification of the eigenstates of the quantum system (\ref{m1}). The low-energy states, with energies in the range $-2J < E < 0$, can be termed {\em regular transporting states}. Besides the energy, they can be labeled by an additional quantum number $\ell$. Example of a regular state, obtained by direct diagonalization of the Hamiltonian matrix, is given in Fig.~\ref{figD1}(a). The physical significance of the quantum number $\ell$ is similar to that of an angular momentum but, strictly speaking, it cannot be equivalent to this latter, because of the absence of rotational symmetry of the Hamiltonian (\ref{m1}). We will come back to this quantum number shortly.

In the energy interval $0<  E<E_{cr}$ we find transporting states associated with either one of the stability island in Fig.~\ref{figD2}(c), as well as chaotic states which have no additional quantum number. Occasionally, we also find hybrid states, that can be viewed as superposition of regular and chaotic states. Example of a hybrid state is given in Fig.~\ref{figD1}(b), where a chaotic state is superimposed to `inner'  and `outer' transporting states, associated with the central and lower/upper stability islands, respectively.

Finally, for $E>E_{cr}$, all states are localized, see  Fig.~\ref{figD1}(d). In regard to this picture, observe that, due to the four--fold lattice symmetry, there exist three other eigenstates at almost the same energy, which look similar to the depicted state. From this set of four exact states one can construct a new set of four {\em approximate} eigenstates, each of which is localized in one segment of the circle. Therefore, a particle with mean energy $E>E_{cr}$, which is initially localized within one of the segments, remains localized in this segment for an exponentially large time.
%#############################################################

Let us now clarify the meaning of the quantum number $\ell$, that also serves to illuminate the topological properties of regular states. Let us consider the response of the quantum system (\ref{m1}) to the local flux
%******************************************************
\begin{equation}
\label{m7}
{\bf A}(x,y) \sim \left(-\frac{y}{x^2+y^2},\frac{x}{x^2+y^2}\right) \;.
\end{equation}
The vector potential (\ref{m7}) corresponds to a magnetic field which is zero everywhere except at the coordinate origin. It is possible to show that the tight-binding counterpart of Eq. (\ref{m7}) is an additional contribution to the Peierls phase, given by
%*****************************************************
\begin{eqnarray}
   \label{m8}
\theta_x(l,m)\equiv \theta(l,m) = \phi \left[ \arctan\left(\frac{2l-1}{2m-1}\right) - \arctan\left(\frac{2l+1}{2m-1}\right) \right]
\end{eqnarray}
in the $x$-direction and  $\theta_y(l,m)=\theta(m,l)$ in the $y$-direction. Here $\phi$ quantifies the inserted flux in units of the magnetic flux quantum. Thus we have
%****************************************************
\begin{eqnarray}
\label{m9}
(\widehat{H}\psi)_{l,m}=
-\frac{J}{2}\left(e^{-i\pi\alpha m} e^{-i\theta(l,m)} \psi_{l+1,m}  +  h.c.\right)
-\frac{J}{2}\left(e^{i\pi\alpha l}e^{i \theta(m,l)} \psi_{l,m+1}+ h.c.\right)
+\frac{\gamma}{2}\left[\left(l-\frac{1}{2}\right)^2+\left(m-\frac{1}{2}\right)^2\right]\psi_{l,m}  \;.
\end{eqnarray}
At difference with (\ref{m1}), we have employed the symmetric gauge for the uniform magnetic field and we have shifted the coordinate origin from the site $(l,m)=(0,0)$ to the center of the plaquette. Figure \ref{figD1}(e-g) shows the low-, mid-, and high-energy parts of the spectrum of (\ref{m9}) as a function of the inserted flux $\phi$. Notice that the energy levels for $\phi=0$ and $\phi=1$ coincide. This fact gives to the spectrum a cylindrical topology, in which an energy level is connected to another according to some continuation rule. The crucial observation here is that regular transporting states are connected to each other by a helical line\footnote{We elect to ignore tiny avoided crossing. Without this convention every energy level of the system (\ref{m9}) is a periodic function of $\phi$.}. Thus the quantum number $\ell$ can be associated with the winding number originated by this connection rule. It is also worth mentioning that the step of helical line is determined by the frequency $\Omega$ (\ref{m5}) but not by the cyclotron frequency $\omega_c$. In this sense the encircling frequency $\Omega$ is the main fundamental frequency of the system, which takes the role of the cyclotron frequency for a plane lattice.

%%%%%%%%%%%%%%%%%%%%%%%%%%%%%%%%%%%%%%%%%%%%%%%
\subsubsection{Wave-packet dynamics}
\label{secD1c}

We can now consider the time--dynamics of the system (\ref{m1}). We shall show that the motion of wave-packets reveals properties of the eigenstates at different energies and allow us to determine two main characteristics of the system -- the encircling frequency $\Omega$ and the critical energy $E_{cr}$ -- from a time-resolved measurement.

\begin{figure}[t]
\includegraphics[width=8.5cm]{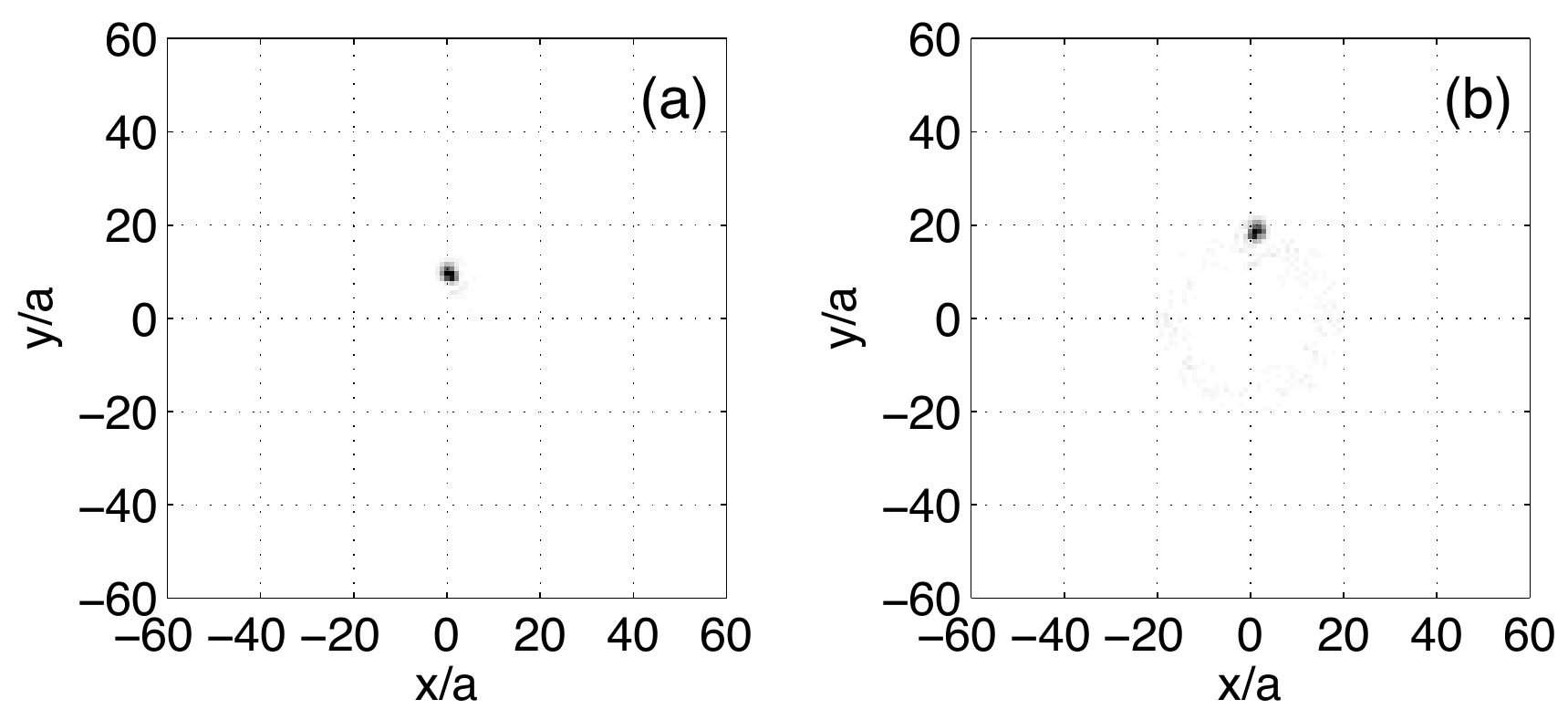}
\includegraphics[width=8.5cm]{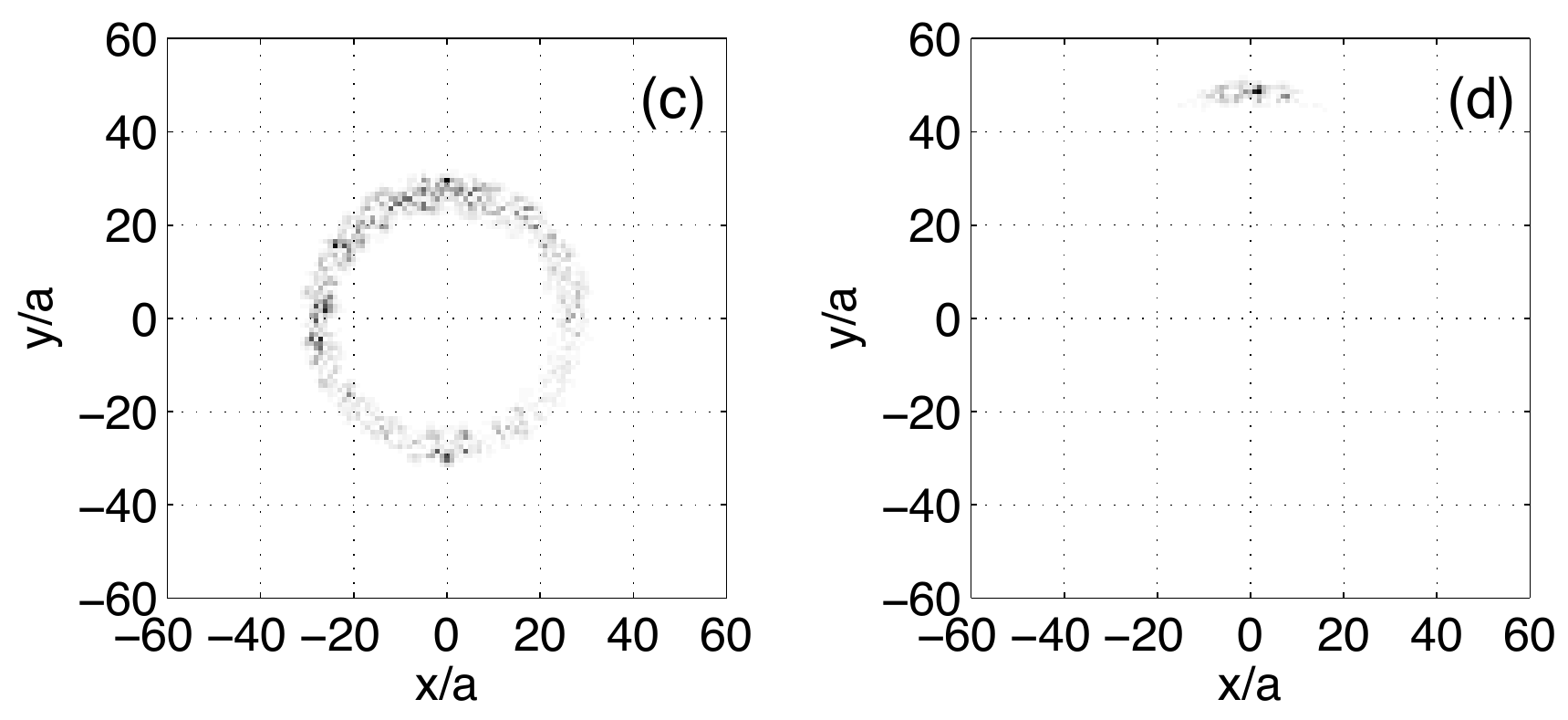}
\caption{Population of the lattice sites at $t=2\pi/\Omega$. The initial wave packet corresponds to a narrow Gaussian shifted by 10 (a), 20 (b), 30 (c), and 50 (d) sites from the lattice origin. The system parameters are $J=1$, $\alpha=0.1$, and $\gamma=0.02$.}
\label{figD5}
\end{figure}

Our numerical experiment follow the same scheme of laboratory experiments on dipole oscillations of cold atoms in parabolic lattices. This protocol involves a sudden shift of the origin of the harmonic potential by a distance $r_0$, so that the atomic cloud appears on the slope of the parabolic lattice, where it has an energy $E\approx\gamma r_0^2/2$. Successively, the system  is let free to evolve for a time span $t$, which is followed by a (destructive) measurement of the atomic density. The result of this numerical experiment is shown in Fig.~\ref{figD5}(a-d), where we have chosen a narrow Gaussian packet as initial condition. In case (a), which corresponds to the regular regime in the classical approach, the packet encircles clockwise the lattice origin and returns periodically to its initial position, with period $T_\Omega=2\pi/\Omega$. The regular regime remains dominant for $r_0=20$, with a minor contribution of chaotic states, seen as a faint circle. With a further increase of the shift ($r_0=30$), chaotic and regular states become equally important. The wave-packet now moves (spreads) in both clockwise and counterclockwise directions. Finally, for $r_0=50$, the mean energy of the packet is larger than the critical energy (\ref{m6}) and we observe localization of the packet in a narrow segment of the circle.
%#############################################################

%%%%%%%%%%%%%%%%%%%%%%%%%%%%%%%%%%%%%%%%%%%
\subsection{Open boundaries}
\label{secD2}

\begin{figure}[t]
\includegraphics[width=8.5cm]{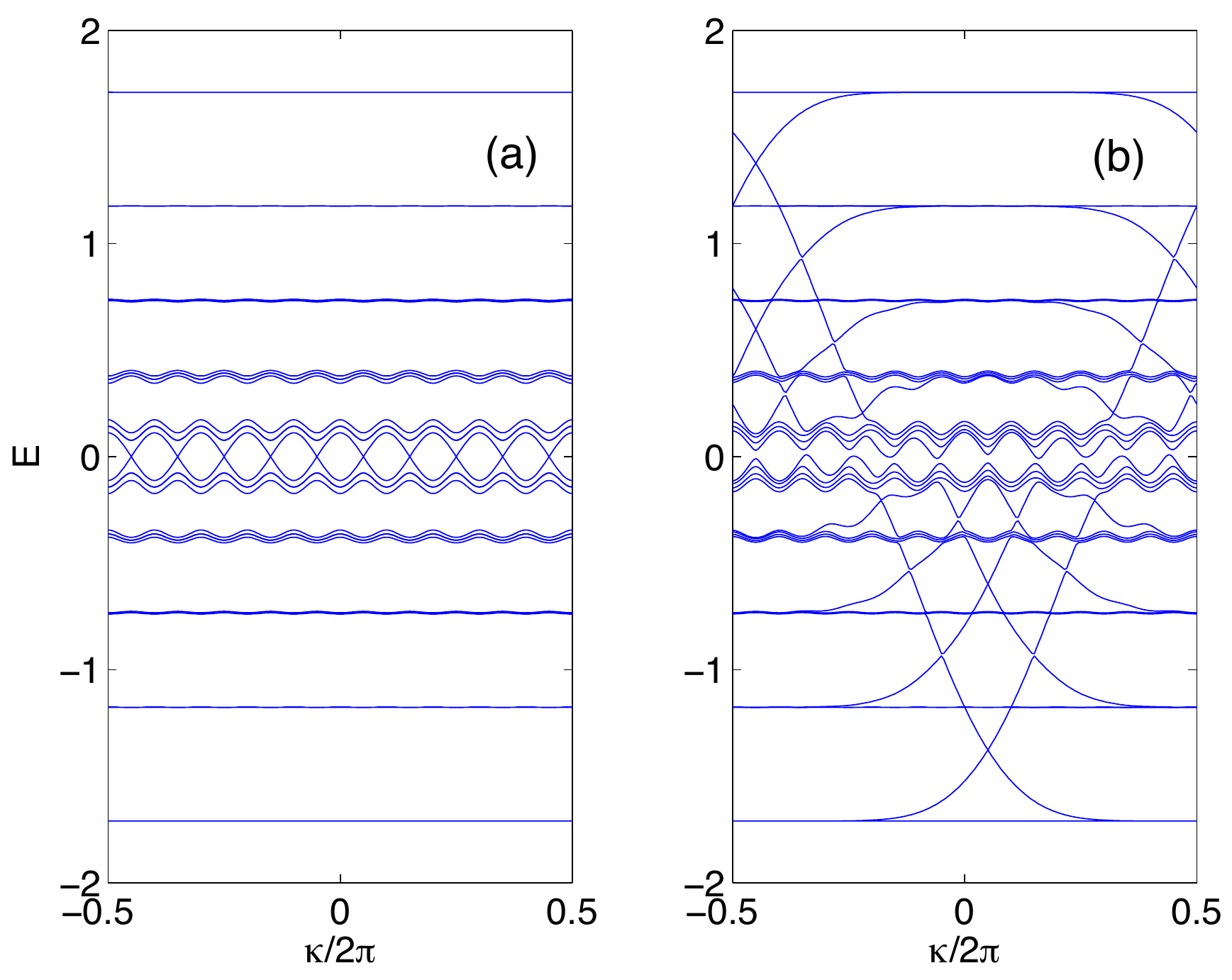}
\caption{Energy spectrum of the system (\ref{b3}) for vanishing electric field and periodic (a) and Dirichlet (b) BCs. The system parameters are $\alpha=1/10$, $J_x=J_y=1$,  $F_x=F_y=0$, the lattice size is $L_x=40$ and $L_y=\infty$.}
\label{figE1}
\end{figure}

The case of Dirichlet BC has attracted much attention in the physical literature because here the system supports edge states, which can carry a non-vanishing current even for vanishing electric field.  The standard set-up for studying these states is given by the Hamiltonian (\ref{b3}) with $F=0$, in which the index $l$ is restricted to a finite interval, $1\le  l \le L_x$, and periodic BC with period $L_y$ (which  eventually tends to infinity) characterize the $y$ direction.  This set-up is usually  opposed to  periodic BC,
%******************************************************
\begin{equation}
   \label{p1}
\Psi_{1,m}=\Psi_{L+1,m} \;,
\end{equation}
where $L\equiv L_x$ is assumed to be multiple of the denominator $q$ of the Peierls phase $\alpha=r/q$ which, in turn, is assumed to be a rational number. In both set-ups, the substitution (\ref{d1}) leads to the Harper equation (\ref{d2}), which is parameterized by the quasimomentum $\kappa_y$. For future reference, Fig.~\ref{figE1} compares the spectrum of the Harper equation for $\alpha=1/10$ with periodic and Dirichlet BCs. In the latter case, eigenenergies inside the energy gaps appear clearly. These are associated, in configuration space, to states localized at the left or the right edge of the lattice, depending on the sign of the group velocity. The total number of edge states in a gap is given by the difference between the Chern numbers of the nearby magnetic bands \cite{Hats93}, a theorem known as bulk-edge correspondence.
%\footnote{We stress that the notion of magnetic bands implies infinite lattices. Thus,  knowing the Chern numbers of magnetic bands of the infinite lattice one can find the number of edge states for the finite lattice and, vice versa, the number of edge states in every gap yields Chern numbers.}.
Finally recall that the group velocity $v_g$ also determines the mean value of the current operator:
%******************************************************
\begin{equation}
   \label{p2}
v_g\equiv  \frac{\partial E_\nu(\kappa)}{\partial \kappa}= \langle\Psi^{(\nu,\kappa)} |\hat{v}_y | \Psi^{(\nu,\kappa)}\rangle \equiv v \;,
\end{equation}
where $\kappa\equiv\kappa_y$ and the discrete index $\nu$ labels the solutions of the finite-size Harper equation. Since the dispersion relations for low-energy bulk states are almost flat, only edge states and bulk states in the center of the Bloch band can carry non-negligible currents.
%#############################################

%%%%%%%%%%%%%%%%%%%%%%%%%%%%%%%%%%%%%%%%%%
\subsubsection{Landau-Stark states}
\label{secD2a}

\begin{figure}[t]
\includegraphics[width=8.5cm]{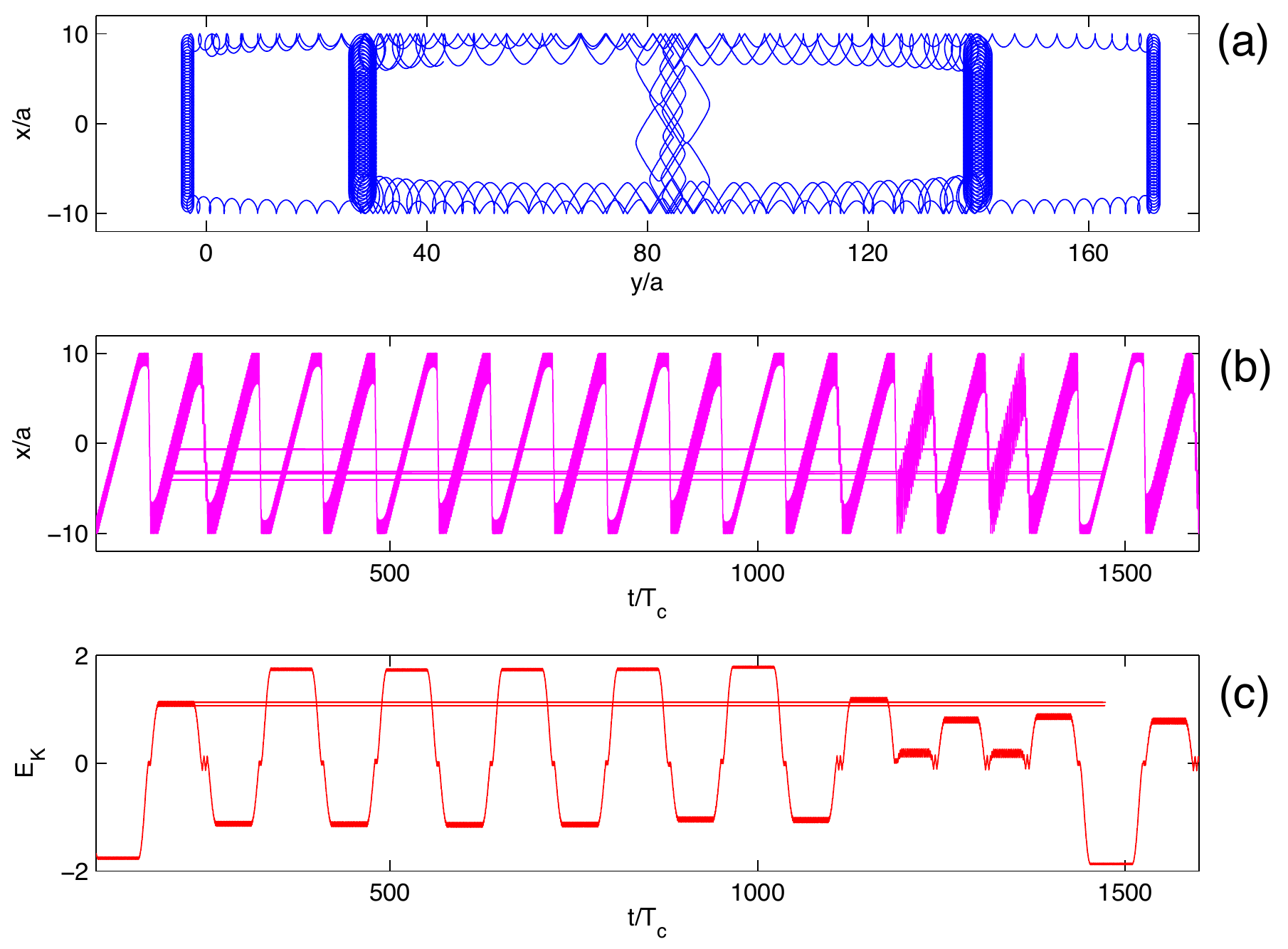}
\caption{A classical trajectory in the $(x,y)$ plane (a); coordinate $x$ (b) and kinetic energy $E_K$ (c) as functions of time. The Peierls phase is $\alpha=1/10$, the electric field magnitude is $F=0.02$, and the initial kinetic energy is $E_K=-2J+\omega_c/2$. Time is measured in units of the cyclotron period $T_c=2\pi/\omega_c$. In the upper panel the trajectory is shown only for the time interval $[0,400T_c]$.}
\label{figE2}
\end{figure}

We are now prepared to discuss LS-states for finite lattices.  As before, one gets a useful insight into the problem by the classical dynamics of the system. Let us assume for the moment that electric field is aligned with the $y$ direction, {\em i.e.}, $F_x=0$. Then, the classical Hamiltonian  reads
%*******************************************
\begin{equation}
\label{p3}
H_{cl}=-J_x\cos(p_x)-J_y\cos(p_y+2\pi\alpha x)+V(x) +Fy
\end{equation}
where $V(x)$ is a box potential confining the system to size $L_x$. A typical trajectory of the system (\ref{p3}) is shown in the upper panel of Fig.~\ref{figE2}. For $F=0$ the low-energy dynamics of the system consists of cyclotron oscillations, where the particle moves along circular orbit with the cyclotron frequency. If $F\ne 0$ the center of the orbits moves in the $x$ direction at the drift velocity (\ref{k1}), until the particle hits the right wall of the potential box. From this moment onward, it moves along the wall, where it is accelerated by the electric field. After approximately one half of the Bloch period, the kinetic energy takes the value $E_K\approx0$ and the particle is scattered to the opposite (left) wall, where it is decelerated by the electric field to lower energies. A second dynamical possibility is that the kinetic energy continues to grow, that for $E_K>0$ is equivalent to deceleration  of a particle with negative mass. As a result of this deceleration, the trajectory eventually leaves the left wall and the process is repeated. In summary, we meet here a kind of BO where the particle is accelerated only at the edges. It is worth remarking that the discussed classical Bloch oscillations are actually chaotic, so that a small change in the initial condition is amplificated in the course of time, and yields a significantly different trajectory. However, the dynamics remains globally the same, being made of time spans $T_v\approx L_x/v^*$, in which the particle moves across the sample, alternated by time intervals in which the particle is accelerated (or decelerated) along the edges, as in Fig.~\ref{figE2}(b,c).
%#############################################

Quantum states of the system inherit the global features of classical trajectories, see Fig.~\ref{figE3}(a,b).  To analyze them, it suffices to find $L_x$ LS-states in the fundamental energy interval $|E| \le F/2$. The other LS-states can be obtained by translating these states in the $y$ direction and imprinting a phase according to Eq.~(\ref{e2}). Thus every LS-state can be labeled by the ladder index $n$, $-\infty<n<\infty$, and by the transverse index $\nu$, $1\le \nu \le L_x$.

As mentioned above, the classical system (\ref{p3}) belongs to the class of chaotic systems. On the quantum level this is manifested in a high sensitivity of eigenvalues and eigenstates to variations of the system parameters: in particular, to the electric field magnitude $F$. If we plot the energies of LS-states versus $F$, we observe the typical `level spaghetti' of a quantum chaotic system. Moreover, the distribution function for the spacings between nearest-neighbor levels, that is the simplest and earlier test of quantum non-integrability, was shown in \cite{99} to coincide with the Wigner-Dyson distribution for random matrices.

Although fine features of individual LS-states are sensitive to variation of the system parameters, their global structure is stable. Out of many possible global characteristics one can consider the spatial density
%*******************************************
\begin{equation}
\label{p4}
\rho_{l,m}^{(n)}=\frac{1}{L_x}\sum_{\nu=1}^{L_x} |\Psi_{l,m}^{(\nu,n)}|^2 \;.
\end{equation}
The density (\ref{p4}) is shown in Fig.~\ref{figE3}(c). Remarkably, this figure reproduces the magnetic bands structure of Fig.~\ref{figE1}(a).

%%%%%%%%%%%%%%%%%%%%%%%%%%%%%%%%%%%%%%%%%%%
\subsubsection{Enhanced Landau-Zener tunneling}
\label{secD2b}

\begin{figure}[t]
\includegraphics[width=8.5cm]{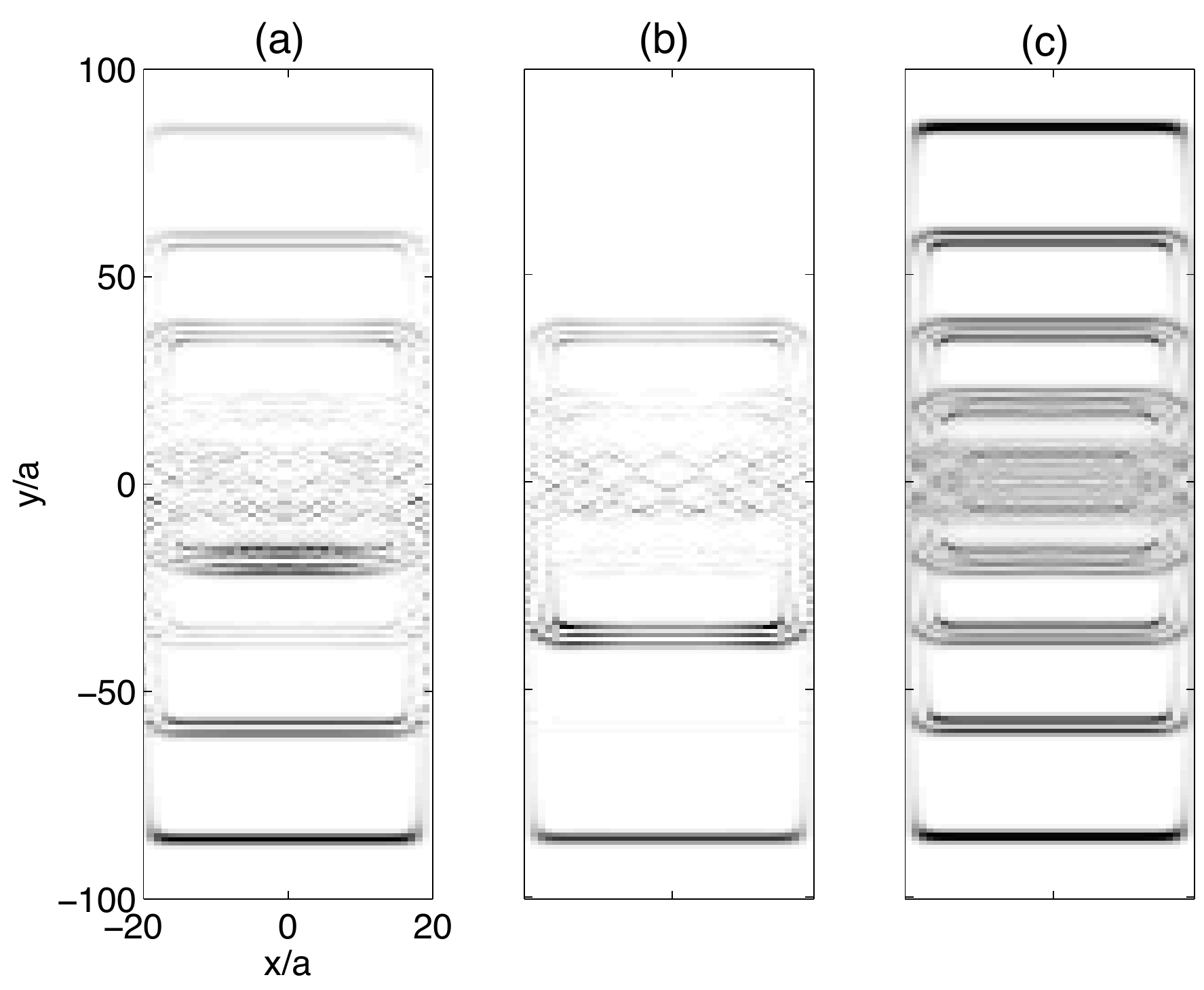}
\caption{Examples of  LS-states (a-b) and spatial density (\ref{p4}) (c) for $F=0.02$, $\alpha=1/10$, $J_x=J_y=1$, and $L_x=40$.}
\label{figE3}
\end{figure}

The  analysis just developed shows that a finite electric field mixes the bulk states of the systems through the edge states. This leads to the appearance of localized LS-states, with the characteristic structure depicted in Fig.~\ref{figE3}. This phenomenon can be also  explained using a different set of arguments. In fact, assume that the initial state of the system belongs to the ground magnetic band, and is characterized by the quasimomentum $\kappa_y$. The effect of the electric field on this state is twofold: firstly, it changes the quasimomentum $\kappa_y$ linearly in time (Bloch acceleration theorem); secondly, it induces interband transitions to higher magnetic bands (LZ-tunneling).  In the absence of edge states -- that is, in the case of periodic BC -- the rate of LZ-tunneling decreases exponentially when $F$ decreases. To the contrary, edge states connect magnetic bands directly, and this path fundamentally modifies the Landau-Zener result \cite{99}: depletion of the ground band is now linear in time, with the rate
%*******************************************
\begin{equation}
\label{p5}
\tau^{-1}=v^*/L_x \sim F \;.
\end{equation}
%
%Eq.~(\ref{p5}) can be proved as follows. Let us consider non-interacting fermions with the Fermi just above the ground magnetic band. Classically this quantum states corresponds to ensemble of particles with the energy $E=-2J+\omega_c/2$ uniformly distributed over the sample. When electric field is switched on all particles start to move to the right edge of the sample with the drift velocity, where they get accelerated and, hence, gain the energy.  As soon as the last particle reaches the right edge, the ground magnetic band becomes completely depleted.
The enhanced rate (\ref{p5}) of interband transitions ensures that all magnetic bands are coupled, as soon as $F$ deviates from zero.

Finally, let us comment on the effect of the orientation of the electric field ${\bf F}$, when it is different from the case considered so far, $\beta=F_x/F_y=0$. Recall that in the case of infinite lattices, it was shown in Sec.~\ref{secA4} that the parameter $\beta$ determines whether LS-states are localized with discrete spectrum (irrational $\beta$) or extended states with continuous spectrum (rational $\beta$). For finite lattices, however, the spectrum is always discrete and the previous distinction is less relevant. Moreover, in the limit of small $F$, we can always satisfy the condition that the localization length  be larger than the system size $L_x$. Then, the main effect of non-zero $\beta$, rational or irrational, is to change the geometry of LS-states from a parallelepiped-like structure to a rhomb-like structure.
%#############################################

%%%%%%%%%%%%%%%%%%%%%%%%%%%%%%%%%%%%%%%%%%%
\section{Beyond the single-band approximation}
\label{secE}
%%%%%%%%%%%%%%%%%%%%%%%%%%%%%%%%%%%%%%%%%%%

The analysis of LS-states and cyclotron-Bloch oscillations would be incomplete without a discussion of the validity of the single-band approximation and of the physical effects neglected in doing this approximation.  As it is well known, for zero magnetic field, the main effect beyond the single-band analysis is LZ-tunneling across the energy gap $\Delta$, which separates the ground Bloch band from the rest of the spectrum. This tunneling causes the decay of any state  supported by ground Bloch or Wannier functions, including WS-states (\ref{c3}). Because of this, these states must be rigorously described as quantum resonances, with complex energies ${\cal E}_{n,k}=E_{n,k}-i\Gamma/2$ \cite{51}. Naturally, one expects that a similar effect holds also when $B\ne0$. We discuss this problem in the next two subsections, which analyze LZ-tunneling across the energy gap $\Delta$ in the presence of a magnetic field.

%%%%%%%%%%%%%%%%%%%%%%%%%%%%%%%%%%%%%%%%%%%
\subsection{Numerical approach}
\label{secE1}

There is a straightforward numerical approach to check the validity of single-band and tight-binding approximations.  One must solve the time-dependent Schr\"odinger equation with the Hamiltonian (\ref{b1}) and project its solution $\Psi(x,y,t)$ on the basis of the Wannier states:
%********************************************************
\begin{equation}
\label{q1}
\tilde{\psi}_{l,m}(t)=\int  w_{l,m}(x,y) \Psi(x,y,t) \; {\rm d}x{\rm d}y \;.
\end{equation}
Then, the amplitudes $\tilde{\psi}_{l,m}(t)$ must be compared with the amplitudes $\psi_{l,m}(t)$ obtained by solving the Schr\"odinger equation with the tight-binding Hamiltonian (\ref{b3}). To complete the analogy, the hopping matrix elements $J_x$ and $J_y$ in the last Hamiltonian obviously depend on the explicit form of the periodic potential $V({\bf r})$ and can be extracted from the Bloch band spectrum of the continuous Hamiltonian (\ref{b1}) for $B,F=0$. In this subsection we present results obtained with the separable periodic potential
%********************************************************
\begin{equation}
   \label{q2}
V(x,y)=V_x\cos (2\pi x/a)+V_y\cos(2\pi y/a)  \;.
\end{equation}
The parameters $V_x$ and $V_y$ are conventionally measured in units of the recoil energy $E_R=h^2/Ma^2$, and we have set $V_x=V_y=0.5 \; E_R$. In this case the ground Bloch band is well approximated by the cosine dispersion relation (\ref{b5}), with $J_x=J_y=0.043 \; E_R$. It is separated from the rest of the spectrum by the gap $\Delta=0.5 \; E_R$. As initial condition we choose $\psi(x,y,t=0)=w_{0,0}(x,y)$ which means $\tilde{\psi}_{l,m}(t=0)=\psi_{l,m}(t=0)=\delta_{l,0}\delta_{m,0}$.

For vanishing electric and magnetic fields the dynamics of the system is ballistic spreading and we observe full agreement between the amplitudes (\ref{q1}) and the tight-binding amplitudes $\psi_{l,m}(t)$. A finite magnetic field spoils this perfect correspondence. We now observe depopulation of the ground Bloch band, that is reflected in deviation of the normalization
%****************************************************
\begin{equation}
\label{q3}
N(t)=\sum_{l,m} |\tilde{\psi}_{l,m}(t)|^2
\end{equation}
from unity. In our simulations the amplitude $B$ of the magnetic field corresponds to the Peierls phase $\alpha=1/8$. At this value of the magnetic flux the normalization (\ref{q3}) rapidly drops to $N\approx 0.85$ and then oscillates around this value. This oscillatory behavior indicates that one can again obtain a nice correspondence between continuous and discrete systems by taking into account a finite number of Bloch bands, {\em i.e.}, by extending the single-band tight-binding model into two- or three-band tight-binding models \footnote{See \cite{Parr13}, for example. A second option is to consider smaller values of $\alpha$, where one can tolerate the deviation of the normalization from unity.}. The case changes dramatically when the electric field is non-null: here, we observe a monotonic decrease of $N(t)$ according to the exponential law,
%****************************************************
\begin{equation}
\label{q4}
N(t)=\exp(-\Gamma t) \;,
\end{equation}
for the whole span of numerical simulations. This means that, similar to the Wannier-Stark system ($B=0$), there is a flow of probability to high-energy states.
%#############################################################
\begin{figure}[t]
\includegraphics[width=7.5cm]{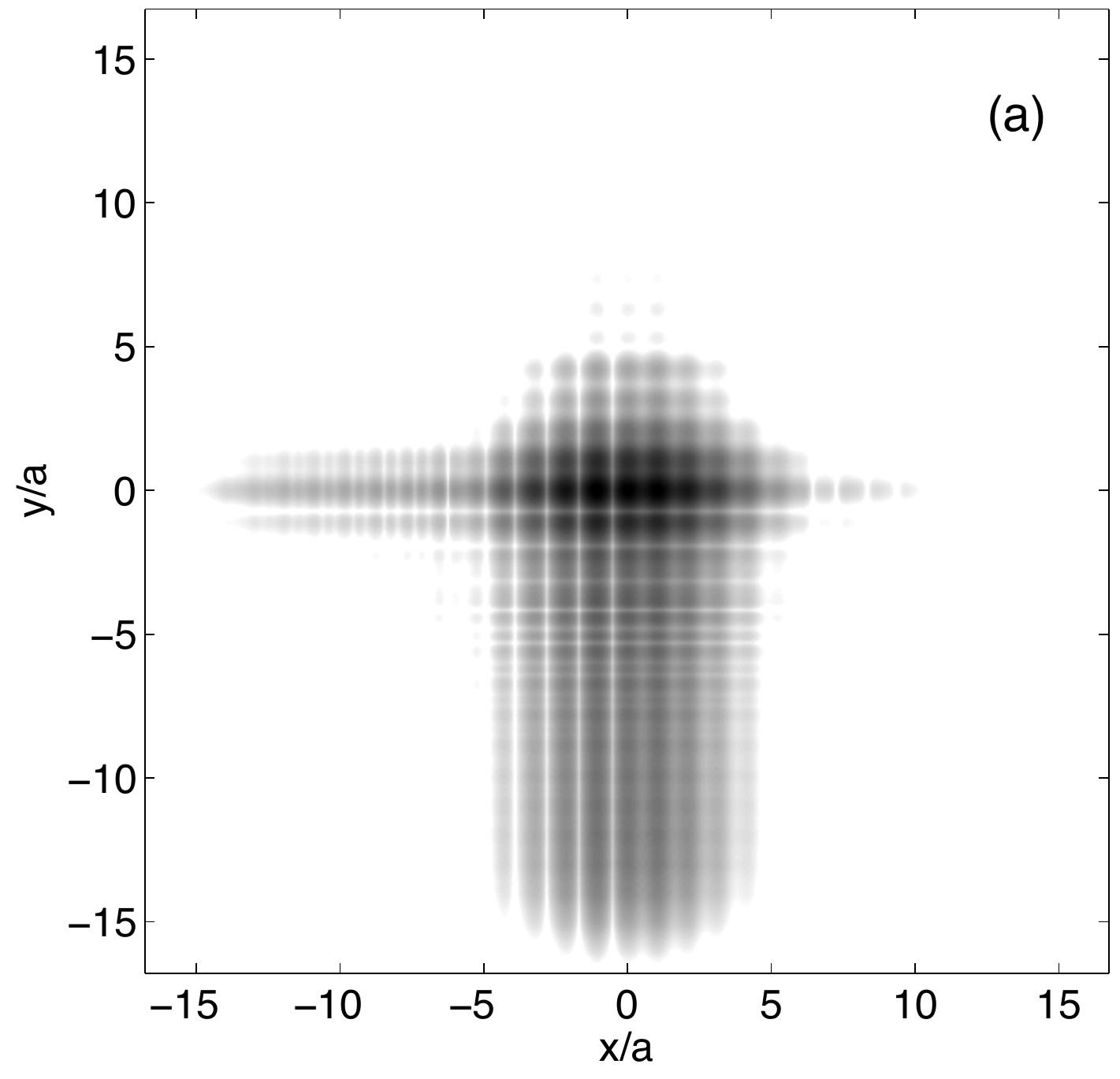}
\includegraphics[width=7.5cm]{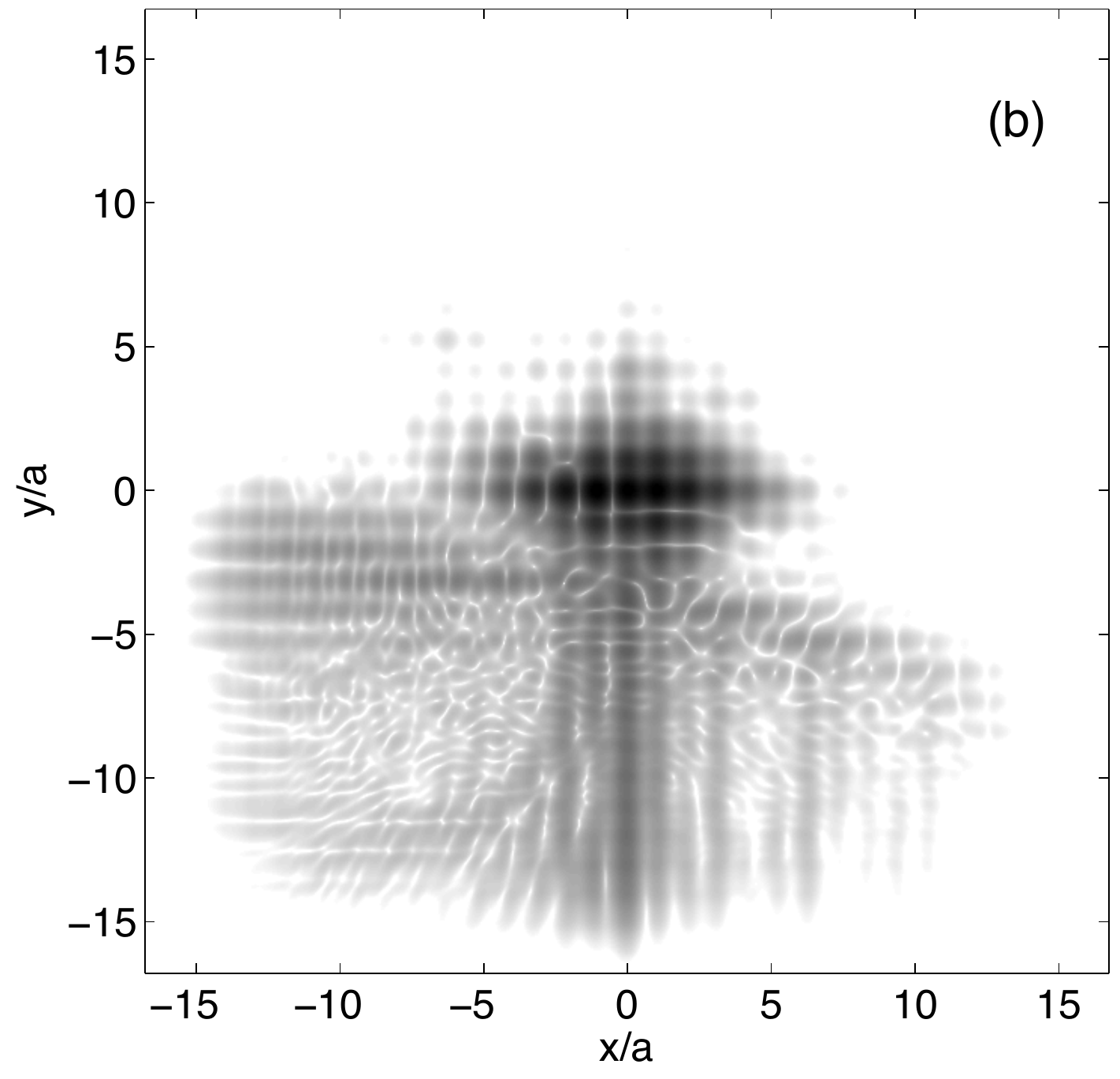}
\caption{Logarithm of $|\psi(x,y,t)|^2$ at the final time of numerical simulations for $\alpha=0$ (a) and $\alpha=1/8$ (b). The other parameters are $V_x=V_y=0.5 \; E_R$,  $eaF=0.015 \; E_R$, and $F_x/F_y=(\sqrt{5}-1)/4$.}
\label{figG1}
\end{figure}

One gets additional insight into this probability leakage by analyzing the wave-packet dynamics. Figure \ref{figG1}(a) shows the logarithm of the squared wave function at the final time  of numerical simulations, for $eaF=0.015 \; E_R$, $\beta=F_x/F_y=(\sqrt{5}-1)/4\approx1/3$ and zero magnetic field. Notice that in the numerical simulations we use absorbing boundary conditions \cite{Osko08} at $x/a=\pm 16$ and $y/a=\pm16$, to prevent wave packet reflection. This ensures rapid relaxation to a quasi-stationary regime with the characteristic pattern seen in Fig.~\ref{figG1}(a). Analyzing this pattern we conclude that  the electric field accelerates the particle only along the crystallographic axes of the lattice. Moreover, when the particle is captured in the minima of the 1D potential $V_x\cos (2\pi x/a)$, it is accelerated in the $y$ direction; when capture happens in the minima of the 1D potential $V_y\cos (2\pi y/a)$, it is accelerated in the $x$ direction \footnote{This effect has a simple classical interpretation, discussed in detail in \cite{55}, for both separable and not separable 2D periodic potentials.}. This acceleration is constant: the particle's kinetic energy grows quadratically in time and tends to infinity in the infinite time limit.
% acquires an infinite amount of kinetic energy in an infinitely long time.

The case of non-zero magnetic field, shown in Fig.~\ref{figG1}(b), is more involved.  The particle is only temporally captured in the minima of the 1D periodic potentials, where it is accelerated by the electric field. This results in an upper boundary for the kinetic energy that can be gained during capture \cite{94},
%****************************************************
\begin{equation}
   \label{q5}
E_{max}\sim E_R\left(\frac{V_0}{\alpha}\right)^2 \;,
\end{equation}
where $V_0\sim V_x/E_R\sim V_y/E_R$ is the depth of the 2D periodic potential, in units of the recoil energy. In the case of small $\alpha$ this upper boundary is a huge, yet finite energy. It is an open problem whether finiteness of $E_{max}$ may affect the decay law (\ref{q4}) for large times.

%%%%%%%%%%%%%%%%%%%%%%%%%%%%%%%%%%%%%%%%%%%
\subsection{Analytical approach to decay rates}

In this subsection we describe an analytical approach to compute the decay rate $\Gamma$  in Eq.~(\ref{q4}). At difference with the previous subsection, we shall consider a new initial state of the system: this is now assumed to belong to a subspace of the Hilbert space associated with a given magnetic band. Thus, for  rational $\alpha=r/q$, we find $q$ decay rates $\Gamma_j$ which, of course, are functions of the electric field magnitude.

\subsubsection{Magnetic bands}

To study LZ-tunneling, we need to compute $q$ ground magnetic bands, and also the spectrum above the energy gap $\Delta$. We now describe a method % for calculating magnetic bands which
that accomplishes this task. To simplify equations we use from now on dimensionless variables, where length is measured in units of $a/2\pi$, energy in units of $E_R=h^2/Ma^2$ and time in units of $h/E_R$. In these units the Hamiltonian (\ref{b1}) takes the form
%****************************************************
\begin{equation}
\label{a1}
\widehat{H}=\frac{1}{2}\left[\hat{p}_x^2 + (\hat{p}_y - B x)^2\right] +
V_x\cos x + V_y\cos y+Fy \;, \quad B=\frac{\alpha}{2\pi}  \;,
\end{equation}
where the dimensionless electric field is given by the Stark energy for one lattice period divided by the recoil energy. We also set $F_x=0$ and, hence, $F_y\equiv F$.

\begin{figure}
\includegraphics[width=8.5cm]{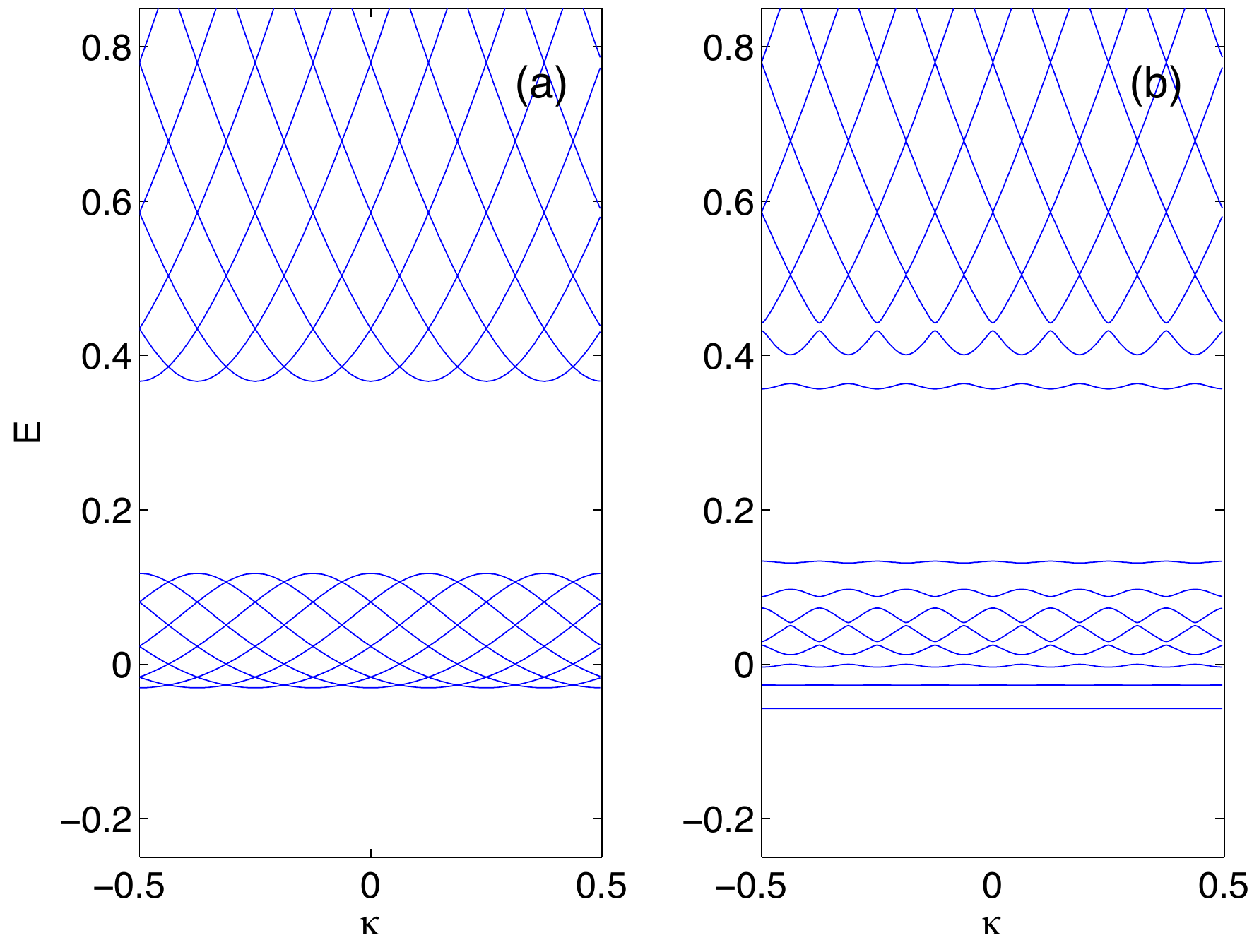}
\caption{Magnetic bands for $\alpha=1/8$ and $(J_x,V_y)=(0.0,0.25)$, left panel,  and  $(J_x,V_y)=(0.0431,0.25)$, right panel.}
\label{figG2}
\end{figure}

Let us assume that $V_x$ is large enough to justify the tight-binding approximation in the $x$ direction. Then, we can use the ansatz
%*************************************************
\begin{equation}
\label{a3}
\Psi(x,y)=\sum_{l=-\infty}^\infty \psi^{(l)}(y) w_l(x) \;,
\end{equation}
where $w_l(x)$ are the Wannier functions associated with the ground Bloch band of the 1D Hamiltonian $\widehat{H}_x=\hat{p}_x^2/2 +
V_x\cos x $. Substituting (\ref{a3}) into the stationary Schr\"odinger equation with the Hamiltonian  (\ref{a1}), in which we have temporarily set $F=0$, we obtain
%**********************************************
\begin{equation}
\label{a4}
E_0\psi^{(l)}(y) -\frac{J_x}{2}\left[\psi^{(l+1)}(y) + \psi^{(l-1)}(y)\right]
+\widehat{H}^{(l)}_y \psi^{(l)}(y)=E\psi^{(l)}(y) \;,
\end{equation}
where
%********************************************
\begin{equation}
\label{a5}
\widehat{H}^{(l)}_y =\frac{1}{2}\left(\hat{p}_y - \alpha l\right)^2 + V_y\cos y \;.
\end{equation}
The eigenfunctions of the Hamiltonians (\ref{a5}) are Bloch waves with a shifted dispersion relation. Namely, if $E(\kappa_y)$ is the Bloch spectrum of the Hamiltonian $\widehat{H}^{(0)}_y$, then for $l \ne 0$ we have
%*******************************************
\begin{equation}
   \label{a6}
E^{(l)}(\kappa_y)=E^{(0)}(\kappa_y+\alpha l)  \;,
\end{equation}
see Fig.~\ref{figG2}(a). Next, using the Fourier expansion of Bloch waves,
%********************************************
\begin{equation}
   \label{a7}
\psi^{(l)}(y)=\exp(i\kappa_y y)  \sum_{n=-\infty}^\infty  c^{(l)}_n(\kappa_y) \exp(in y)  \;,
\end{equation}
the system of partial differential equations (\ref{a4}) can be reduced to a system of algebraic equations for the coefficients $c^{(l)}_n$:
%**********************************************
\begin{equation}
\label{a8}
-\frac{J_x}{2}\left[c^{(l+1)}_n + c^{(l-1)}_n\right] + \frac{1}{2}(n+\kappa_y+\alpha l)^2 c^{(l)}_n
+\frac{V_y}{2}\left[c^{(l)}_{n+1} + c^{(l)}_{n-1}\right]=E c^{(l)}_n \;.
\end{equation}
In the general case of arbitrary $\alpha$, the index $l$ in (\ref{a8}) ranges from minus to plus infinity. However, when $\alpha=r/q$ is a rational number, we can restrict $l$ to a magnetic period, $1\le l \le q$. In this case Eq.~(\ref{a8}) must be supplemented by the periodic boundary condition
%****************************************************
\begin{equation}
\label{a9}
c^{(q+1)}_n=\exp(i2\pi\kappa_x) c^{(1)}_{n-r}  \;,
\end{equation}
where the dimensionless quasimomentum $\kappa_x$ in the Bloch phase factor belongs to the reduced Brillouin zone $|\kappa_x|\le1/2q$. The system of algebraic equations (\ref{a8}), together with the boundary condition (\ref{a9}), provide an alternative method for calculating the ground magnetic bands.
%#############################################################

As an example of this calculation, the right panel in Fig.~\ref{figG2} shows the solution of Eqs.~(\ref{a8},\ref{a9}) for $\alpha=1/8$, $V_y=0.25$, and $J_x=0.0431$, which corresponds to $V_x=0.5$. Magnetic bands are plotted as functions of the quasimomentum $\kappa_y$, for a single value of the quasimomentum $\kappa_x=0$. This figure must be compared with Fig.~\ref{figA1}(a) that shows the magnetic bands in the tight-binding approximation. Two differences are to be noted. The first is that the presence of high-energy Bloch bands breaks the symmetry between magnetic bands associated with the bottom and the top of the ground Bloch band. This is, however, a minor correction. More importantly, in the complete theory, a segment of the energy spectrum appears above the energy gap. This allows us to find the decay rates $\Gamma_j$ of the ground magnetic bands when $F\ne0$.

%%%%%%%%%%%%%%%%%%%%%%%%%%%%%%%%%%%%%%%%%%%%
\subsubsection{Decay rates}

To find the decay rates  $\Gamma=\Gamma_j(F)$ of the individual magnetic bands we employ the truncated Floquet  matrix method of \cite{51}, adapted to the present problem. The main steps of this method are as follows. First, we consider the time-dependent counterpart of the stationary Schr\"odinger equation (\ref{a8}), in which the quasimomentum $\kappa_y$ changes linearly in time:  $\kappa_y\rightarrow \kappa_y-Ft$. Next, we compute the (formally infinite) matrix of the evolution operator over one Bloch period, $T_B=1/F$, and truncate it to a finite size. Notice that when calculating the Floquet matrix we explicitly use the periodic boundary condition (\ref{a9}). In so doing, the matrix is truncated only with respect to the index $n$, $|n|\le n_{max}$. The method rapidly converges when $n_{max}$ is increased: in our calculations we use $n_{max}=7$. The eigenvalues $\lambda_j$ of the truncated Floquet matrix are known to be the complex poles of the scattering matrix. We can therefore conclude: the individual decay rates $\Gamma_j$ of ground magnetic bands are found from the equation $|\lambda_j|^2=\exp(-\Gamma_j T_B)$, where $\lambda_j$ are the first $q$ eigenvalues which are closest to the unit circle.

\begin{figure}
\includegraphics[width=8.5cm, clip]{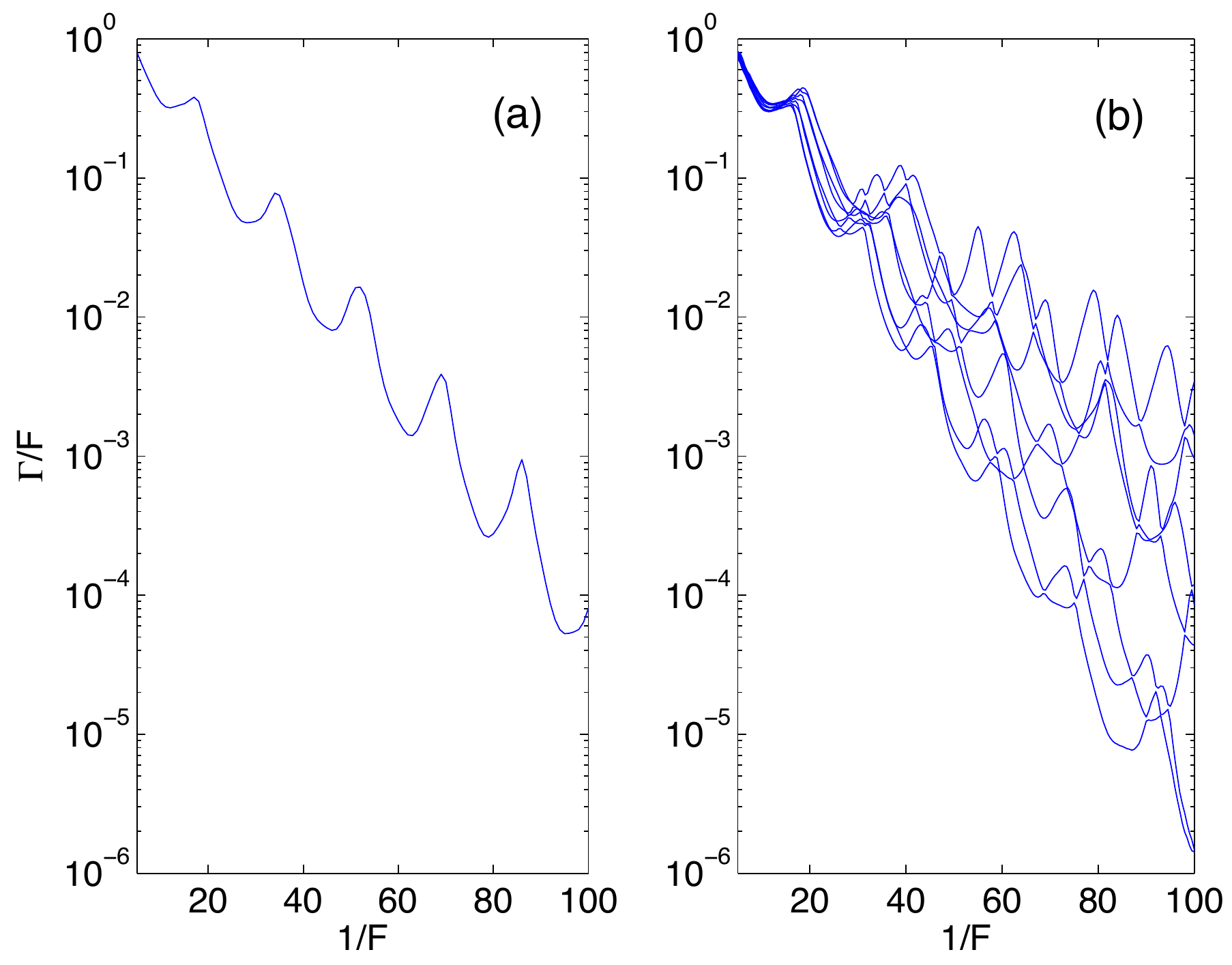}
\caption{Decay rate $\Gamma$ of the ground Bloch band ($\alpha=0$), left panel, and decay rates $\Gamma_j$ of the ground magnetic bands for $\alpha=1/8$, right panel, as functions of the inverse (scaled) electric field magnitude. The other parameters are the same as in Fig.~\ref{figG2}.}
\label{figG3}
\end{figure}

The decay rates $\Gamma_j$ for $\alpha=1/8$ are shown in Fig.~\ref{figG3}(b) versus the inverse (scaled) electric field magnitude.  Compare the right panel of this figure with the left panel, which shows the decay rate of the ground Bloch band for $B=0$. In first approximation the functional dependence of the decay rate is given by the Landau-Zener formula
%************************************************
\begin{equation}
\label{a10}
\bar{\Gamma}(F) \sim F\exp\left(-\frac{b}{F}\right) \;,
\end{equation}
where the coefficient $b$ is proportional to the square of the energy gap $\Delta$. Deviations from (\ref{a10}) are due to the phenomenon of resonant tunneling \cite{43}. It is seen in Fig.~\ref{figG3}(b) that this process also takes place when $B\ne0$. We can thus decompose $\Gamma_j$ into two terms,
\begin{equation}
   \label{a13}
\Gamma_j(F)=\bar{\Gamma}_j(F)+\Gamma^R_j(F)
\end{equation}
where $\bar{\Gamma}_j(F)\sim F\exp(-b_j/F)$ and $\Gamma^R_j(F)$ is an oscillating part. By a careful analysis of the coefficients $b_j$ we can conclude that LZ-tunneling is suppressed for lower magnetic bands, $j\ll q$, but it is enhanced for higher bands, $j\approx q$. Notice that this effect is well pronounced only for weak electric fields, while in the strong field regime the decay rates $\Gamma_j$ approximately coincide with that for $B=0$.
%#############################################################

In conclusion, the main result of this analysis is the existence of an additional critical magnitude, $F_{LZ}$, of the electric field, besides $F_{cr}$ given in Eq.~(\ref{j9}). The resulting physical picture of LZ-tunneling is the following:  there exist three different regimes of inter-band transitions. Suppose that only the lowest magnetic band is initially populated. When $F<F_{cr}$, the probability to find the system in the lowest band remains close to unity for an exponentially long time. As soon as $F$ exceeds $F_{cr}$, all $q$ ground magnetic bands become involved in the dynamics while their populations add up to unity to high accuracy.  Finally, when the second threshold is surpassed, $F>F_{LZ}$, a rapid decrease in the total population of ground magnetic bands takes place.

%%%%%%%%%%%%%%%%%%%%%%%%%%%%%%%%%%%%%%%%%%%
\section{Open problems}
\label{secEE}

In this review we have addressed the spectral and dynamical properties of a quantum particle in a 2D lattice, subject to potential and gauge fields---for example, electric and magnetic fields acting on a charged particle. When the gauge field is not present there exist two complementary approaches to the problem. The first, dynamical approach, is based on the notion of Bloch band spectrum and Bloch waves associated with this spectrum. The potential field is known to change the quasimomentum of Bloch waves linearly in time, leading  to Bloch oscillations of the quantum particle. Simultaneously, it induces Landau-Zener transitions between Bloch bands. Thus the main ingredient of this theory is a thorough analysis of the interband LZ-tunneling.

The second, spectral approach, does not need knowledge of the Bloch spectrum. It requires instead to find the eigenstates of the full Hamiltonian, known nowadays as Wannier-Stark states. These states were first introduced by Wannier \cite{Wann60} by using the single band approximation, where they appear as stationary states of the system. This result, however, conflicts with the finite rates of LZ-tunneling across the energy gaps. A natural way to resolve this contradiction is to assume that WS-states are metastable Gamov states. However, it took a decade to prove that WS-states are indeed quantum resonances and several decades to find them explicitly \cite{53}.

In the present work we generalize the theory of WS-states to the case of non-zero gauge field.  As in the former problem, we use in parallel dynamical and spectral techniques. The dynamical approach is now based on the notion of magnetic bands, which emerge from Bloch bands.
Unfortunately, magnetic bands are well defined only for rational values of the magnetic flux through the elementary cell (rational Peierls's phase) and this somehow restricts the applicability of this technique.

The spectral approach does not suffer from this drawback: since it does not rely on magnetic bands, it is equally suited for rational and irrational values of the magnetic flux. This approach requires the computation of the eigenstates of the full Hamiltonian, which we termed in this review Landau-Stark states.  Essentially, we have accomplished a first step in this approach: we have found LS-states in the single band (more exactly, the tight-binding) approximation. Within this approximation, LS-states are stationary states of the system, with discrete or continuous spectrum, depending on the orientation of the potential field relative to the primary axes of the lattice. The results of Sec.~\ref{secE}, however, indicate that these states are subject to decay, at least, for a finite time. It is an open problem whether the true ({\em i.e.}, beyond the tight-binding approximation) LS-states are stationary states,  metastable state similar to WS-states, or a new kind of states `in between'.

Besides this fundamental question, a number of other points, perhaps less fundamental yet not less important for applications require further investigation. In particular, in this review we restricted ourselves to the case of a simple square lattice and only briefly we have touched upon the case of staggered magnetic fields. As it is well known, combination of nontrivial lattice geometry with a staggered field may result in interesting topological systems like the Haldane lattice \cite{Hald91}. It would be interesting to study LS-states for such lattices and to clarify the question how the transition between topologically trivial and non-trivial lattices \footnote{In present day experiments with cold atoms this transition is tuned, for example, by changing the amplitudes of the laser beams creating the optical lattice \cite{Tarr12}.} is reflected in the properties of LS-states.

\begin{acknowledgements}
The authors would like to express their gratitude to H.~J.~Korsch, I.~Guarneri and J. Bellissard for useful discussions and acknowledge contribution of our coauthors in the original papers \cite{94,96,98}, the results of which were partially presented in this review.
\end{acknowledgements}
%%%%%%%%%%%%%%%%%%%%%%%%%%%%%%%%%%%%%%%%%%%%%
%\include{references}

%%%%%%%%%%%%%%%%%%%%%%%%%%%%%%%%%%%%%%%%%%%%%%

\end{document}